\documentclass[11pt]{article}
\usepackage[top=.75in, bottom=.75in]{geometry}

\usepackage{multirow} 
\usepackage{booktabs} 
\usepackage{tabularx} 
\usepackage{xfrac} 
\usepackage{subcaption} 
\usepackage{color,soul}
\usepackage[flushleft]{threeparttable}
\usepackage{setspace} 
\usepackage{lineno} 
\usepackage[T1]{fontenc} 
\usepackage[bitstream-charter]{mathdesign}
\title{Discovering and interpreting transcriptomic drivers of imaging traits using neural networks}
\author{Nova F. Smedley$^{1,2,3}$, Suzie El-Saden$^{1,2}$, and William Hsu$^{1,2,3,4}$}
\date{}

\begin{document}

\maketitle
\begin{enumerate}\setlength{\itemsep}{0pt}
\item Medical \& Imaging Informatics, University of California Los Angeles, USA
\item Department of Radiological Sciences, University of California Los Angeles, USA
\item Department of Bioengineering, University of California Los Angeles, USA
\item Bioinformatics IDP, University of California Los Angeles, USA
\end{enumerate}

\begin{abstract}

\textbf{Motivation:} Cancer heterogeneity is observed at multiple biological levels. To improve our understanding of these differences and their relevance in medicine, approaches to link organ- and tissue-level information from diagnostic images and cellular-level information from genomics are needed. However, these ``radiogenomic'' studies often use linear, shallow models, depend on feature selection, or consider one gene at a time to map images to genes. Moreover, no study has systematically attempted to understand the molecular basis of imaging traits based on the interpretation of what the neural network has learned. These current studies are thus limited in their ability to understand the transcriptomic drivers of imaging traits, which could provide additional context for determining clinical traits, such as prognosis. 

\textbf{Results:} We present an approach based on neural networks that takes high-dimensional gene expressions as input and performs nonlinear mapping to an imaging trait. To interpret the models, we propose gene masking and gene saliency to extract learned relationships from radiogenomic neural networks. In glioblastoma patients, our models outperform comparable classifiers (>0.10 AUC) and our interpretation methods were validated using a similar model to identify known relationships between genes and molecular subtypes. We found that imaging traits had specific transcription patterns, e.g., edema and genes related to cellular invasion, and 15 radiogenomic associations were predictive of survival. We demonstrate that neural networks can model transcriptomic heterogeneity to reflect differences in imaging and can be used to derive radiogenomic associations with clinical value.

\textbf{Availability and implementation:} {\small\texttt{https://github.com/novasmedley/deepRa}} {\small\texttt{diogenomics}}.

\textbf{Contact:} whsu@mednet.ucla.edu


%

\end{abstract}
%
%
\doublespacing
%
\section{Introduction}

Radiogenomic mapping is the integration of traits observed on medical images and traits found at the molecular level, such as gene expression profiling \cite{Segal2007, Diehn2008}. As such, the study of radiogenomics plays a role in precision medicine, where associations can describe prognosis or therapy response \cite{Pope2008, Naeini2013, Chang2018}. A common approach to radiogenomic mapping involves dimensionality reduction and pairwise comparisons \cite{Diehn2008, Zinn2011, Gutman2013, Aerts2014, Gevaert2014, Jamshidi2014, Grossmann2017} or predictive models, such as decision trees \cite{Segal2007, Hu2016, Kickingereder2016, Zhang2017, Gevaert2017} or linear regression \cite{Gevaert2012, Zhu2015, Guo2015, Yamashita2016, Grossmann2017}. Markedly, these approaches often require feature selection; assume linearity; and/or depend on pairwise associations, limiting their capacity to represent complex biological relationships. 

Neural networks, with their ability to automatically learn nonlinear, hierarchical representations of large input spaces \cite{LeCun2015, Goodfellow2016book}, are alternate approaches for radiogenomics \cite{Chang2018, Korfiatis2017, Li2018, Ha2019, Chen2016}. However, current applications of neural networks have focused on diagnostic images as inputs to predict a single gene status and have excluded gene interactions \cite{Chang2018, Korfiatis2017, Ha2019}. To the best of our knowledge, no prior studies have interpreted the neural networks to ascertain what radiogenomic associations are learned. Towards understanding the biological basis of imaging traits, we thus present an approach using the representational power of neural networks to model tumor transcriptomes and nonlinearly map genes to tumor imaging traits. No \textit{a priori} selection is used on the transcriptome. 

Importantly, we provide approaches for understanding radiogenomic neural networks based on visualization techniques, such as input masking \cite{Zeiler2014} and saliency maps \cite{Simonyan2014}. Although neural networks are considered ``black boxes'', these methods probe the trained neural networks to understand relationships between gene expressions and imaging traits. The methods can identify cohort-level imaging-transcriptomic associations, which we refer to as \textit{radiogenomic associations}, and patient-level radiogenomic associations, which we refer to as \textit{radiogenomic traits}. We validated the associations and traits generated by a network in a classification task with known relationships, i.e., gene expressions (model input) and molecular subtypes (model output) \cite{Verhaak2010}.

\begin{figure}[h!]
	\caption{Examples of phenotypic differences observed in glioblastoma patients. Shown are single, axial images of pre-operative MRI scans from the TCGA-GBM cohort. Four MRI sequences were used to annotate imaging traits: T1-weighted (T1W), T1W with gadolinium contrast (T1W+Gd), and T2-weighted (T2W), and fluid-attenuated inversion recovery (FLAIR) images.  MRI traits included enhancing (enhan.), non-contrast enhancing tumor (nCET), necrosis (necro.), edema, infiltrative (infil.), and focal, where class labels were indicated by black (proportions $< \sfrac{1}{3}$, expansive, focal) or gray (proportions $\geq \sfrac{1}{3}$, infiltrative, non-focal) blocks.}
	\label{fig_mri_traits}
	\centering
	\includegraphics[width=\textwidth]{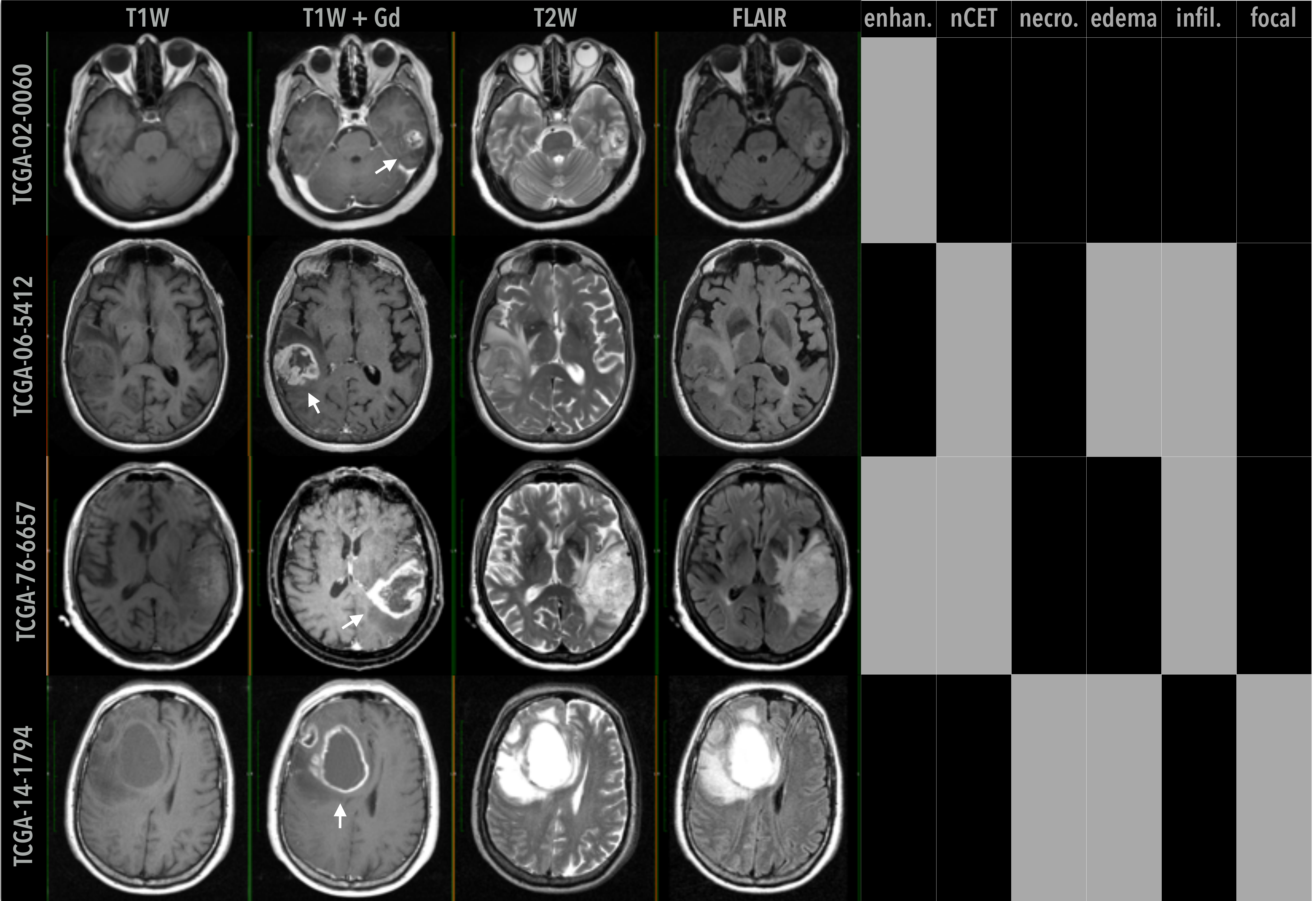}
\end{figure}

As a use case, we study radiogenomic associations related to human-understandable imaging traits in magnetic resonance imaging (MRI) scans from patients with glioblastoma (GBM). GBM is a grade IV malignant brain tumor with poor prognosis, and for which imaging is heavily used in diagnosis, prognosis, and treatment assessment. MRI traits like tumor enhancement, non-contrast enhancement, edema, and necrosis in Fig. \ref{fig_mri_traits}, describe some of the visual, phenotypic variations between patients as they are diagnosed and treated. We present here extracted associations found by the radiogenomic neural networks using our approach, compare against previous work to show both new and consistent findings, and establish the radiogenomic associations' clinical value in estimating patient survival over clinical or imaging traits alone.

%
\section{Materials and methods}

\begin{figure}[h]
	\caption{An overview of the study's approaches to radiogenomic neural network
		(\textbf{a}) training and
		(\textbf{b}) interpretation, i.e., gene masking and gene saliency to extract radiogenomic associations and radiogenomic traits.}
	\label{fig:pipeline}
	\centering
	\includegraphics[width=\textwidth]{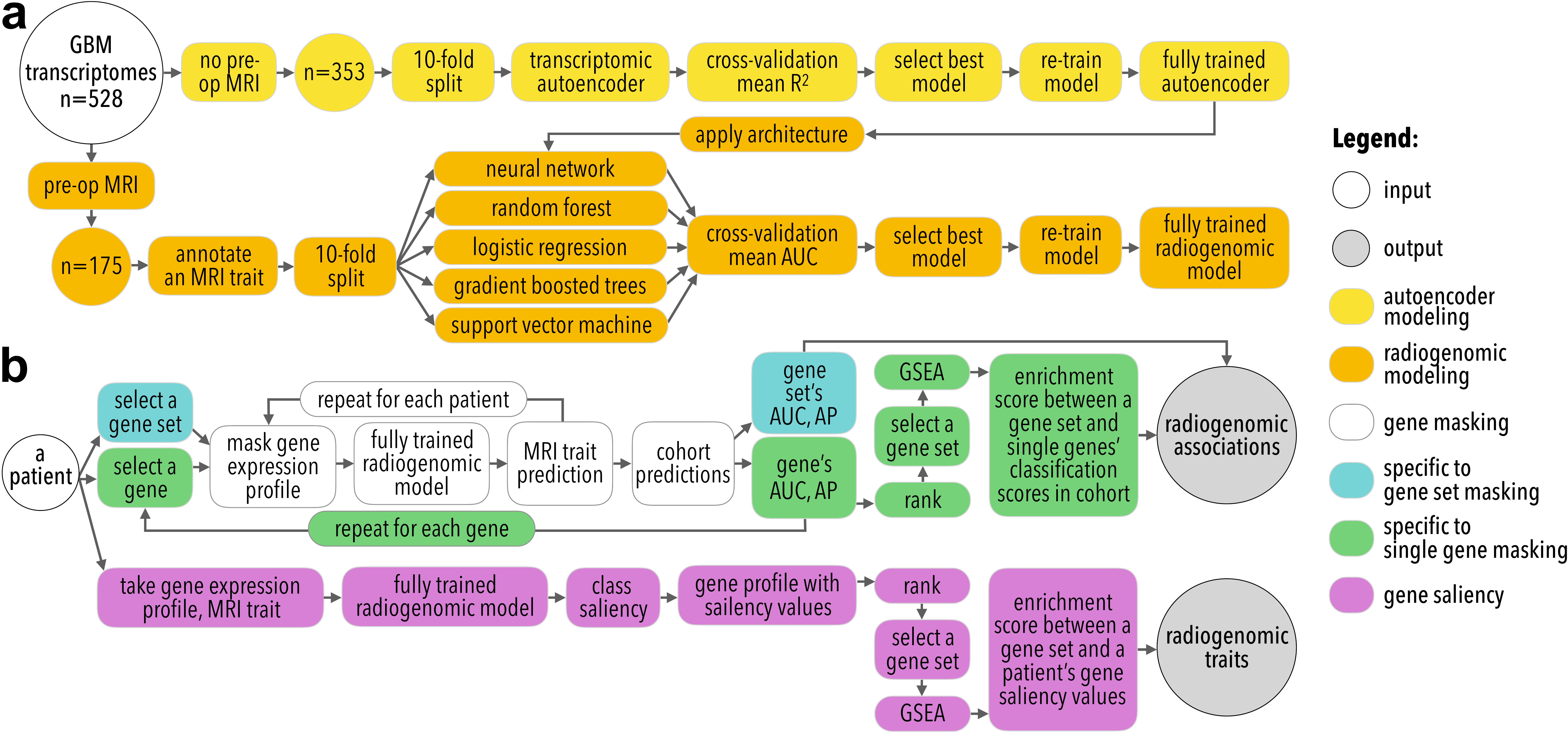}
\end{figure}

\subsection{Gene expression}
Transcriptomes were available for 528 GBM patients as part of The Cancer Genome Atlas (TCGA), see Supp. Table \ref{tab:demo}. Samples were primary tumors, untreated, and had $\geq$80\% cancer and $\leq$50\% necrotic cells \cite{McLendon2008}. Samples were analyzed by the Broad Institute on Affymetrix arrays, quantile normalized, and background corrected. Level 3 data were downloaded from the Genomic Data Commons at {\small\texttt{https://gdc.cancer.gov/}}. Each profile had 12,042 genes.

\subsection{Magnetic resonance imaging}
Medical images for 262 GBM patients were downloaded from The Cancer Imaging Archive (TCIA) \cite{tcia_gbm}. Patients were matched based on barcodes shared by TCGA and TCIA. A board-certified neuroradiologist, Dr. El-Saden (26 years of experience), evaluated images using the Osirix medical image viewer. An electronic form was used to record MRI traits according to the Visually Accessible Rembrandt Images (VASARI) feature guide \cite{Vasari}. 175 patients had pre-operative (pre-op) MRIs and transcriptomes, see Supp. Table \ref{tab:demo}. Six MRI traits were annotated from the pre-op studies. Traits were modeled as binary labels given the small sample sizes. Supp. Table \ref{tab:labels} describes all labels and their percentages in the cohort.

\subsection{Radiogenomic modeling}
\begin{figure}[h]
	\caption{ The radiogenomic neural network's
		(\textbf{a}) architecture,
		(\textbf{b}) transfer learning with a deep transcriptomic autoencoder,
		and interpretation methods through 
		(\textbf{c}) gene masking and
		(\textbf{d}) gene saliency.
		Pretrained weights learned in the autoencoder were transferred to a radiogenomic model, where weights were frozen (non-trainable, long red arrows) and/or fine-tuned (trainable, dashed red arrow) during radiogenomic training.
	}
	\label{fig_neural_net_methods}
	\centering
	\includegraphics[width=.9\textwidth]{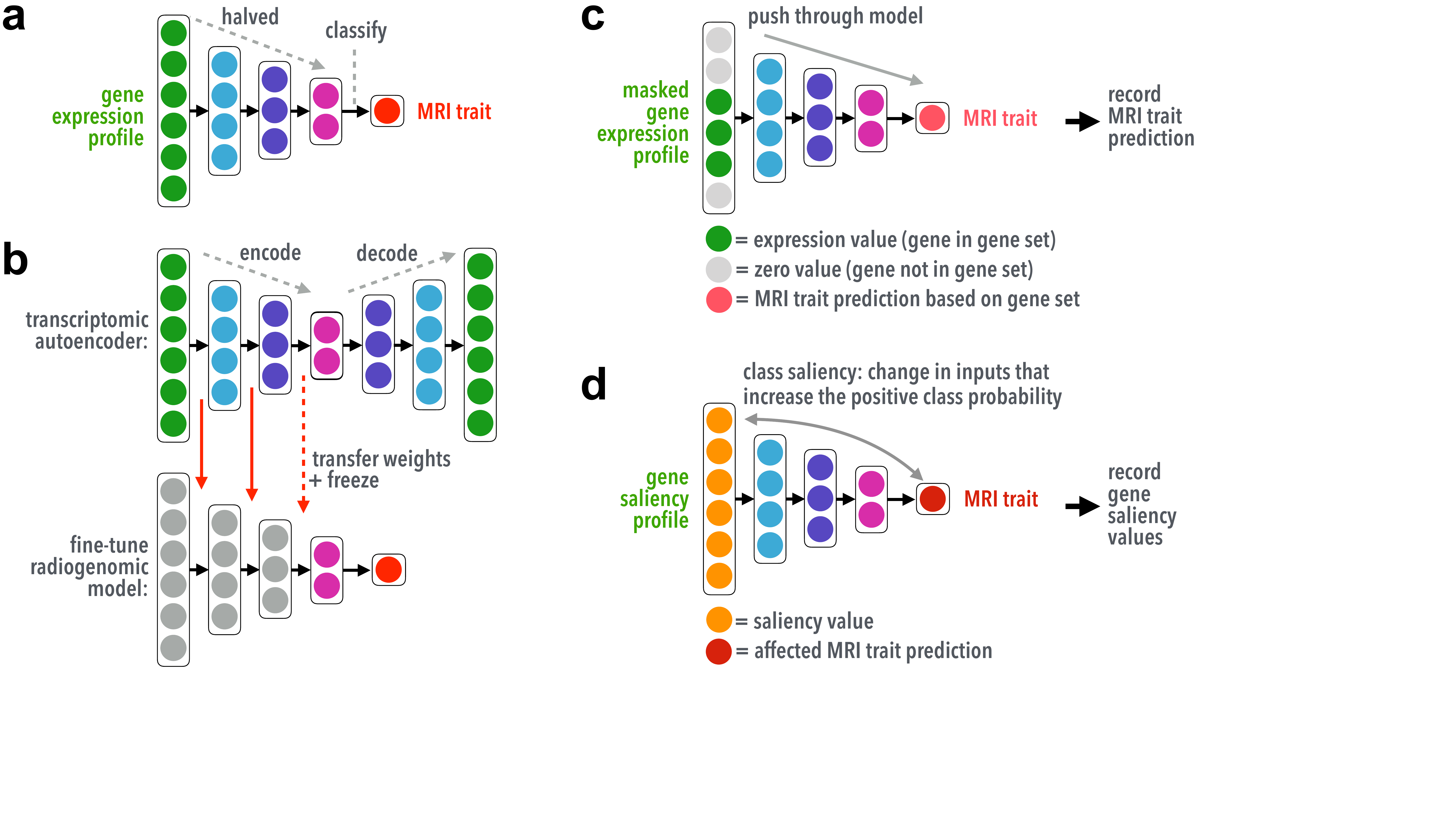}
\end{figure}

To map relationships between gene expression profiles and MRI traits, feed-forward neural networks \cite{Goodfellow2016book} were used, see Fig. \ref{fig_neural_net_methods}a--b. Each MRI trait was a binary classification task, where the positive class was the least frequent label (Supp. Table \ref{tab:labels}). Models were provided all 12,042 gene expressions as input vectors to classify each imaging trait, resulting in one model per trait. During training, early stopping with a patience of 200 epochs was used while monitoring the receiver operating characteristic curve, and the area under the curve (AUC) calculated. To help learning in the radiogenomic model, an autoencoder was used since many more patients had transcriptomic data. The radiogenomic neural networks were pretrained using weights transferred from a deep transcriptomic autoencoder trained on a separate subset of 353 patients. The transcriptome dataset consisted of TCGA-GBM patients with transcriptomes, but no pre-op MRIs and thus excluded the radiogenomic dataset.

The transcriptomic autoencoder takes a gene expression profile as input, compresses the information through three encoding layers, and then decodes the information to reconstruct the transcriptome. Early stopping was used in training and monitored the mean coefficient of determination ($R^2$) of each gene. The trained autoencoder weights, along with the gene pre-processing parameters, were then used as non-random weight initialization (i.e., weights \textit{can} be fine-tuned during training) and/or frozen weights (i.e., weights \textit{cannot} be fine-tuned during training) in the radiogenomic models. 

Performance was estimated with 10-fold cross-validation with sample weighting and stratified fold splitting. Each gene expression was mean subtracted and divided by its standard deviation, a process performed at each fold split. Hyperparameters were optimized via grid search. An illustration of the overall methods for radiogenomic model training is shown in Fig. \ref{fig:pipeline}a.  The autoencoder achieved a mean validation of 0.45 $R^2$ in cross-validation and 0.61 $R^2$ in retraining, see Supp. Fig. \ref{supfig:ae_perf}. Radiogenomic models were then constructed to have the same encoding architecture, activation function, and optimizer as the best performing autoencoder.

Neural networks were trained on NVIDIA Tesla K80 and V100 GPUs through Amazon Web Services using \texttt{Python 3.6}, \texttt{Keras 2.2.4} \cite{chollet2015keras}, and \texttt{TensorFlow 1.12.0} on a Ubuntu 16.04 machine. Other classifiers were implemented via \texttt{XgBoost 0.80} \cite{ChenGuestrin2016} and \texttt{sklearn 0.20.0} \cite{Pedregosa2011}. For more modeling details, see Supp. Table \ref{tab:models}.

\subsection{Bootstrapping}
The best performing models of each model type (see Supp. Tables \ref{tab:models}) were evaluated on bootstrapped datasets to measure classification performance variability within and between model types. For each bootstrap, a radiogenomic dataset was split for 10-fold cross-validation using a different seed. In each split, training and validation samples were separately resampled with replacement to obtain bootstrapped sets. Sets were resampled if not all classes were observed. Each model was trained and compared on the same bootstrapped dataset using the same procedure in the aforementioned methods. This process was repeated 100 times for each MRI trait.

\subsection{Molecular subtype modeling}
Molecular subtypes and their gene sets were downloaded from \cite{Verhaak2010}, which contained 840 genes to describe four GBM subtypes, i.e., classical, mesenchymal, proneural, and neural. Of the 528 patients, 171 had subtype labels and Affymetrix profiles, see Supp. Tables \ref{tab:demo} and \ref{tab:labels}. A neural network was trained to predict subtypes using gene expression profiles. The four subtypes were modeled as multi-class classification task via one-hot-encoded labels. Gene expressions pre-processing and model training were performed in the same manner as the radiogenomic models, mainly, 10-fold cross-validation and hyperparameter grid search, see Supp. Tables \ref{tab:subtype_hyp}.

\subsection{Gene masking}
Masking is a sensitivity analysis where the actual value of one or more components of the input are kept while all others are replaced with zeros. The goal is to determine the impact that the kept input components have on the end classification; this procedure was previously described in \cite{Zeiler2014}. Here, we define ``gene masking'' to extract radiogenomic associations from a neural network, see Fig. \ref{fig_neural_net_methods}c. For each individual, the gene expression values for a particular gene set were kept while all other expressions were replaced with zeros. The masked profiles were pushed through a fully trained neural network and the output, i.e., a class probability based on using genes from the gene set, was recorded. After repeating this process for each patient, classification performance was calculated. Each gene set was evaluated by AUC and average precision (AP) to measure the strength of a radiogenomic association. As such, gene masking reported radiogenomic associations that were generally predictive of an MRI trait in the entire cohort.

In \textit{single gene} masking, each gene was additionally used in gene set enrichment analysis (GSEA) \cite{Subramanian2005} (ranked by AP or AUC). We also used \textit{gene set} masking, where predefined gene sets smaller than 500 genes  (see GSEA methods) from the Broad Institute's molecular signatures database (MSigDB, v6.2) \cite{Liberzon2011}, molecular subtypes \cite{Verhaak2010}, and brain cell types and phenotypes \cite{Patel2014, Darmanis2015, Zhang2016neuron} taken from \cite{Puchalski2018} were used. MSigDB was also queried for gene sets that include the 22 genes characterized as potential contributors of GBM tumorigenesis \cite{McLendon2008, Parsons2008}, see Supp. information. The top performing genes or gene sets for each MRI trait were visualized together by clustering classification scores using \texttt{pheatmap} in R. 

\subsection{Gene saliency}
Class saliency is a visualization technique used to compute the gradient an output class prediction with respect to an input via back-propagation \cite{Simonyan2014, Kotikalapudi2017}. Thus, class saliency identifies the relevant input components whose values would affect the positive class probability in a neural network. Here, we define ``gene saliency'' as the genes whose change in expression would increase the model's belief of the positive class label, see Fig. \ref{fig_neural_net_methods}d. In each model, salient genes are derived for each patient, ranked, and used in GSEA to determine if a gene set is relevant to predicting his/her MRI trait. Subsequently, positive enrichment between a single patient's salient genes and a gene set is defined as a ``radiogenomic trait.'' For example, the edema model was probed to identify a single patient's salient genes. The most salient genes were the genes that increased the probability of edema being $\geq \sfrac{1}{3}$. If GSEA found the salient genes were enriched by a gene set, then the prediction of the patient's edema was related to the gene set, i.e., the patient has the radiogenomic trait between the gene set and $\geq \sfrac{1}{3}$ edema. Saliency was implemented using \texttt{keras-vis 0.4.1} \cite{Kotikalapudi2017}. The input range was determined by the gene expression range in the dataset, and \texttt{guided} was used as the \texttt{backprop\_modifier}; other parameters were set to default. In the subtype neural network, gene saliency was repeated for each class as one-versus-others. Fig. \ref{fig:pipeline}b depicts the overall process for gene masking compared to gene saliency.

\subsection{Gene set enrichment analysis}
Pre-ranked GSEA \cite{Subramanian2005} was implemented using \texttt{fgsea} \cite{Sergushichev2016}. Gene sets were parameterized by the recommended maximum size of 500 genes in a gene set, a minimum size of 15, and 10,000 permutations. Genes were ranked via single gene masking classification scores or gene saliency values for a patient. Enrichments were significant at an adjusted p-value $<0.05$ \cite{Sergushichev2016}. Correlation between a gene expression and an imaging trait was used for comparison. Clustering of enrichment scores was performed as previously defined.

\subsection{Survival}
Clinical data from TCGA was used to define patient outcomes. Patient covariates included gender, race (binned as white and non-white), and age at initial diagnosis (binned above or below the median value in all 528 patients). Overall survival (OS) and time-to-death events were in the \texttt{patient} file. Progression-free survival (PFS) outcomes were defined by the \texttt{followup} file, where all event types with days-to-event data were considered, but only the earliest event was used (see also Supp. information).

Cox proportional hazards models were used to estimate univariate and adjusted hazard ratios (HR). Radiogenomic traits enriched in at least five patients were kept. Adjusted HRs were estimated by a Cox model with backward feature selection and forced to keep all three patient covariates, while free to choose any MRI or radiogenomic trait. 

Survival analyses were done in R using the \texttt{survival} and \texttt{survminer}. Kaplan-Meier estimates were obtained with \texttt{survfit}, \texttt{ggsurvplot}, and \texttt{survdiff}. Patients with missing information were removed and the log-rank test was used to test for significant differences between groups. Cox models were obtained with \texttt{coxph} and feature selection performed with \texttt{step}.
%
%
%
%
%
\section{Results}

\subsection{Neural networks achieve best performance in classifying MRI traits}
\begin{figure}
	\caption{Radiogenomic models performances. (\textbf{a}) Observed 10-fold cross-validation performances. (\textbf{b}) Performance differences between a neural network and another model in 100 bootstrapped datasets. Notation: neural network (nn), gradient boosted trees (gbt), random forest (rf), support vector machines (svm), logistic regression (logit).}	\label{fig:performance}	\centering	\includegraphics[width=\textwidth]{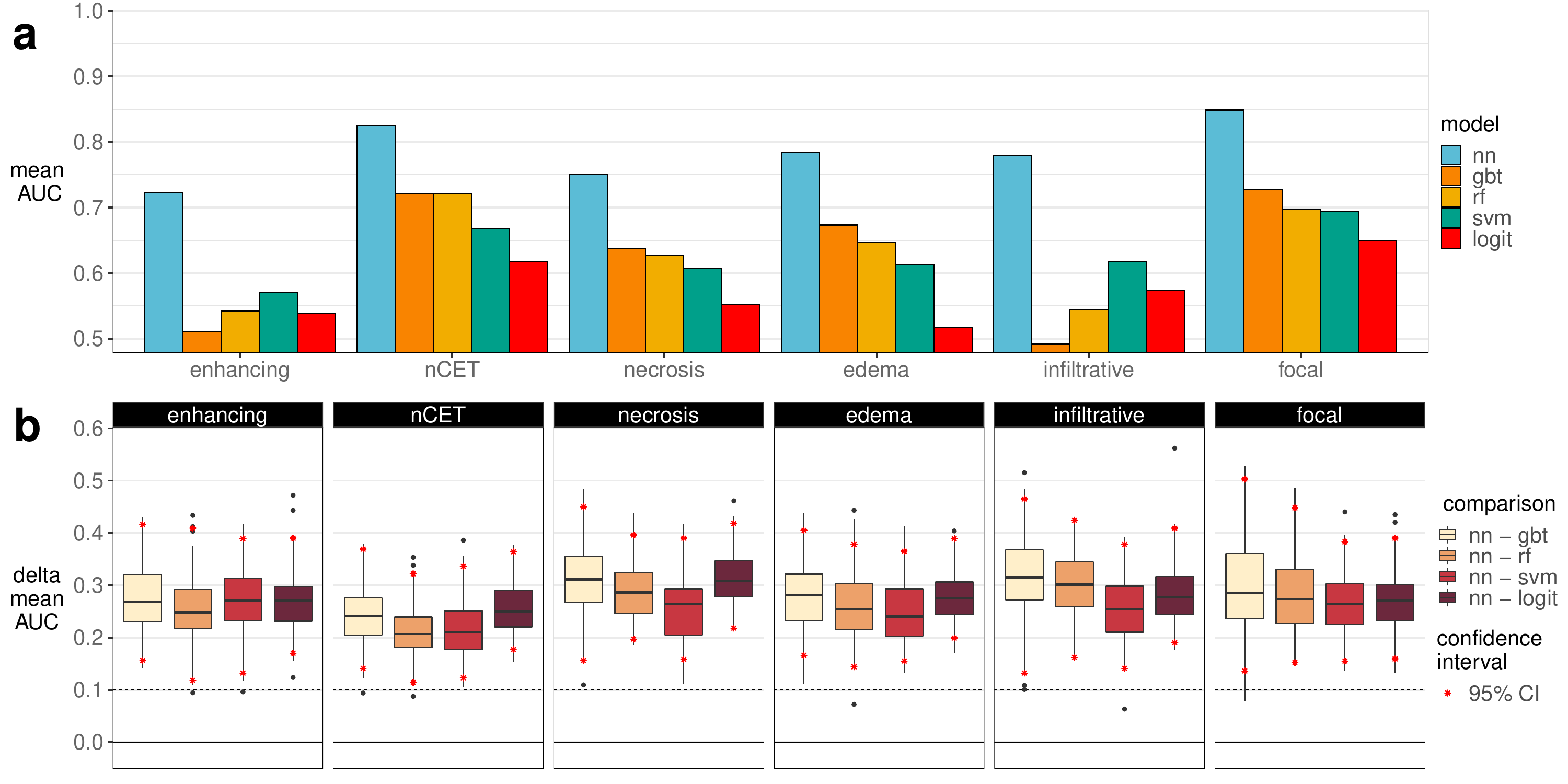}
\end{figure}

Neural networks were better at estimating MRI traits than all other classifiers, see Fig. \ref{fig:performance}a. In predicting proportions of nCET, necrosis, and edema, the first hidden layer used frozen pretrained weights and only the last two hidden layers were used for fine-tuning, see Supp. Table \ref{tab:nn_selected_hyps}. Bootstrapping showed neural networks had higher performances, where 95\% confidence intervals (CI) in Fig. \ref{fig:performance}b indicate neural networks outperformed other models by more than 0.10 AUC. Supp. Table  \ref{tab:model_perf} and Figs \ref{supfig:boot_comparison}--\ref{supfig:boot_nn_metrics} have further modeling results.

{\setlength{\tabcolsep}{7pt}
	
	\begin{table}
		\centering
		\scriptsize
		\begin{threeparttable}	
			\caption{Neural network hyperparameters.  Layers refers to the depth of hidden layers in the radiogenomic model that used pretrained weights from the autoencoder (AE), e.g., two AE layers indicate the first two layers used pretrained weights. Retrain refers to models trained on the full dataset.}
			\label{tab:nn_selected_hyps}	
			\begin{tabular}{lll lll lll ll} 
				\toprule
				& \multicolumn{2}{c}{CV means} & \multicolumn{2}{c}{retrain} && &&  \\
				\cmidrule(lr){2-3} \cmidrule(lr){4-5}
				\textbf{label} 		& \textbf{$R^2$} &  \textbf{eph} & \textbf{$R^2$} &  \textbf{eph}&  \textbf{architecture} & \textbf{opt}	& \textbf{act}	& \textbf{drop} &	&\\
				\midrule
				transcriptome		& 0.45 & 467 & 0.61 & 486  & 4000, 2000, 1000  & Adadelta  & tanh & 0.0 &&\\
				
				& 	&	&					&& 		& 						&		 &	& \multicolumn{2}{c}{layers} \\
				\cmidrule(lr){10-11}
				\textbf{label} 		& \textbf{AUC} &  \textbf{eph} & \textbf{AUC} &  \textbf{eph}	& \textbf{architecture} & \textbf{opt}	& \textbf{act}	& \textbf{drop} & \textbf{AE}  &\textbf{frozen} \\
				\midrule
				enhancing			& 0.72 & 38 	& 1.00	& 14	& 4000, 2000, 1000  & Adadelta  & tanh & 0.6 & 3 & 0	 	   \\
				nCET				& 0.83 & 38  	& 1.00	& 11	& 4000, 2000, 1000  & Adadelta 	& tanh & 0.0 & 1 & 1 	  \\
				necrosis 			& 0.75 & 44 	& 1.00	& 11 	& 4000, 2000, 1000  & Adadelta  & tanh & 0.0 & 1 & 1 	  \\
				edema 				& 0.78 & 109 	& 1.00	& 16 	& 4000, 2000, 1000	& Adadelta	& tanh & 0.0 & 1 & 1 	  \\
				infiltrative 		& 0.78 & 70  	& 1.00	& 12 	& 4000, 2000, 1000	& Adadelta	& tanh & 0.0 & 2 & 1		  \\
				focal				& 0.85 & 44 	& 1.00	& 12 	& 4000, 2000, 1000	& Adadelta	& tanh & 0.6 & 3 & 0	 \\
				subtype				& 0.994 & 14 	& 0.998 & 66 	& 3000, 1500, 750	& Nadam		& sigmoid & 0.4 & - & -		 \\
				\bottomrule	
			\end{tabular}
			\begin{tablenotes}
				\scriptsize
				\item[] eph (epoch), opt (optimizer), act (activation), drop (dropout).
			\end{tablenotes}
		\end{threeparttable}
	\end{table}
}

\subsection{Neural networks correctly learned known associations between gene expressions and molecular subtypes}

\begin{figure}[h!]
	\centering
	\caption{Gene masking in the fully trained subtype neural network. The model's 
		(\textbf{a}) estimated subtype probabilities, where each row was a patient and grouped by their true subtype; and 
		(\textbf{b}) classification performance measured by AP in gene set masking, where each row was a gene set and each column was the subtype prediction. The random gene set excluded any gene included in a subtype set.  For visualization purposes, rows were sorted by the model's estimated mesenchymal probability. Notation: classical (CL), mesenchymal (MES), neural (NL), proneural (PN), coverage (percent of gene set that exist in gene expression profiles).  For more gene masking, see Supp. Figs. \ref{supfig:subtype_verhaak}--\ref{supfig:subtype_hallmark}.}
	\label{fig:subtype_subtype_pert}
	\includegraphics[width=.7\textwidth]{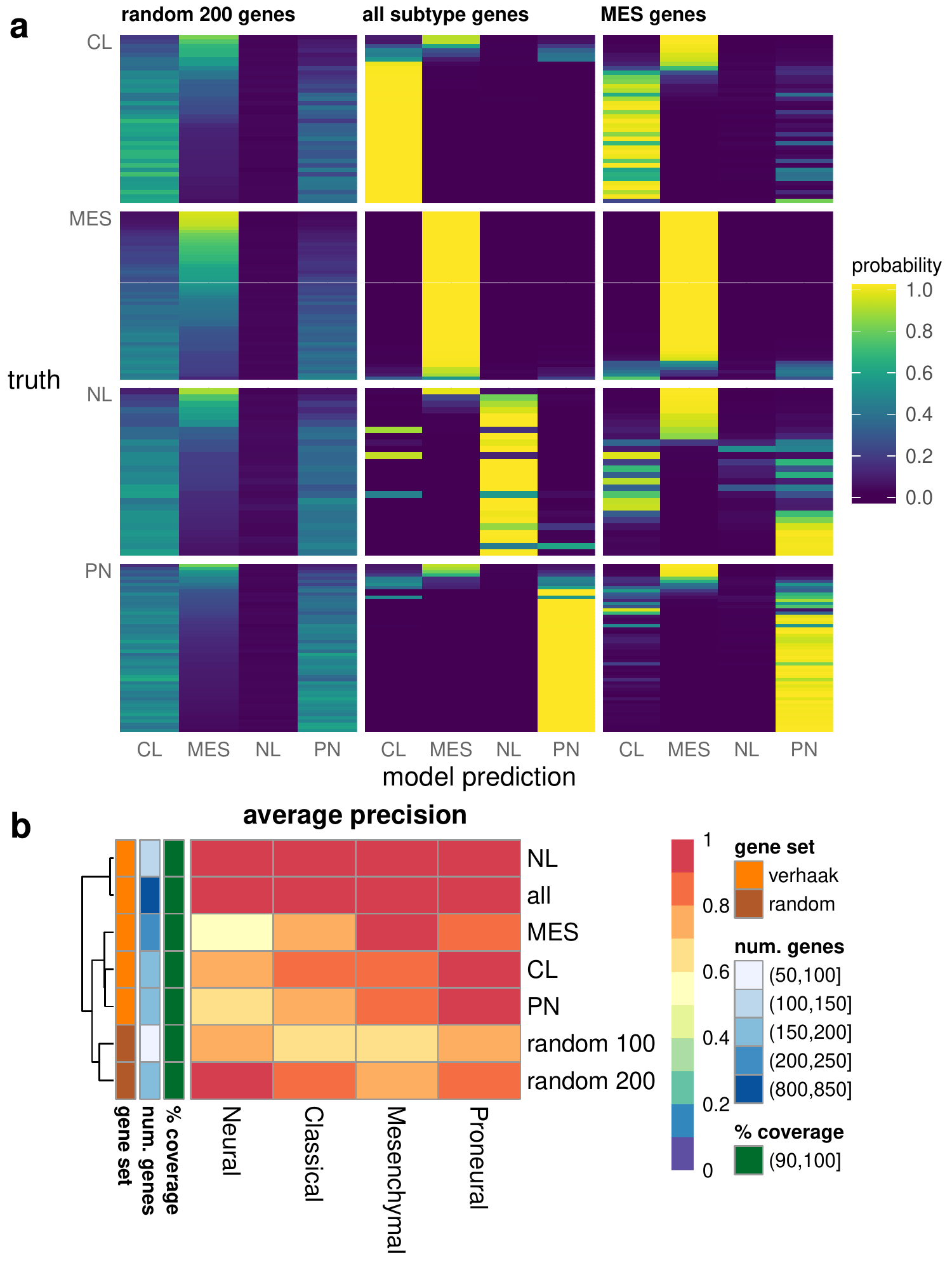}
\end{figure}

To verify the relationships learned in the neural network's layers, a molecular subtype model was trained. The model achieved a micro-averaged AUC of 0.994 in 171 patients, see Supp. Table \ref{tab:suptype_best_cv}. Gene masking with subtypes gene sets showed the neural network was able to predict subtype classes with high certainty, see  Fig. \ref{fig:subtype_subtype_pert}. For example, when the model was given the expressions of the 216 mesenchymal genes, subtype probabilities approached 1 or 0 and often corresponded to the correct subtype (0.90 AUC, 0.89 AP). Performance improved when the model was given all 840 subtype genes (0.99 AUC, 0.98 AP). Conversely, given the expressions of 200 random, non-subtype genes, the model was less certain (probabilities away from 1 or 0) and a fully trained performance of 1.0 AP dropped to 0.68 AP, see Supp. Table \ref{tab:subtype_subtype_score}.

\begin{figure}[h!]
	\centering
	\caption{Single gene masking in the subytpe model: 
		(\textbf{a}) the top 20 genes used to predict each subtype;
		(\textbf{b}) the percent of subtype genes covered in the top N genes; 
		(\textbf{c}) GSEA with genes ranked by average precision, where positive enrichment indicated the subtype gene set was correlated with high average precision and vice versa; and 
		(\textbf{d}) GSEA with genes ranked by correlation with label, where positive enrichment indicated the subtype gene set was correlated with subtype and vice versa.}
	\label{fig:subtype_single}
	\includegraphics[width=\textwidth]{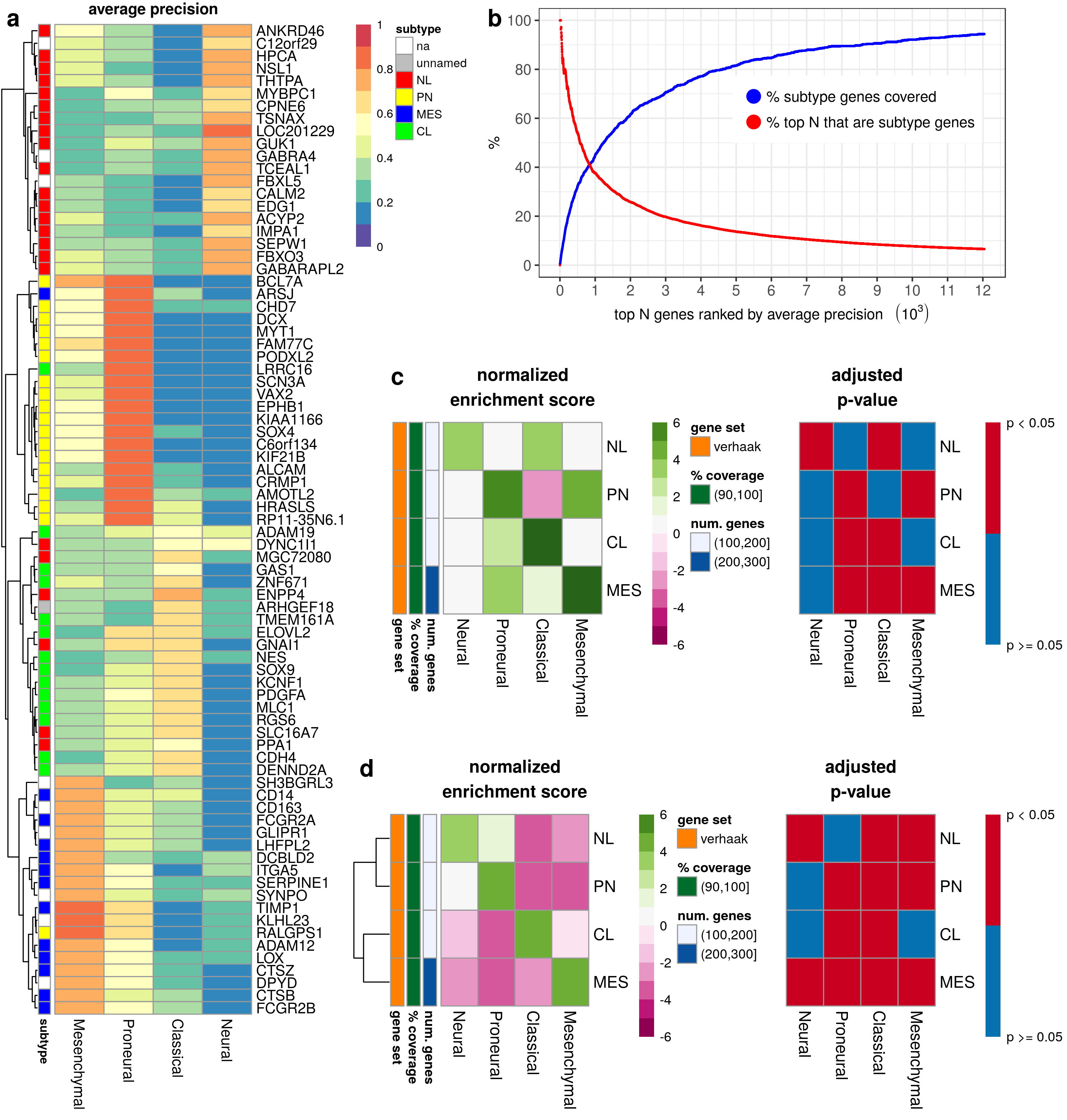}
\end{figure}

The majority of the top 20 predictive genes in each subtype class belonged to the original subtype class definition, see Fig. \ref{fig:subtype_single}a. For example, 18 of the top 20 genes for predicting the proneural subtype were a part of the proneural gene set and each had least 0.80 AP and a 0.80 AUC. Of the top 500 genes ranked by AP, 270 genes (54\%) were subtype genes; this represented 32\% of all subtype genes, see Fig. \ref{fig:subtype_single}b. As expected, subtype genes were predictive of more than one subtype. Similarly, GSEA showed the most predictive single genes for each subtype prediction were significantly (adjusted p-value $<$ 0.05) and positively enriched by the corresponding subtype gene set, see  Fig. \ref{fig:subtype_single}c. This observation was corroborated in GSEA using ranked genes based on the correlation, see Fig. \ref{fig:subtype_single}d.

\subsection{Genes driving the prediction of MRI traits}

\subsubsection{Enhancing tumor}
Tumor enhancement was measured on T1W+Gd images. Low grade or well-differentiated brain tumors tend to generate blood vessels with intact blood-brain barriers (BBB) and do not enhance. Poorly differentiated or more aggressive tumors, like GBM, generate leaky blood vessels without an intact BBB and enhance on T1W+Gd images. 

\begin{figure}[h!]
	\captionsetup{font=small}
	\caption{Gene masking of the radiogenomic models with the MSigDB hallmark gene sets  \cite{Liberzon2015}. 
		(\textbf{a}) Model performance in gene set masking. Shown are the top five gene sets ranked by average precision in each MRI trait, see also Supp. Fig. \ref{supfig:vasari_hallmark_auc}. 
		(\textbf{b}) Enrichment among genes ranked by average precision in single gene masking. Positive enrichment indicated gene sets were predictive of an MRI trait and negative enrichment indicated the opposite. Shown are hallmarks with at least one significant enrichment.}
	\label{fig:vasari_hallmark}
	\includegraphics[width=\textwidth]{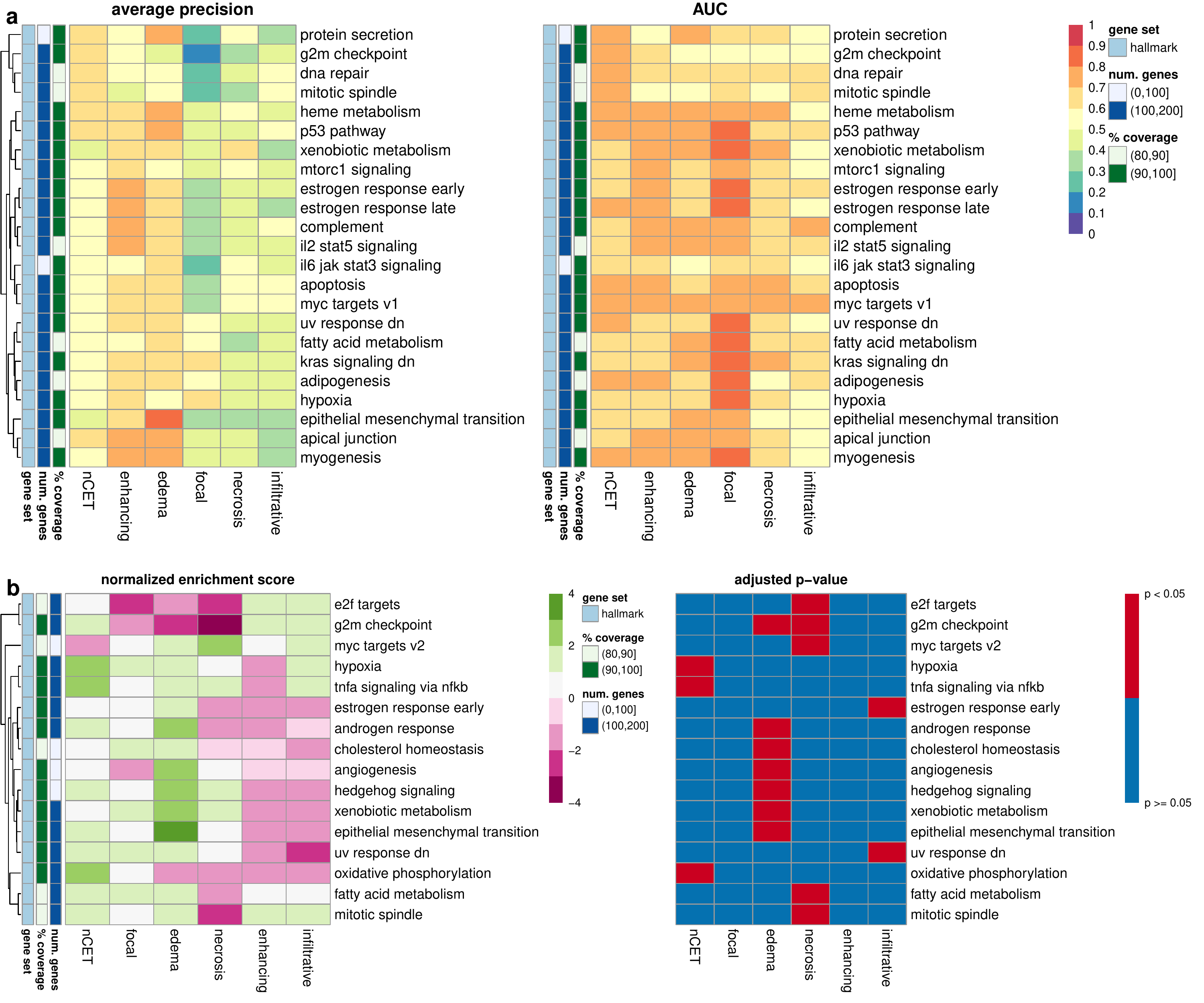}
	\centering
\end{figure}

Enhancement was found to be associated with growth, immune responses, hormones, the extracellular matrix (ECM), vasculature, and kinase activity in gene masking, see Table \ref{tab_summary_masking1}. The ECM association included gene expressions of ECM-related proteins \cite{Naba2012}, see Supp. Fig. \ref{supfig:vasari_canonical}. Of the MSigDB hallmark gene sets, apical junction (cellular components, adherens and tight junctions), IL2/STAT5 signaling immune response activation), complement system, early and late responses to estrogen (associated with ESR1 expression), and heme metabolism (erythroid differentiation, STAT5 activation) \cite{Liberzon2015} were most predictive of enhancement, see Fig. \ref{fig:vasari_hallmark}a. Gene Ontology (GO) gene sets related to GBM-abnormalities support the association of growth, immune system, and hormones with enhancement.

In single gene masking, enhancement was best predicted by \textit{SNTB1} (0.67 AUC, 0.58 AP) and \textit{B4GALT6} (0.64 AUC, 0.60 AP), see Supp. Fig. \ref{supfig:vasari_single}. SNTB1, a cytoskeletal protein, was down-regulated in a GBM cells study \cite{Mongiardi2016} and a potential binder to PTPRZ, a protein contributing to glioma cell growth \cite{Bourgonje2014}.


In previous GBM radiogenomic studies, enhancement was associated with hypoxia, ECM, angiogenesis in 22 patients \cite{Diehn2008}; Biocarta pathways and genes, \textit{C1orf172}, \textit{CAMSAP2}, \textit{KCNK3}, and \textit{LTBP1} in 23 patients \cite{Jamshidi2014}; and \textit{EGFR} copy number amplification in 75 patients \cite{Gutman2013}. Gene sets involving the ECM, \textit{EGFR}, \textit{C1orf172}, \textit{KCNK3}, and \textit{LTBP1} were confirmed to have performances greater than 0.70 in both AUC and AP, see Supp. Fig. \ref{supfig:vasari_lit_enhan}a.

\begin{flushright}
	{\setlength{\tabcolsep}{3.4pt}
		
		\begin{table}[h!]
			\centering
			\scriptsize
			\begin{threeparttable}	
				\caption{Summary of transcriptomic drivers of MRI traits in GBM patients. Shown are the top five hallmark gene sets ranked by AUC or AP and compared against gene sets related to prior GBM work, see Supp. Figs \ref{fig:vasari_gene_query} (gene abnormalities \cite{McLendon2008, Parsons2008}) and \ref{supfig:vasari_lit_enhan}--\ref{supfig:vasari_lit_f10} (radiogenomics). Note: common themes can comprise of different gene sets.}
				\label{tab_summary_masking1}	
				\begin{tabular}{lll lll lll ll} 
					\toprule
					& \multicolumn{2}{c}{\textbf{transcriptomic drivers}} \\
					\cmidrule{2-3}
					\textbf{MRI trait} & theme &gene set (collection*, query$\dagger$) & \textbf{AUC} & \textbf{AP} & \textbf{see also}\\
					
					\midrule
					enhancing	
					&growth/death				 & growth (GO, \textit{PTEN})& 0.86& 0.84\\
					&							& sensory organ development (GO, \textit{EGFR, KCNK3}) & 0.85 & 0.84 & \cite{Jamshidi2014, Gutman2013}\\
					&immune system  & IL2/STAT5 signaling (H) & 0.77& 0.76 & \\
					&							& complement system (H) & 0.79 & 0.75 \\
					&							& activation of immune response (GO, \textit{PTEN}) & 0.90 & 0.89 \\
					&							& leukocyte \& lymphocyte activation (GO, \textit{PIK3R1}) & 0.86, 0.85 & 0.85, 0.83\\
					&							& immune effector process (GO, \textit{PIK3CA}) & 0.87 & 0.84 \\
					&hormones	& early \& late responses to estrogen (H) & 0.79, 0.78 & 0.73, 0.73\\
					&					 & response to steroid hormone (GO, \textit{RBI}) & 0.88 & 0.88 \\
					&					 & regulation of hormone levels (GO, \textit{PARK2})  & 0.87 & 0.84\\
					&ECM 	& related to ECM proteins   (C, ECM) &0.77--0.84 & 0.73--0.76 & \cite{Diehn2008}\\
					&			&  apical junction (H) & 0.80 & 0.75 \\
					&vasculature & heme metabolism (H) & 0.77 & 0.65\\
					&						& vasculature \& heart development (GO, \textit{LTBP1}) & 0.81, 0.78 & 0.80, 0.77 & \cite{Jamshidi2014}\\  
					&kinases activity 	& multiple (GO, \textit{EGFR, LTBP1, KCNK3}) & all 0.87 & all 0.85 & \cite{Jamshidi2014, Gutman2013}\\
					
					%
					edema	
					& EMT		& EMT (H) & 0.80 & 0.80 \\
					&				& positive regulation of locomotion (GO, \textit{EGFR, POSTN}) & 0.80 & 0.81 & \cite{Zinn2011} \\
					& 				& taxis (GO, \textit{CXCL12, KIF5C}) & 0.80 & 0.80 & \cite{Zinn2011}\\
					& 				& apical junction (H) & 0.77 & 0.74 \\
					&			  	& related to cell adhesion (GO, \textit{CDKN2A, EGFR, CTNNA2}) &  0.75--0.83 & 0.78--0.82 & \cite{Zinn2011}\\
					& growth/death & p53 pathway (H) & 0.77 & 0.77 \\
					&			  & autophagy (GO, \textit{CDKN2A}) & 0.75 & 076 \\
					&			  & myogenesis (H) & 0.75 & 076 \\
					&			& urogenital system development (GO, \textit{PTEN}) & 0.81& 0.81\\
					&			 & muscle structure development (GO, \textit{COL6A3}) & 0.80 & 0.80 & \cite{Zinn2011}\\
					&			 & response to growth factor (GO, \textit{EGFR, POSTN}) & 0.76 & 0.81 & \cite{Zinn2011}\\
					& vasculature 				& heme metabolism (H) & 0.77 & 0.73 \\
					& differentiation 	& central nervous system neuron differentiation (GO, \textit{PTEN}) & 0.79 & 0.81 \\
					& 						  	&  cell differentiation (GO, \textit{MET}) & 0.79 & 0.81\\
					&						 	&  stem cell differentiation (GO, \textit{CDK6}) & 0.80 & 0.80\\
					& immune system & regulation of cell activation (GO, \textit{CDKN2A}) & 0.83 & 0.82\\ 
					&								& neg. regulation of immune system (GO, \textit{CDKN2A, CXCL12}) & 0.82 & 0.81 & \cite{Zinn2011}\\
					&								& immune effector process (GO, \textit{PIK3CA}) & 0.80 & 0.81\\
					& other				& glycolysis (H) & 0.76 & 0.70 \\
					nCET
					& cell cycle & mitotic spindle (H) & 0.78 & 0.70\\
					& 				 & DNA repair (H) & 0.77 & 0.66\\
					&  				& G2M checkpoint (H) & 0.72 & 0.63\\
					&				& regulation of mitotic cell cycle (GO, \textit{TP53}) & 0.78 & 0.71 \\
					& growth/death  & p53 pathway (H) & 0.76 & 0.65\\
					&							& urogenital \& vasculature development (GO, \textit{PTEN}) & 0.81, 0.80 & 0.76, 0.71 \\
					&							& neg. regulation of cell cycle (GO, \textit{TP53}) & 0.80 & 0.74 \\
					&							& reproductive system development (GO, \textit{EGFR}) & 0.80 & 0.72 \\
					& UV response	& UV response down (H) & 0.78 & 0.59 \\
					&	 						& response to radiation (GO, \textit{TP53})	& 0.83 & 0.73 \\
					& other	& glycerphospholip metabolism process (GO, \textit{EGFR}) & 0.75 & 0.74		 \\
					&								& small molecule catabolic process 	(GO, \textit{PTEN}) & 0.78 & 0.73 \\
					\bottomrule	
				\end{tabular}
				\begin{tablenotes}
					\scriptsize
					\item[] * an MSigDB collection, where H=hallmarks, GO=Gene Ontology, and C=Canonical. $\dagger$ Gene sets queried from MSigDB using gene names or functions reported by previous work as keyword(s). neg.=negative. 
				\end{tablenotes}
			\end{threeparttable}
		\end{table}
	}
\end{flushright}
\begin{flushright}
	{\setlength{\tabcolsep}{2pt}
		\begin{table}[h!]
			\centering
			\scriptsize
			\begin{threeparttable}	
				\caption{Summary of transcriptomic drivers of MRI traits in GBM patients, continued.}
				\label{tab_summary_masking2}	
				\begin{tabular}{lll lll lll ll} 
					\toprule
					& \multicolumn{2}{c}{\textbf{transcriptomic drivers}} \\
					\cmidrule{2-3}
					\textbf{MRI trait} & theme &gene set (collection*, query$\dagger$) & \textbf{AUC} & \textbf{AP} & \textbf{see also}\\
					
					\midrule
					necrosis 
					& vasculature & heme metabolism (H) & 0.72 & 0.61 \\	
					& growth/death &  apoptosis (H) & 0.71 & 0.58 \\
					&						 & apoptotic signaling pathway (GO, \textit{IL4, TP53}) & 0.72 & 0.63 & \cite{Gevaert2014} \\
					&						 & related to \textit{TP53} & 0.75--0.76 & 0.63--0.68\\
					&						 & gland development (GO, \textit{EGFR}) & 0.76 & 0.65 \\

					& immune system  & IL6/JAK/STAT3 signaling (H) & 0.67 & 0.56 \\
					& 								& leukocyte cell cell adhesion (GO, \textit{IL4, ITGA5})	& 0.76 & 0.59 & \cite{Gevaert2014, Jamshidi2014} \\
					& 								& regulation of leukocyte proliferation (GO, \textit{CDKN2A}) & 0.77 & 0.61 & \cite{Gutman2013} \\
					& others & xenobiotic metabolism (H) & 0.76 & 0.65 \\
					&  & related to \textit{PTEN} & 0.73--0.78 & 0.63--0.67 \\
					&  & regulation of homeostatic process (GO, \textit{NF1}) & 0.78 & 0.65\\
					&  & glycolysis (H) & 0.76 & 0.56 \\
					
					%
					focal
					& growth/death & regulation of anatomical structure size (GO, \textit{PTEN}) & 0.96 & 0.88  \\
					&						 & response to growth factor (GO, \textit{EGFR}) & 0.92 & 0.83\\
					& transport & secretion by cell (GO, \textit{NF1}) & 0.95 & 0.88 \\
					&					& neg. regulation of transport (GO, \textit{PTEN}) & 0.96 & 0.87 \\
					&					& regulation of cytoplasmic transport  (GO, \textit{TP53}) & 0.95 & 0.85 \\
					& 					& monovalent inorganic cation transport (GO, \textit{PARK2}) & 0.95 & 0.81 \\
					& response to & steroid hormone, lipid, \& organic cyclic compound (GO, \textit{RB1}) &0.94--0.96 & 0.81--0.87\\
					& vasculature & vasculature development (GO, \textit{PTEN}) & 0.93 & 0.84 \\
					& 					& muscle \& circulatory system process (GO, \textit{PIK3CA}) & 0.94, 0.94& 0.82, 0.83\\
					& oxygen & hypoxia (H)  & 0.85 & 0.61\\
					& others & genes down-regulated by \textit{KRAS} (H) & 0.88 & 0.64 \\
					& 	& protein heterodimerization activity (GO, \textit{TP53}) & 0.97 & 0.85 \\
					&  & neg. regulation of intracellular signaling transduction (GO, \textit{PTEN}) & 0.96 & 0.84 \\
					&  & synaptic singaling (GO, \textit{PTEN}) & 0.92 & 0.82 \\
					
					%
					infiltrative
					& oxygen & reactive oxygen species pathway (H) & 0.71 & 0.50 \\
					&				& response to oxygen levels (GO, \textit{TP53}) &  0.67 & 0.60 \\
					& transport & neg. regulation of transport (GO, \textit{PTEN, NFKBIA}) & 0.74 & 0.64 &  \cite{Colen2014}\\
					& healing & wound healing (GO, \textit{NF1}) & 0.80 & 0.64 \\
					&  			    & hemostasis (GO, \textit{PIK3CA}) & 0.78 & 0.59 \\
					
					& growth/death & developmental growth (GO, \textit{PTEN}) & 0.74 & 0.62\\
					&							& spinal cord development (GO, \textit{NF1}) & 0.66 & 0.60\\
					& response to & response to drug (GO, \textit{MDM2 , MYC}) & 0.75 & 0.60 \\
					&   				 & response to inorganic substance (GO, \textit{PTEN}) & 0.73 & 0.61 & \cite{Colen2014}\\
					& others & DNA repair  (H) & 0.70 & 0.58 \\
					&  			 & ligase activity (GO, \textit{MDM2}) & 0.75 & 0.64 \\
					& 			 & ubiquitin like protein transferase  activity (GO, \textit{MDM2}) & 0.70 & 0.59 \\
					& 			 & ubiquitin like protein ligase binding (GO,  \textit{NFKBIA}) & 0.67 &  0.57 & \cite{Colen2014}\\
					&  & related to protein \& transcription factor complex (GO, \textit{TP53}) & 0.74--0.75& 0.60--0.63 \\
					&  & WNT signaling pathway (GO, \textit{PTEN, MYC}) & 0.73 & 0.62 & \cite{Colen2014}\\
					
					\bottomrule	
				\end{tabular}
				\begin{tablenotes}
					\scriptsize
					\item[] * an MSigDB collection, where H=hallmarks, and GO=Gene Ontology. $\dagger$ Gene sets queried from MSigDB using gene names or functions reported by previous work as keyword(s). neg.=negative. 
				\end{tablenotes}
			\end{threeparttable}
		\end{table}
	}
\end{flushright}

\subsubsection{Edema}
Tumor edema was identified as abnormal hyperintensity on FLAIR or T2W images. Edema often co-occurs with enhancement, implying a more aggressive tumor and does not typically occur in low grade brain tumors. Edema also results from leaky capillaries and usually surrounds the tumor, spreading within the white matter. Edema suggests an inflammatory and/or immune response to a malignant tumor, which is essentially a foreign body when highly dedifferentiated. 

Edema was associated with epithelial mesenchymal transition (EMT, metastasis and invasion), cell differentiation, and growth, see Table \ref{tab_summary_masking1}. p53 pathway (cell cycle, death), myogenesis, apical junction, heme metabolism, and glycolysis (cell metabolism) were the top hallmark gene sets, see Fig. \ref{fig:vasari_hallmark}a. GO terms relating to cell differentiation, death and adhesion with $\geq$ 0.80 AP, see Supp. Fig. \ref{fig:vasari_gene_query}b. Similar to the enhancement model, vasculature, immune system and \textit{EGFR}-related processes (albeit through different GO terms) were apart of the most predictive gene sets.

Growth and metastasis was also found to be predictive of edema in single gene masking. \textit{RAI2}, \textit{ANXA2}, \textit{POSTN} genes, all related to cell growth, were the top three single most predictive gene with 0.68--0.70 AUC and 0.60--0.64 AP. The top three ranked by AP were \textit{MTSS1}, \textit{LAMA5}, and \textit{KLHDC3}, with 0.65--0.66 AP and 0.63--0.67 AUC, see Supp. Fig. \ref{supfig:vasari_single}. Both \textit{MTSSI} and \textit{LAMA5} were associated with metastasis \cite{Agarwala2018}. 

GSEA showed significant enrichment in EMT, angiogenesis, androgen response (hormone), hedgehog signaling (including \textit{MTSSI}), and xenobiotic metabolism (drug metabolism), see Fig. \ref{fig:vasari_hallmark}b. The appearance of drug metabolism could be due to the use of symptomatic relief drugs  prior to surgery, such as corticosteroids for patients with neurologic symptoms caused by edema \cite{Omuro2013, Pitter2016}.

Previously, \textit{POSTN} was associated with edema in 78 GBM patients; the authors suggested \textit{POSTN} was regulated by miR-219 and contributed to cell migration or invasion \cite{Zinn2011}.  
GO gene sets related to the study's top five upregulated genes and microRNAs were found to be predictive of edema, see Supp. Figs. \ref{supfig:vasari_lit_edema}. Table \ref{tab_summary_masking1} shows an overlap of gene set patterns between the study's findings and gene masking of the edema model. In particular, gene sets associated \textit{POSTN}, cell taxis, and cell adhesion added to the association between edema and EMT.

\subsubsection{Non-contrast enhancing tumor}
Non-contrast enhancing tumor (nCET) was best identified on contrast-enhanced T1W and FLAIR or T2W images. nCET is typically lower grade tumor (better cellular differentiation, more closely resembling normal brain tissue), generating vessels with an intact BBB, and absent of contrast enhancement. While the abnormality on images is a mass-like neoplastic tissue, it is not rapidly dividing or aggressively dedifferentiating.  

Cell cycle, growth, and radiation response were themes among the most predictive gene sets for nCET.
Mitotic spindle, UV response down (genes down-regulated in response to ultraviolet radiation), DNA repair, and p53 pathway were predictive of nCET, see Fig. \ref{fig:vasari_hallmark}a. Of the gene sets related to GBM genomic alterations, the nCET model had a mix of ones found in the enhancing and edema model, see Fig. \ref{fig:vasari_gene_query}c, and supported transcription patterns found in hallmark gene sets in Table \ref{tab_summary_masking1}.  

\textit{SGPL1} (0.73 AUC, 0.56 AP) and \textit{DDR1} (0.66, 0.61 AP) were the top performing genes in single gene masking of the nCET model, see Fig. \ref{supfig:vasari_single}, where hypoxia, TNFA signaling via NFKB, and oxidative phosphorylation were significantly enriched, see Fig. \ref{fig:vasari_hallmark}b. The latter two gene sets were also identified in gene set masking, see Supp. Fig. \ref{supfig:vasari_hallmark_auc}.

\subsubsection{Necrosis}
Tumor necrosis was evaluated as the area of fluid signal intensity on T1W+Gd images. As tumors proliferate, they create new blood supply (angiogenesis) and/or expands to recruit blood from adjacent tissue. Subsequently, necrosis occurs, typically within the central portions of an aggressive tumor as the outer rim of enhancing surviving cells can be observed on MR images. 

Vasculature, apoptosis, immune system, and homeostasis was associated with necrosis, see Table \ref{tab_summary_masking2}. Predictive GO terms for necrosis included several \textit{TP53} and \textit{PTEN} related processes. Gene masking found some gene sets related to \textit{IL4}, \textit{WWTR1}, \textit{RUNX3}, \textit{ITGA5}, and \textit{CDKN2A} (found in previous radiogenomic studies \cite{Gevaert2014, Jamshidi2014, Gutman2013}) support the association between necrosis and apoptosis and the immune system. Earlier, drug metabolism was predictive of edema, but was also predictive of necrosis. Besides corticosteroids, antiepilepics can be prescribed patients who experience seizures from tumors \cite{Omuro2013}.

In single gene masking, \textit{CACNB2} (0.67 AUC, 0.51 AP) and \textit{PACSIN3} (0.64 AUC, 0.51 AP) the most predictive genes for necrosis, see Supp. Fig. \ref{supfig:vasari_single}. Notably, the MYC targets hallmark was significantly enriched, see Fig. \ref{fig:vasari_hallmark}b.

\subsubsection{Focal vs. non-focal}
Focal vs. non-focal traits were determined via T1W+Gd and FLAIR or T2W images. Focal tumors appear in one region. Non-focal tumors included those described as multifocal, multicentric, or with gliomatosis cerebri. A \textit{multifocal} tumor is one with separate enhancing regions that appear connected on FLAIR/T2W images with contiguous hyperintensity spreading via white matter tracts. A \textit{multicentric} GBM has multiple enhancing or non-enhancing tumors growing synchronously without contiguity on FLAIR/T2W. Gliomatosis cerebri is a rare, diffusely infiltrating subtype and involve at least three cerebral lobes. 

Focal traits were associated with growth, transport, vasculature, and hypoxia, see Table \ref{tab_summary_masking2}. Several of these involved \textit{PTEN} and \textit{RB1}. Secretion by cell includes genes potentially overlapped by others, e.g., \textit{NF1}, \textit{EGF}, and \textit{VEGF}. 

In general, focal traits were better predicted by GO gene sets related to GBM genes (Supp Fig. \ref{fig:vasari_gene_query}e) than hallmark gene sets (Fig. \ref{fig:vasari_hallmark}) or single genes (Supp. Fig. \ref{supfig:vasari_single}). These highly predictive GO gene sets ($\geq$ 0.90 AUC and $\geq$ 0.80 AP) indicated broad tumor characteristics, e.g., proliferation (growth, response to growth factors, secretion of growth factors) resulting in the need for angiogenesis (vasculature development), were used by the focal model to determine focal vs. non-focal tumors.

\subsubsection{Expansive vs. infiltrative}
Expansive vs. infiltrative was measured as the ratio of T1/FLAIR abnormality on T1W and FLAIR or T2W images. Expansive tumors have similar distribution on T1W as on FLAIR/T2W; the closer the two, the better defined the tumor margins and the better for surgical resection. Infiltrative tumors have FLAIR/T2W abnormality that is large compared to T1W abnormality, where the tumor is spreading through white matter tracts to cause large edema relative to the core tumor mass. Infiltrative traits indicate ill-defined tumor margins, less successful surgical debulking, and worse prognosis.

Infiltrative traits were best predicted by gene sets related to oxygen, transport, healing, and growth, see Table \ref{tab_summary_masking2}. Gene masking showed that GO gene sets were more predictive than hallmark gene sets (Fig. \ref{fig:vasari_hallmark}). Of the top GO gene sets, would healing and hemostasis were the most predictive and several included \textit{TP53}, \textit{MDM2}, and \textit{PTEN}. Notably, \textit{MDM2} transcription is regulated by \textit{TP53}.

Previous radiogenomic studies related to expansiveness or infiltrative traits found associations with \textit{MYC}, NFKBIA, and immune cell gene modules \cite{Colen2014}. The infiltrative model was masked with related gene sets and was able to predict infiltrative tumors with 0.50--0.70 AP, see Supp. Fig. \ref{supfig:vasari_lit_f10}. 

In single gene masking, \textit{ZBTB48} (0.68 AUC, 0.53 AP) and \textit{PRTN3} (0.67 AUC, 0.57 AP) as the best single gene predictors in Supp. Fig. \ref{supfig:vasari_single}.  For more gene masking, see Supp. Figs \ref{supfig:vasari_verhaak}--\ref{supfig:vasari_onco}. 


\subsection{Radiogenomic traits: Patient-specific radiogenomic associations}

Gene masking was used to identify cohort-level radiogenomic associations as genes were ranked by their overall classification performance among all tumors. In contrast, gene saliency was measured for each patient and identified patient-level radiogenomic associations, termed 'radiogenomic traits.' For salient genes in the subtype model, see Supp. Fig. \ref{supfig:subtype_subtype_saliency}. 

\begin{figure}[h!]
	\caption{Radiogenomic traits: results of gene saliency applied to the neural networks. Patients were considered enriched for a gene set at an adjusted p-value $< 0.05$. Gene sets with at least 10 enriched patients in a model were shown. All enriched patients were positively enriched. Positive enrichment indicated that the gene set was among the most salient genes for predicting a single patient's imaging trait.}
	\label{fig:vasari_saliency}
	\centering
	\includegraphics[width=\textwidth]{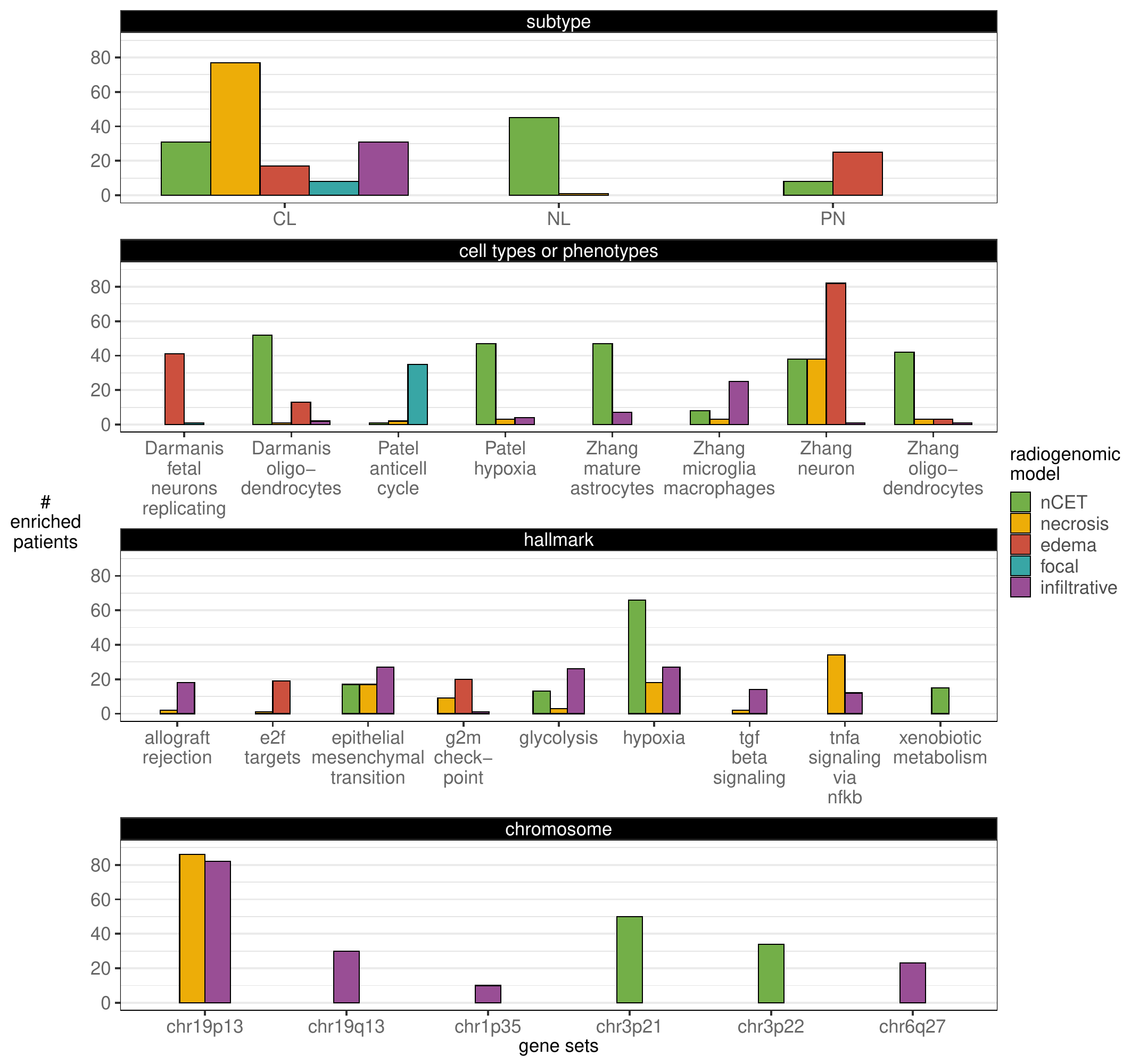}
\end{figure}

Classical subtype genes were salient in predicting larger proportions of necrosis, where 77 patients were enriched with classical genes in the necrosis model, see Fig. \ref{fig:vasari_saliency}. Similarly, neural and proneural gene sets were associated with greater edema and nCET proportions, respectively.

The nCET model showed more than 40 patients had salient genes enriched by oligodendrocytes, mature astrocytes, and hypoxia gene sets. Larger edema proportions were associated with neurons and replicating fetal neurons genes. The anti-cell cycle genes (negatively correlated with the cell cycles genes, some of which were a part of the hypoxia gene set in \cite{Patel2014}) were associated with prediction of the non-focal class (35 patients enriched). Patients were not significantly enriched with cell type or phenotype in the enhancing model, possibly reflecting more tumor heterogeneity in patients with more enhancing and aggressive tumor.

The hypoxia and nCET association (66 enriched patients) was consistent with the aforementioned anti-cell and hypoxia findings and with gene masking analysis, where vasculature development was predictive of nCET at the cohort-level. The association between hypoxia and greater proportions of nCET may be linked to lower-grade cells in the beginning stages of aggressive tumor growth and therefore responding to the beginning stages of hypoxic conditions and driving angiogenesis. Xenobiotic metabolism was enriched in 15 patients for predicting larger nCET, suggesting an increase in tumor size will result in an increased dosage of drugs administered prior to surgical resection or biopsy \cite{Pitter2016}.

Interestingly, the edema model showed only two hallmarks enriched by more than 10 patients. Although gene masking showed the edema model had high overall performance with the EMT gene set, other genes that are not associated with a predefined gene set may have been more influential in predicting each individual patient's edema proportions. In fact, the EMT hallmark was more associated with the radiogenomic model's belief of an infiltrative tumor in 27 patients. This subset of patients support the hypothesis that tumor cells with alterations in EMT-related genes are driving the observation of higher edema proportions than tumor cell proportions. Glycolysis and hypoxia hallmarks were also moderately ($<$ 25 patients) associated with infiltrative tumors. TNFA signaling via NFKB was associated with larger proportions of necrosis in 34 patients. 

Chromosomal aberrations have been reported in GBM \cite{Ohgaki2007, Verhaak2010}. There were 86 and 82 patients who were enriched by the chr19p13 gene set in predicting their necrosis and infiltrative traits, respectively. Genes in chr19q13, chr1p35, and chr6q27 were also salient to infiltrative tumors. chr3p21 and chr3p22 gene sets were also salient for greater nCET proportions.

\subsubsection{Radiogenomic traits with survival implications}

Of the 175 patients with radiogenomic data, 127 had all six MRI traits labeled and outcomes data. Given that only a subset of these individuals had clinical, imaging, and transcriptomic data to perform a survival analysis, we tested whether this subset of patients had any differences in outcomes compared to the transcriptome cohort (n=528). There were no overall survival (OS) or progression-free-survival (PFS) differences between the transcriptome cohort, the subset of patients with MRI traits, and the subset of patients with all six MRI traits, see Supp. Fig. \ref{supfig_cohort_sv_km}. In building the multivariable Cox models, three clinical traits, six MRI traits, and 54 radiogenomic traits were considered, see Supp. Fig. \ref{supfig_rg_traits}.

\begin{figure}[h!]
	\caption{Overall survival (OS) and progression-free survival (PFS) dichomotized by 
		(\textbf{a}) imaging traits compared to 
		(\textbf{b}) radiogenomic traits.
		Patients split based on the association between the nCET trait and neural (NL) subtype genes had a median PFS of 0.96 years vs. 0.52 years (161 day difference). The median OS was 1.19 years vs. 0.91 years (101 day difference) when split by the nCET and chr3p21 trait, 1.18 years vs. 1.14 years (15 day difference) split by the infiltrative and chr19p13 trait, and 1.19 years vs. 0.85 years (125 day difference) split by the infiltrative and chr1p35 trait.}
	\label{fig_km}
	\centering
	\includegraphics[width=.75\textwidth]{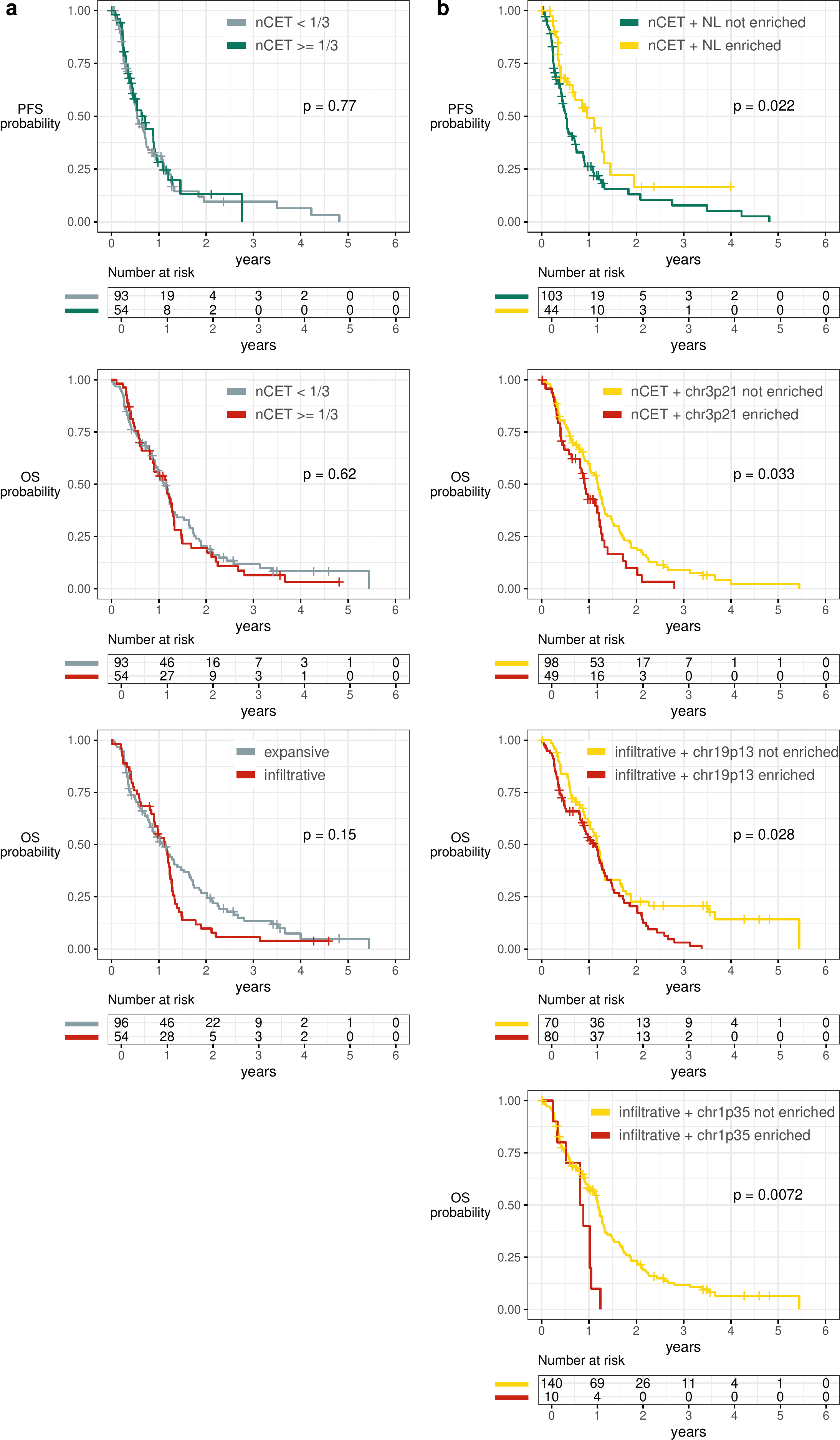}
\end{figure}

Figure \ref{fig_km} show patients had significant differences when dichotomized by their radiogenomic traits. Patients with neural genes among the most salient genes for predicting larger nCET had \textit{better} PFS compared to those who did not. Likewise, patients had significantly \textit{worse} OS when chr3p21 genes were important in predicting nCET proportion was $\geq \sfrac{1}{3}$, and when chr19p13 or chr1p35 genes were important in predicting infiltration. In contrast, dichotomizing patients based solely on individual MRI traits had no OS or PFS differences, except in the counterintuitive case of expansive vs. infiltrative, see Supp. Fig. \ref{supfig_img_split_km}. Infiltrative tumors had a univariate HR of 0.92 when estimating PFS, see Table \ref{tab_cox}. However, after adjusting for patient covariates and radiogenomic traits, infiltrative tumors had an adjusted HR of 1.61 and correctly follows the intuition that infiltrative tumors would have a higher probability of progression than expansive tumors. 

Males and non-white races had better OS and PFS, while patients diagnosed below the median age had better OS, but worse PFS. The final Cox model consisted of six significant traits, five radiogenomic traits and race when estimating OS, see Table \ref{tab_cox}. The final PFS model had five significant radiogenomic traits. In comparison, a Cox model with clinical and imaging traits had no significant factors in estimating OS or PFS. The survival analyses suggest that radiogenomic traits extracted from neural network models have prognostic value.

		
{\renewcommand{\arraystretch}{0.7}
	\setlength{\tabcolsep}{5pt}
	
	\begin{table}
		\scriptsize
		\centering
		\caption{Cox regression analysis of traits associated with overall survival and progression-free survival. Both OS and PFS multivariable models had p<0.001 in the likelihood ratio test.}
		\label{tab_cox}	
		\begin{threeparttable}
		\begin{tabular}{lllllll} 
			\toprule	
			  & \textbf{univariate HR} 	& \multicolumn{2}{c}{\textbf{adjusted HR}}& \\	
			  \cmidrule{3-4}
			  & \textbf{(95\%CI)} 	& \textbf{(95\%CI) } & \textbf{p-value}\\		
			\midrule
			Overall survival (n=127, deaths=107) \\
			\midrule
			
			\textbf{clinical} \\
			gender is male							 & 	0.95 (0.64--1.41)	& 0.80 (0.53--1.20)  & 0.280 \\
			race is white				  				& 1.69 (0.85--3.37)  	& 2.30 (1.09--4.85)& 0.029* \\
			diagnosis age is below median &  0.82 (0.56--1.21)	&  0.79 (0.53--1.20)&  0.273\\
			\addlinespace[0.4em]

			\textbf{radiogenomic} \\
			infiltrative $+$ chr1p35                                      	& 2.06  (0.98--4.31)&	2.06 (0.95--4.43) 		&  0.066\\
			edema $+$ endothelial$^{\dagger}$             &  2.07 (0.76--5.67) &  4.36 (1.47--12.89) &  0.008*\\
			necrosis $+$ GBM core astrocytes$^{\dagger}$       & 0.46 (0.14--1.49) &    0.11 (0.03--0.45) &     0.002*\\
			necrosis $+$ epithelial mesenchymal transition & 1.40 (0.80--2.43) &   3.45 (1.75--6.82) &    <0.001*\\
			nCET $+$ myogenesis                          		  & 2.85 (0.89--9.07) &	10.73 (2.48--46.51) &    0.002*\\
			necrosis $+$ MYC targets (v2)                      		& 0.76 (0.31--1.87) & 0.31 (0.10--0.98) &       0.045*\\
			infiltrative $+$ mTORC1 signaling                      & 1.87 (0.76--4.61) &	2.38 (0.93--6.10) &        0.071\\
			
			\midrule
			Progression-free survival (n=127, progressions=88) \\
			\midrule
			
			\textbf{clinical} \\
			gender is male							 & 1.03 ( 0.65--1.62)	& 0.80 (0.48--1.34)  & 0.394\\
			race is white				  				& 1.28 (0.62--2.65)	& 1.96 (0.89--4.30) & 0.094\\
			diagnosis age is below median & 0.98 ( 0.64--1.50)	&  1.25 (0.77--2.04)  &  0.373\\
			\addlinespace[0.4em]
		
			\textbf{imaging} \\
			tumor was infiltrative                            & 0.92 (0.58--1.45)		&       1.61 (0.96--2.71) & 0.072\\
			\addlinespace[0.4em]
			
			\textbf{radiogenomic} \\
			infiltrative $+$ chr1p35                                      		& 	2.06 (0.98--4.31)	&       4.20 (1.89--9.33) & <0.001* \\
			infiltrative $+$ epithelial mesenchymal transition  &   2.57 (1.23--5.39)		&      2.24 (1.27--3.96) & 0.006*\\
			necrosis $+$ MYC targets (v2)                      			& 	0.21 (0.03--1.49)	&       0.06 (0.01--0.47) & 0.008*\\
			edema $+$ fetal neurons replicating$^{\dagger\dagger}$  & 	1.48 (0.93--2.35)	&            2.50 (1.44--4.33)   & 0.001*\\
			infiltrative $+$ TGF-$\beta$ signaling                  & 	1.51 (0.80--2.86)	&        1.91 (0.94--3.92) & 0.076\\
			edema $+$ chr18p11                                     	& 	2.22 (0.80--6.15)	&        5.79 (1.87--17.99) & 0.002* \\
			edema $+$ G2M checkpoint                      		& 	0.75 (0.36--1.56)	&       0.46 (0.20--1.05) & 0.066\\
			nCET $+$ chr22q13                                     	& 	 0.35 (0.09--1.42)	&         0.36 (0.08--1.60) & 0.180\\
			infiltrative $+$ chr6q27                                      	& 	1.33  (0.76--2.33)	&        1.79 (0.98--3.30) & 0.060\\
			necrosis $+$ p53 pathway							         & 	1.06 (0.39--2.91)	&         2.44 (0.81--7.39) & 0.114\\
			
			\bottomrule	
		\end{tabular}
		\begin{tablenotes}
		\scriptsize
		\item[] *p<0.05, $^{\dagger}$gene set from \cite{Zhang2016neuron}, $^{\dagger\dagger}$gene set \cite{Darmanis2015} 
	\end{tablenotes}
	\end{threeparttable}
	\end{table}
}

\section{Discussion}

We demonstrate how deep neural networks can be used to discover radiogenomic associations. \textit{First}, we predict imaging traits using gene expression profiles, showing that our neural network-based approaches outperforms other classifiers. We also illustrate the benefit of transfer learning to train radiogenomic neural networks using a transcriptomic autoencoder modeled on a much larger cohort  to address the impedance of relatively small radiogenomic datasets. \textit{Second}, we present methods based on input masking and class saliency that facilitate interpretation of radiogenomic associations, providing a way to understand the results of an otherwise ``black box" method, which is the main criticism against neural networks. \textit{Third}, using our network analysis techniques, we identify pertinent gene expressions that may act as transcriptomic drivers for each imaging trait. We put forth a set of potential imaging surrogates that provide a clearer biological basis of commonly assessed imaging phenotypes in GBM and relate them to trends in patients' overall and progression-free survival.

Gene masking identifies cohort-level radiogenomic associations, where the strength of association was measured by the model's classification when only a subset or one gene was used. Radiogenomic associations have common themes related to major GBM candidate driver genes and terms, e.g., cell growth and vasculature. However, each MRI trait also show specificity towards components of these general themes, e.g., different  functionalities of \textit{EGFR} or cell death by autophagy in the edema model compared to apoptosis in the necrosis model. We identify unique associations between imaging traits and different themes: edema is associated with cell invasion and differentiation; enhancement is associated with immune system processes and hormones; nCET is associated with cell cycle and UV response; necrosis is associated with apoptosis; and focal is associated with cell transport and response to certain compounds.

Prior radiogenomic studies have mainly reported cohort-level associations. In reality, multiple gene expression profiles, when influenced by different environmental factors, may lead to the same observed imaging trait. Towards this end, gene saliency was used to identify patient-level radiogenomic traits.

With gene saliency, each patient has his/her own list of relevant genes for each imaging trait; it is then determined if the patient's salient genes are significantly associated with a gene set. We describe subsets of patients with common radiogenomic associations that are not apparent in gene masking, such as the association between infiltrative traits and epithelial-mesenchymal transition genes or larger nCET proportions and drug metabolism. Some of these radiogenomic traits are significant factors in predicting patient survival.

Furthermore, we validate our modeling approach by training a neural network to predict molecular subtypes. We report an experiment that evaluates model's ability to learn meaningful relationships. Not only does the subtype model achieve near-perfect classification, the model is able to select genes relevant to each subtype among 12,042 genes. In the radiogenomic models, we validate our radiogenomic associations with prior GBM  studies in radiogenomics and genomics and found corresponding relationships. We also identify new findings that have not been widely reported in radiogenomics due to the ability of gene saliency to provide patient-specific radiogenomic traits and the inclusion of the entire gene profile in our models. These results support the neural network's abilities to identify associations with existing domain knowledge and to suggest potential starting points for further investigation.

We recognize the limits of radiogenomic analysis, particularly in terms of small sample size, limited tissue sampling of a heterogeneous tumor, and limited follow-up information. Sample size is an inherent challenge in radiogenomics. TCGA tends to have the most radiogenomic data, but lacks detailed clinical data. While larger cohorts do exist, tumors are across multiple grades \cite{Aerts2014} and do not use molecular profiling \cite{Chang2018}. These limitations may be addressed as the cost of high-throughput platforms decreases and multiple tumor regions are sampled \cite{Puchalski2018}. With 528 gene expression profiles and a radiogenomic subset of 175, we show that neural networks can model transcriptomic heterogeneity to reflect phenotypic differences in imaging. The VASARI feature set is also limited in that it provides a gross categorization of imaging features and only one experienced reader's annotations of the imaging data is used. A more comprehensive analysis that includes quantitative (radiomic) traits may be warranted. Finally, the radiogenomic associations are only hypothesized and not proven through experimentation, though we attempt to compare our discovered associations with those that have been previously reported in literature. To validate the identified associations, cell and animal studies would allow controlled experiments between genes and imaging phenotypes \cite{Zinn2018}.

\section{Conclusion}
Using a neural network-based approach to radiogenomic mapping, we highlight the representational and discriminative capacity of neural networks to model the high-dimensional, nonlinear, and correlative nature of gene expressions to predict typical GBM imaging traits. We demonstrate the use of neural network interpretation techniques, e.g., input masking and class saliency to understand what the model has learned and to extract relevant radiogenomic associations. The learned radiogenomic associations may point to potential transcriptomic drivers of imaging traits and could further clarify the understanding of the relationship between two often disjoint datasets, gene expression profiling and medical imaging. As such, prognostication and treatment options may be further individualized, where a targeted pathway could be considered in the selection of an appropriately tailored chemotherapeutic agent.

\singlespacing




\section{Acknowledgements}
We are thankful for the feedback from the faculty and students of the Medical \& Imaging Informatics group, discussions about performance differences with biostatisticians, Drs. James Sayre and Audrey Winter, and correspondences with TCIA members to better understand the data.

\section{Funding}
This work was supported in part by the National Institutes of Health [\textbf{F31CA221061} and  \textbf{T32EB016640} to N.F.S., \textbf{R01CA157553} to N.F.S., W.H., S.E.S.]; the National Science Foundation [\#\textbf{1722516} to W.H.]; Amazon Web Services and UCLA Department of Computational Medicine partnership to W.H.; and the Integrated Diagnostics Program, jointly funded by the Department of Radiological Sciences and Pathology \& Laboratory Medicine to N.F.S., W.H.. \textit{Conflict of Interest:} none declared.


\bibliography{library}

\newpage
\nolinenumbers
\setcounter{page}{1}
\setcounter{figure}{0} \renewcommand{\thefigure}{S\arabic{figure}}
\setcounter{table}{0} \renewcommand{\thetable}{S\arabic{table}}

\section*{Supplemental data}

\section*{Dataset}
Supplemental Materials contains the following data used in this study:
\begin{itemize}
	\item \texttt{gene\_expression.txt} - Affymetrix transcriptomes
	\item \texttt{gene\_expression\_ids.txt} - patients' TCGA barcodes
	\item \texttt{TCGA\_clinical\_data} - TCGA-GBM clinical data
	\item \texttt{TCGA\_unified\_CORE\_ClaNC840.txt} - molecular subtype and gene sets from Verhaak et al.
	\item \texttt{vasari\_annotations.csv} - MRI traits
\end{itemize}

Generated data and models can be found in the paper's source code on Github.

{\renewcommand{\arraystretch}{0.7}
	\setlength{\tabcolsep}{5pt}
	
	\begin{table}[h!]
		\small
		\centering
		\caption{Patient characteristics with transcriptomes.  Transcriptomes were analyzed with
			Affymetrix HT Human Genome U133 Arrays by the Broad Institute. The quantile normalized
			and background corrected (GenePattern platform; Level 3) data were downloaded from the
			Genomic Data Commons using the legacy data portal at https://gdc.cancer.gov/ .  The radiogenomic subset is based on patients with any pre-op MRI study. The Verhaak et al. study was based on unified gene expressions from multiple platforms. Here, we took gene expression profiles measured on the same an Affymetrix platform as the radiogenomic models to create the subtype subset. }
		\label{tab:demo}
	\begin{threeparttable}	
		\begin{tabular}{lll lll} 
			\toprule
			& 				&			& \multicolumn{3}{c}{\textbf{modeling datasets}} \\
			\cmidrule(r){4-6}
			&				& \textbf{total} & autoencoder	& radiogenomic & subtype\\
			\midrule
			number of patients	& &	528			& 353			&	175			  & 171 \\
			\\
			diagnosis age (years)	
			& min. 			&	10 			& 10			&	14			  & 17\\
			& mean			&	57.9 		& 57.1			& 	59.4		  & 56.3\\
			& med.			&	59 			& 58			&	60.5		  & 57.5\\
			& max.			&	89 			& 89			&	86			  & 86\\
			\\
			vital status	
			& deceased		& 388 			& 250			& 138			 & 146 \\
			& alive			& 113  			& 85			& 28 			  & 8\\
			& n/a			& 27  			& 18			& 9 			  & 17\\						
			\\
			gender  			
			& female 		& 196 			& 134			&	62			  & 56\\
			& male			& 306			& 202			&	104			  & 98\\
			& n/a			& 26			& 17			&	9			  & 17\\			
			\\
			race				
			& white			& 443 			& 302			& 141 			  & 135\\
			& african		& 29 			& 17			& 12 			  & 6\\
			& asian			& 11			& 7				& 4 			  & 5 \\
			& n/a			& 45 			& 27			& 18			  & 25\\	
			\\
			ethnicity 			
			& hispanic 		& 12			& 9				&	3			  & 5\\
			& not hispanic 	& 415			& 278			&	137			  & 141\\
			& n/a 			& 101 			& 66			&	35			  & 25\\
			\\
			KPS				
			& min. 			& 20  			& 20			& 40			  & 40\\
			& mean 			& 77  			& 76			& 78			  & 82\\
			& med. 			& 80  			& 80			& 80			  & 80\\
			& max. 			& 100  			& 100			& 100			  & 100\\
			
			diagnosis method	
			& biopsy		& 63  			& 46			& 17 			  & 6\\
			& resection		& 435  			& 288			& 147 			  & 147\\
			& other			& 2  			& 1				& 1 			  & 0\\
			& n/a			& 28  			& 18			& 10 			  & 18\\
			\\
			molecular subtype 	
			& classical				& 38			& 23			& 15			  & 38\\
			& mesenchymal	& 54			& 26			& 28			  & 54\\
			& neural				& 26			& 12			& 14			  & 26\\
			& proneural			& 53			& 30			& 23			  & 53\\
			& n/a					& 357			& 262			& 95 			  & 0\\
			\bottomrule	
\end{tabular}
\begin{tablenotes}
	\scriptsize
	\item[] KPS (Karnofsky Performance Score), n/a (not available)
\end{tablenotes}
\end{threeparttable}
\end{table}
}

{\renewcommand{\arraystretch}{0.7}
	\setlength{\tabcolsep}{3pt}
	
	\begin{table}[h!]
		\small
		\centering
		\caption{Classification labels in the radiogenomic and subtype datasets. Based on personal communication with TCIA, the most recent imaging studies without signs of surgery or biopsy were estimated to be images closely associated with time of tissue sampling, as pre-op imaging was necessary for surgical planning. (GBM patients usually undergo surgical resection to remove the bulk of the tumor.) Proportion labels were binarized using $\geq \sfrac{1}{3}$ as the threshold due to small numbers in other categories. Likewise, subcategories of multifocal or multicentric (24 patients) and gliomatosis (2 patients) were combined into one class, non-focal.}
		\label{tab:labels}	
		\begin{tabular}{lllll} 
			\toprule	
			\textbf{n} 	& \textbf{trait} & \textbf{description} 					& \textbf{values}  	& \textbf{\# (\%)}   \\
			\toprule
			
			\multirow[t]{2}{*}{175} & \multirow[t]{2}{*}{surgical} 	& evidence of prior surgery or biopsy in the    	& pre-op	& 175 (100\%) \\
			&							& earliest imaging study							& post-op 	&  0 (0\%) \\
			[1.5ex]
			\multirow[t]{2}{*}{166} & \multirow[t]{2}{*}{f5} 	& proportion of tumor estimated to be enhancing   	& $< \sfrac{1}{3}$   	& 95 (57\%)\\
			&  							&    				& $\geq \sfrac{1}{3}$ 	& 71 (43\%) \\
			[1.5ex]
			\multirow[t]{2}{*}{156} & \multirow[t]{2}{*}{f6}  	& proportion of tumor estimated to be    & $< \sfrac{1}{3}$		& 99 (63\%)\\
			&							& non-contrast enhancing and not edema	& $\geq \sfrac{1}{3}$ 	& 57 (37\%) \\ 
			[1.5ex]
			\multirow[t]{2}{*}{167} & \multirow[t]{2}{*}{f7} 	& proportion of tumor estimated to be necrosis   & $< \sfrac{1}{3}$  		& 116 (70\%) \\
			&							&  					& $\geq \sfrac{1}{3}$		& 51 (30\%) \\
			[1.5ex]
			\multirow[t]{2}{*}{161} & \multirow[t]{2}{*}{f9}  	& lesions outside of main tumor and its edema:  &	& \\
			&							&  a) none 										& focal 	& 135 (84\%)\\
			&							&  b) spread via dissemination 		& non-focal	& \multirow[t]{3}{*}{26 (16\%)} \\
			&							&   or in majority of a hemisphere	&	& \\
			[1.5ex]
			\multirow[t]{2}{*}{159} & \multirow[t]{2}{*}{f10}  	& ratio of abnormality sizes in T1 and FLAIR:		& & \\
			&							&  a) T1 $\approx$ FLAIR 	& expansive 	& 103 (65\%)\\
			&							&  b) T1 $<$ FLAIR  or  T1 $<<$ FLAIR		& infiltrative 		& 56 (35\%) \\
			[1.5ex]
			162 & f14   & proportion of tumor estimated to be edema   	& $< \sfrac{1}{3}$   	& 86 (53\%) \\
			&		&												& $\geq \sfrac{1}{3}$	& 76 (47\%)\\
			[1.5ex]
			171 & subtype &  molecular subtypes  defined by Verhaak et al. & classical 	& 38 (22\%)\\
			&		&														& mesenchymal 	& 54 (32\%)\\
			&		&														& neural	& 26 (15\%)\\
			&		&														& proneural	& 53 (31\%)\\
			\bottomrule	
		\end{tabular}
	\end{table}
}
\clearpage
\begin{figure}[h!]
	\small
	\caption{Label association testing with Fisher's exact test (a) within labels and (b) with clinical traits in R using \texttt{fisher.test}. P-values were adjusted using Bonferroni correction in \texttt{p.adjust}. Multi-class labels do not have estimates, i.e, odds ratios. For binary labels, labels ``{$<$ 1/3}", ``focal", and ``expansive" were class 1 and ``$\geq$ 1/3", ``non-focal", ``infiltrative" were class 2. 
		Rows are the explanatory variable. E.g., the odds of nCET $<$ 1/3 and an expansive tumor were 3.353 times higher than nCET $>$ 1/3 and an expansive tumor. 
		Estimates above 1 indicated positive correlations, and vice versa. The traits infiltrative was positively correlated with the traits focal and nCET proportions. nCET was negatively correlated with edema and necrosis. No other associations were significant.
		Diagnosis age was dichotomized based on the mean diagnosis age of 58, a value calculated from all patients whose diagnosis age was known. Patients with MRI traits had tissue samples obtained from eight different sites (site codes: 02, 06, 08, 14, 19, 27, and 76). Karnofksy performance scores included 40, 60, 80, and 100; the higher the better.}
	\label{supfig:label_fishertest}
	\centering
	\includegraphics[width=\textwidth]{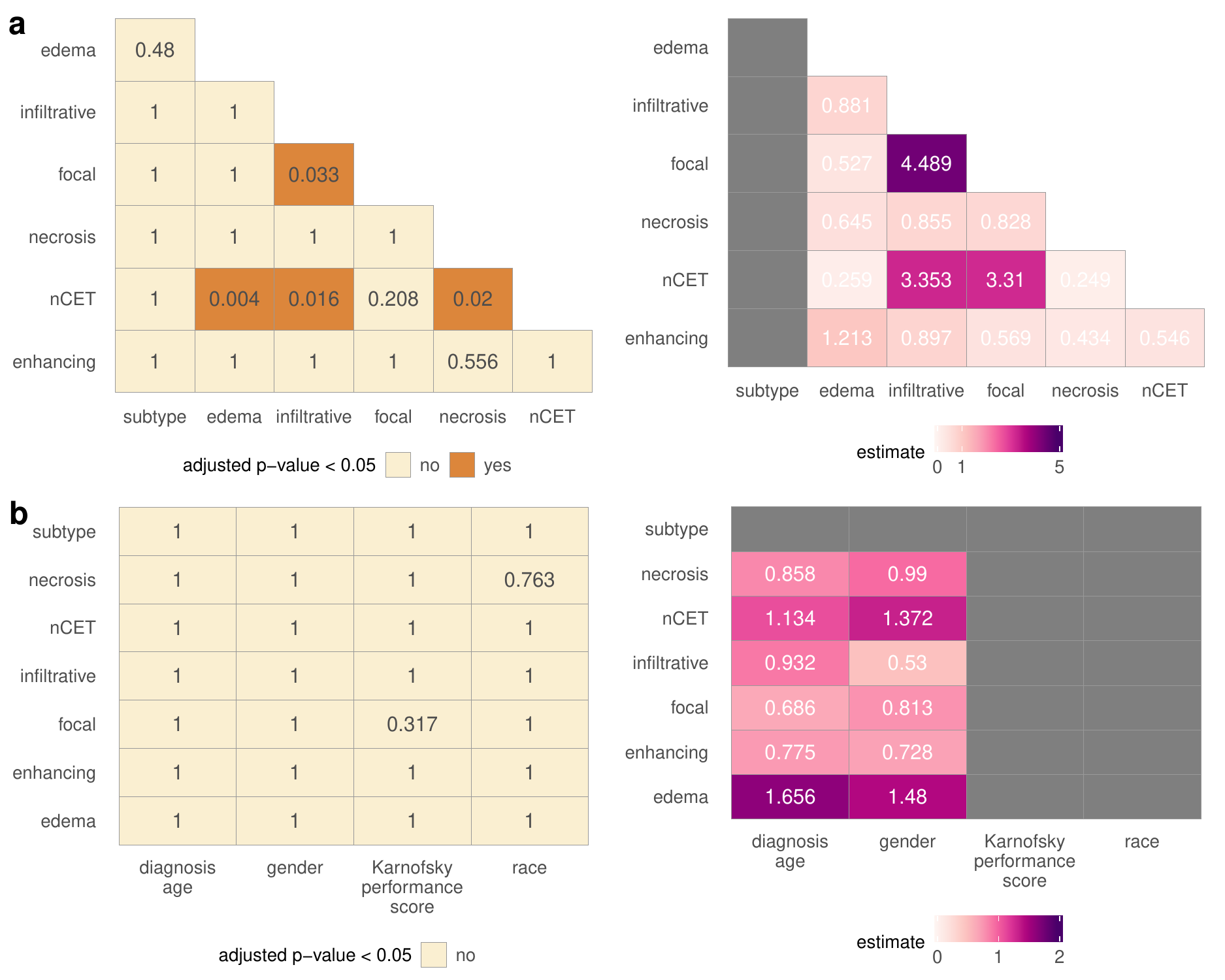}
\end{figure}

\newpage
\section*{Models}

Radiogenomic neural networks were fully connected, had a batch size of ten. Regularization with dropout was used, where the same dropout rate was applied across the input and all hidden layers. For binary classes, 0 and 1 labels, binary cross-entropy loss and sigmoid activation in the prediction layer was used. Since subtypes were multi-classes, the subtype neural networks instead used categorical cross-entropy loss and softmax activation in the prediction layer.

Transcriptomic autoencoders considered five different architectures of three encoding layers: the first hidden layer was either 1000, 2000, 3000, or 4000, where subsequent encoding layers decreased by half each time.  The autoencoder used mean absolute error as the loss function, a batch size of 50, a patience of 200 epochs while monitoring validation $R^2$, and no dropout. 

All neural networks used batch normalization in each hidden layer and a maximum of 500 epochs was set. Due to the inherent class imbalance of imaging traits, sample weighting based on class size and stratified fold splitting were used.

{\setlength{\tabcolsep}{2.8pt}	
	\begin{table}[h!]
		\small
		
		\caption{Radiogenomic models and hyperparameters.}
		\label{tab:models}	
		\begin{threeparttable}
			\begin{tabular}{lllll} 
				
				\toprule
				\textbf{model type} 			& \textbf{no. models }	& \textbf{hyperparameter}	& \textbf{values} \\ 
				\midrule
				
				\multirow[t]{3}{*}{logistic regression} & 4000	&
				penalty type	& L1, L2 \\
				&	& C	penalty		& log(3) - log(1) \\
				& 	& solver		& liblinear, newton-cg,\\
				& 	& 				& lbfgs,  sag, saga \\
				[1.5ex]
				
				\multirow[t]{3}{*}{support vector machines} & 4000	& 
				kernel			& linear, poly, rbf, sigmoid \\
				&	& C	penalty			& log(-6)-log(1)\\
				[1.5ex]
				
				\multirow[t]{3}{*}{random forest} &	1200	&
				trees		& [50: 50: 2000] \\ 
				&	& split criterion		& Gini, entropy \\
				&	& max. features	& $G, \sqrt{G},\log_2(G)$ \\
				&	& max. depth	& None, [1-4]\\
				[1.5ex]
				
				\multirow[t]{3}{*}{gradient boosted trees} & 760	&
				trees	& [50: 50: 1000) \\
				&	& max. depth	& [1-4]\\
				&	& learning rate	& [0.01: 0.05: 0.50] \\  
				[1.5ex]
				
				\multirow[t]{4}{*}{neural network,} & 40 & 	hidden layers 		& 3 \\
				autoencoder 
				& 	& hidden nodes		 	& [4000-250]\\
				& 	& architectures	& 4\\
				&	& optimizer 	& Nadam, Adadelta\\
				&	& activation 	& sigmoid, tanh, relu\\
				&	& dropout		& [0.0:0.2:0.6] \\
				&	& loss				& binary cross-entropy, mean absolute error \\
				&	& epochs		& 200, 500 \\
				&	& patience		& 200 epochs \\
				&	& batch			& 10, 50 \\
				&	& weight initializer		& Glorot normal, autoencoder \\
				&	& no. layers frozen 		& 0, 1, 2 \\
				\bottomrule	
			\end{tabular}
			\begin{tablenotes}
				\scriptsize
				\item[] G: number of genes
			\end{tablenotes}
		\end{threeparttable}
	\end{table}
}

\begin{table}[h!]
	\small
	\centering
	\caption{Subtype neural network hyperparameters.}
	\label{tab:subtype_hyp}	
	\begin{tabular}{lllll} 
		
		\toprule
		\textbf{model} 			& \textbf{no. models }	& \textbf{hyperparameter}	& \textbf{values} \\ 
		\midrule
		
		\multirow[t]{4}{*}{neural network} & 90 & 	hidden layers 		& 3 \\ 
		& 	& hidden nodes		 	& [4000-125]\\
		& 	& architectures	& 5\\
		&	& optimizer 	& Nadam, Adadelta\\
		&	& activation 	& sigmoid, tanh, relu\\
		&	& dropout		& [0.4:0.2:0.8] \\
		&	& loss			& categorical cross-entropy\\
		&	& epochs		& 200 \\
		&	& batch		& 10 \\
		&	& weight initializer		& Glorot normal\\
		\bottomrule	
	\end{tabular}
\end{table}

\clearpage
\section*{Model performance}
\subsection*{Autoencoder}
\begin{figure}[h!]
	\caption{Performance of transcriptomic autoencoder in 
		(\textbf{top}) 10-fold cross-validation grid search, and
		(\textbf{bottom}) $R^2$ distribution after retraining. Architecture refers to the hidden nodes in the three encoding layers.}
	\label{supfig:ae_perf}
	\centering
	\includegraphics[width=.8\textwidth]{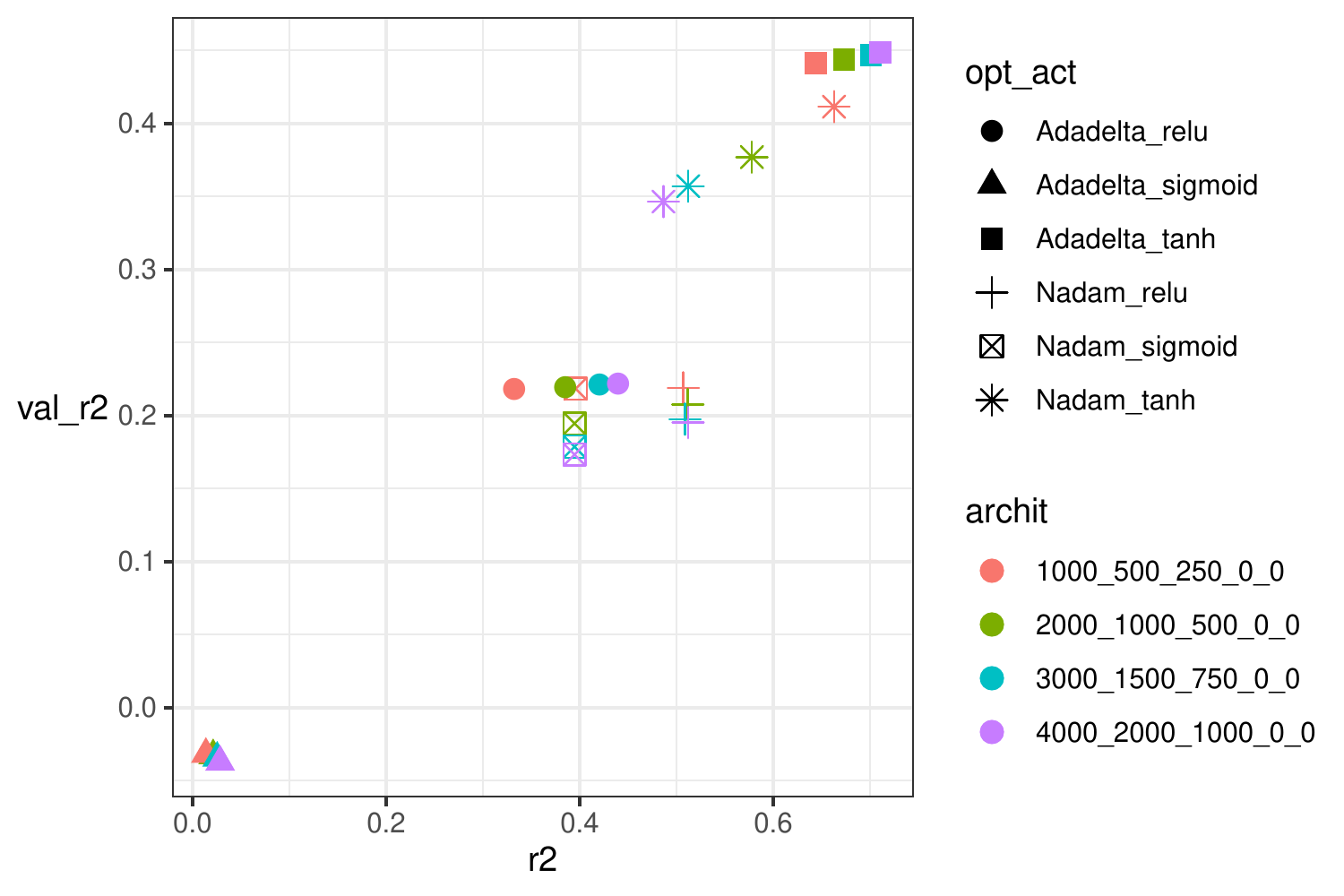}
	\includegraphics[width=.8\textwidth]{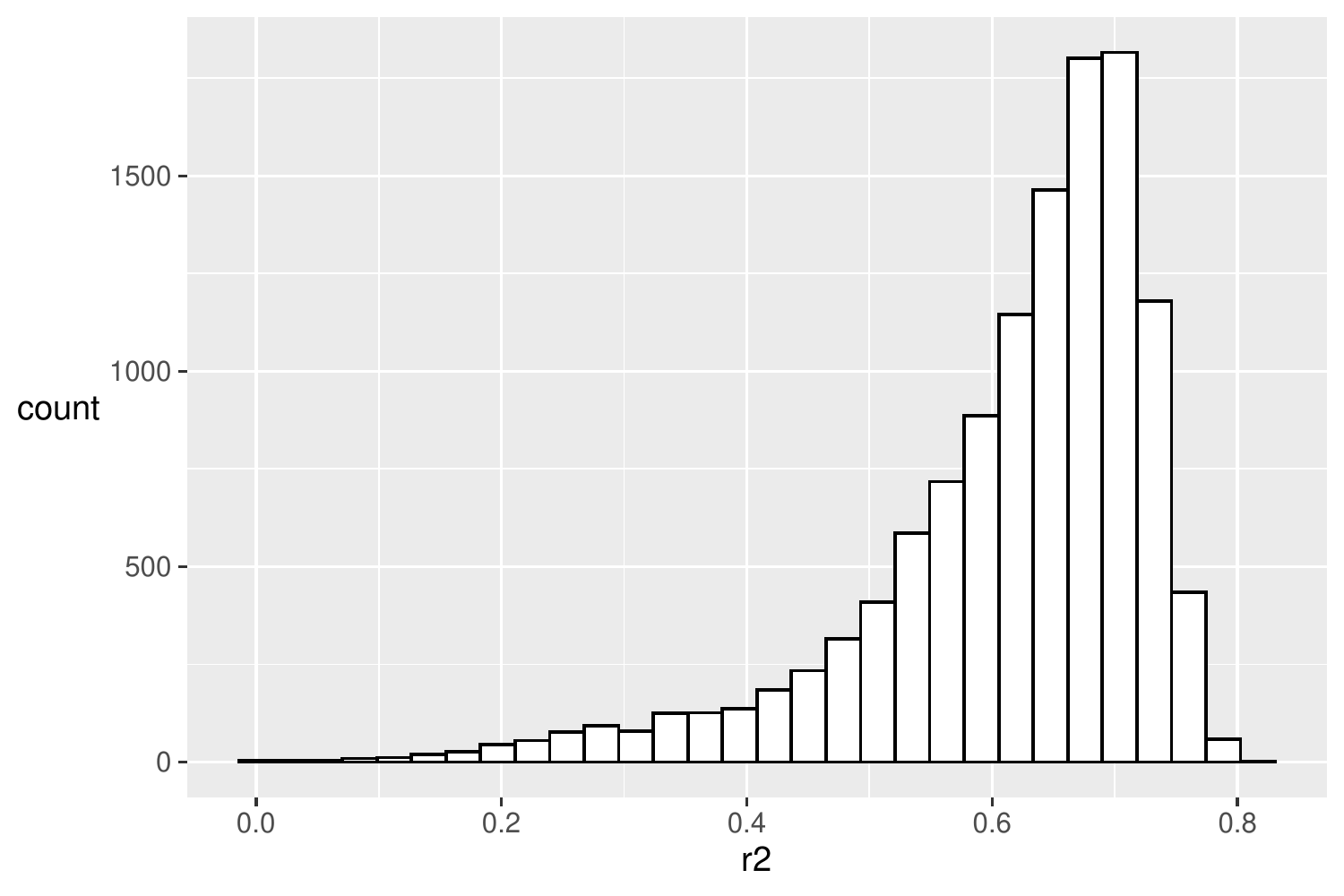}
\end{figure}

\clearpage
\subsection*{Radiogenomic models}

{\setlength{\tabcolsep}{6pt}
	
	\begin{table}[h!]
		\small
		\centering
		\caption{Radiogenomic model performances. For each model, the best performing hyperparameters were selected based on cross-validation AUC, and their results are shown here.}
		\label{tab:model_perf}
		\begin{threeparttable}
			\begin{tabular}{l llll llll  } 
				\toprule
				& \textbf{label}&	\textbf{name}	&	\textbf{nn} & 	\textbf{gbt}&	\textbf{rf}&	\textbf{svm}&	\textbf{logit}\\
				\midrule
				\multirow{6}{*}{\textbf{AUC}}	
				& f5&	enhancing	&	\textbf{0.722}	&	0.511&	0.542&	0.571&	0.538\\
				& f6&	nCET 		&	\textbf{0.826}	&	0.722&	0.721&	0.667&	0.617\\
				& f7&	necrosis	&	\textbf{0.751}	&	0.638&	0.627&	0.607&	0.553\\
				& f14&edema			&	\textbf{0.784}	&	0.673&	0.647&	0.613&	0.517\\
				& f10&infiltrative	&	\textbf{0.780}	&	0.492&	0.545&	0.617&	0.573\\
				& f9&	focal		&	\textbf{0.849}	&	0.728&  0.697&	0.694&	0.650\\		
				\midrule
				\multirow{6}{*}{\textbf{$\Delta$AUC} (nn - another model)}	
				& f5&	enhancing		&	-	&	0.211&	0.180&	0.151&	0.184
				\\
				& f6&	nCET			&	-	&	0.104&	0.105&0.159&	0.209
				\\
				& f7&	necrosis		&	-	&	0.113&	0.124&	0.144&	0.198\\
				& f14&edema				&	-	&	0.111&	0.137&	0.171&	0.267\\
				& f10&infiltrative		&	-	&	0.288&	0.235&	0.163&	0.207\\
				& f9&	focal			&	-	&	0.121&	0.152&	0.155&	0.199\\	
				\bottomrule	
			\end{tabular}
			\begin{tablenotes}
				\scriptsize
				\item[] nCET (non-contrast enhancing tumor), neural network (nn), gradient boosted trees (gbt), random forest (rf), support vector machines (svm), logistic regression (logit)
			\end{tablenotes}
		\end{threeparttable}
	\end{table}
	
}

\clearpage
\subsection*{Subtype neural network}
{\setlength{\tabcolsep}{6pt}
	\begin{table}[h!]
		\small
		\centering
		\caption{Cross-validation results of selected hyperparameters for subtype neural network. Values are in AUC and averaged over 10 folds. Individual subtype AUCs were calculated based on one-versus-others.}
		\label{tab:suptype_best_cv}	
		\begin{tabular}{lll lll lll lll l} 
			\toprule
			& \textbf{classical} & \textbf{mesenchymal} & \textbf{neural} & \textbf{proneural} & \textbf{micro-averaged} & \textbf{macro-averaged} \\
			\midrule
			training	& 0.9965&	0.9980&	1&	0.9974&	0.9964&	0.9983\\
			validation &	0.9841 &	0.9974&	1&	0.9910&	0.9938&	0.9956 \\
			\bottomrule	
		\end{tabular}
	\end{table}
	
	\begin{table}[h!]
		\small
		\centering
		\caption{Performance scores of subtype model in gene masking with various gene sets, corresponds to Fig. \ref{fig:subtype_subtype_pert} in main text. Subtype gene sets were defined by \cite{Verhaak2010}. Random gene sets were obtained by random sampling of genes, excluding subtypes genes. The fully trained subtype model had perfect classification, as expected. The model was able to retain high performance scores when only using subtype genes.}
		\label{tab:subtype_subtype_score}	
		\begin{tabular}{lll ll} 
			\toprule
			\textbf{gene set}	&	\textbf{gene size}& \textbf{AUC}	&	\textbf{f1-score}	& \textbf{average precision}\\
			\toprule
			
			random 			&	100	&	0.741	&	0.485	&	0.528 \\
			random			&	200	&	0.829	&	0.643	&	0.676 \\
			mesenchymal		&	216	&	0.897	&	0.725	&	0.789 \\
			all subtypes	&	840	&	0.994	&	0.930	&	0.984 \\
			
			\bottomrule	
		\end{tabular}
	\end{table}
}

\clearpage
\section*{Bootstrapped performances}
\begin{figure}[h!]
	\caption{Comparison of neural network performance compared to other models in 100 bootstrapped datasets. Points centered below the diagonal line indicate cases where neural networks had better 10-fold cross-validation performance, and was the case for all bootstrapped datasets. For each bootstrap, the difference was equal to the neural network performance minus another model's performance. Notation - gbt: gradient boosted trees, rf: random forest, svm: support vector machines, logit: logistic regression.}
	\label{supfig:boot_comparison}
	\centering
	\includegraphics[width=\textwidth]{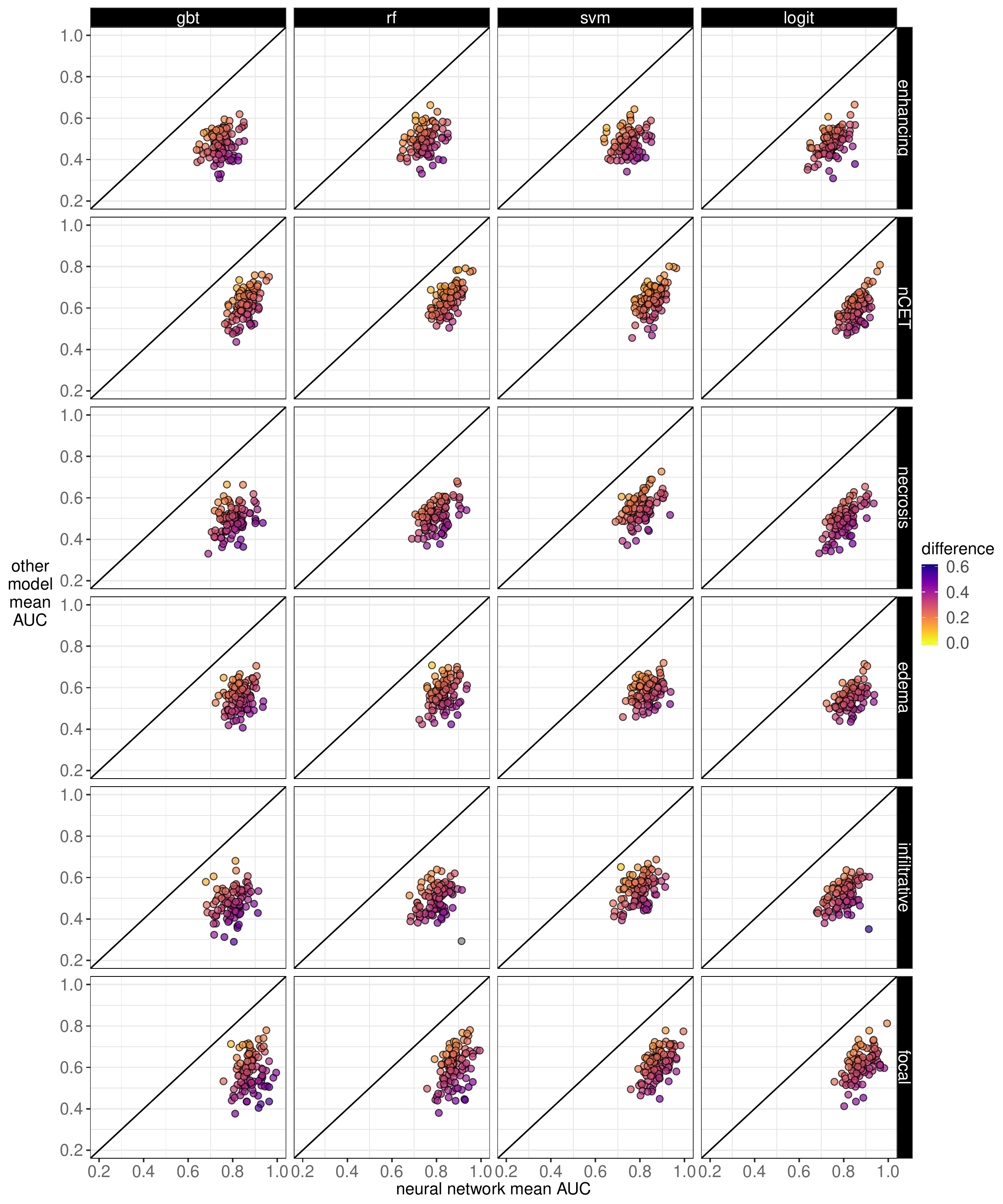}
\end{figure}

\begin{figure}[h!]
	\caption{Distribution of 10-fold cross-validation performances in 100 bootstrapped datasets across models. Dashed vertical lines represent 95\% confidence intervals (CI). Solid vertical lines indicate 0.5 AUC - classification was as good as random. Confidence intervals did not overlap between neural networks and any other model and suggested neural networks were better models.}
	\label{supfig:boot_individual}
	\centering
	\includegraphics[width=\textwidth]{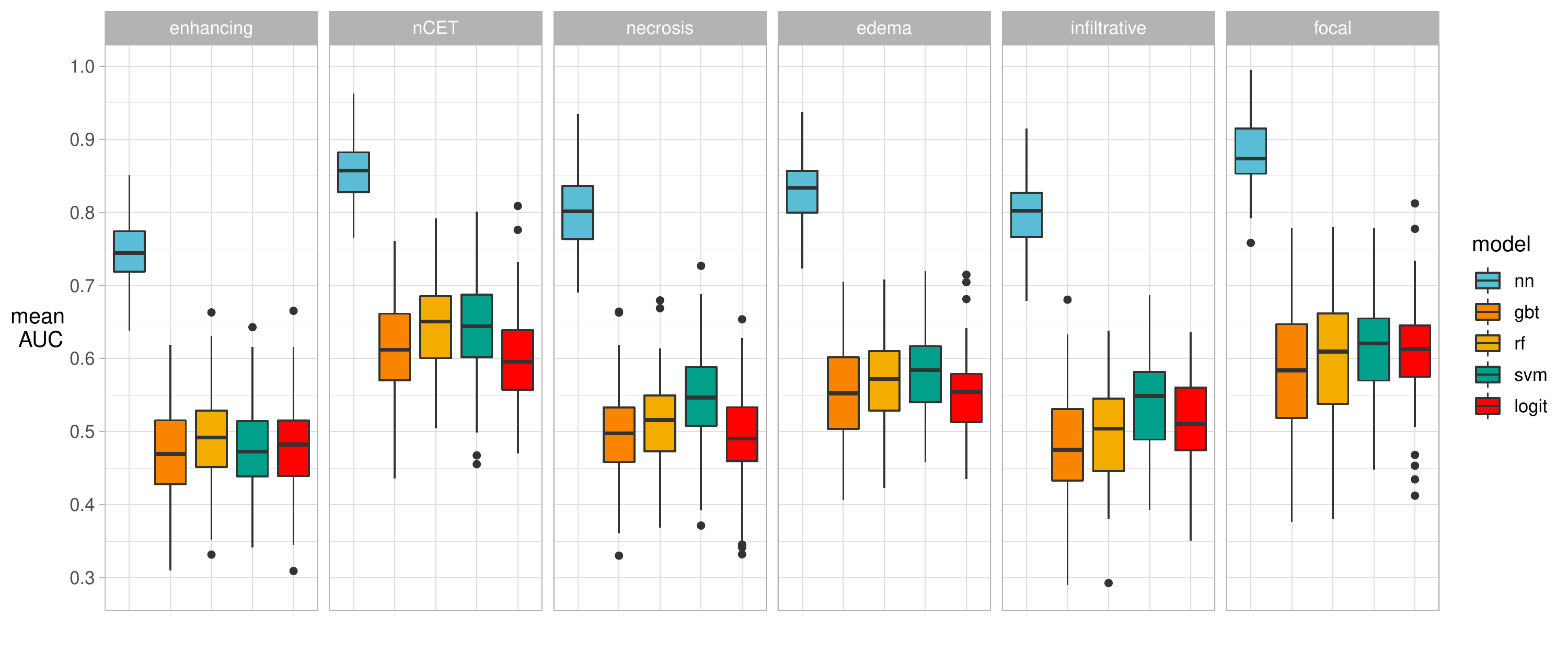}
	\includegraphics[width=\textwidth]{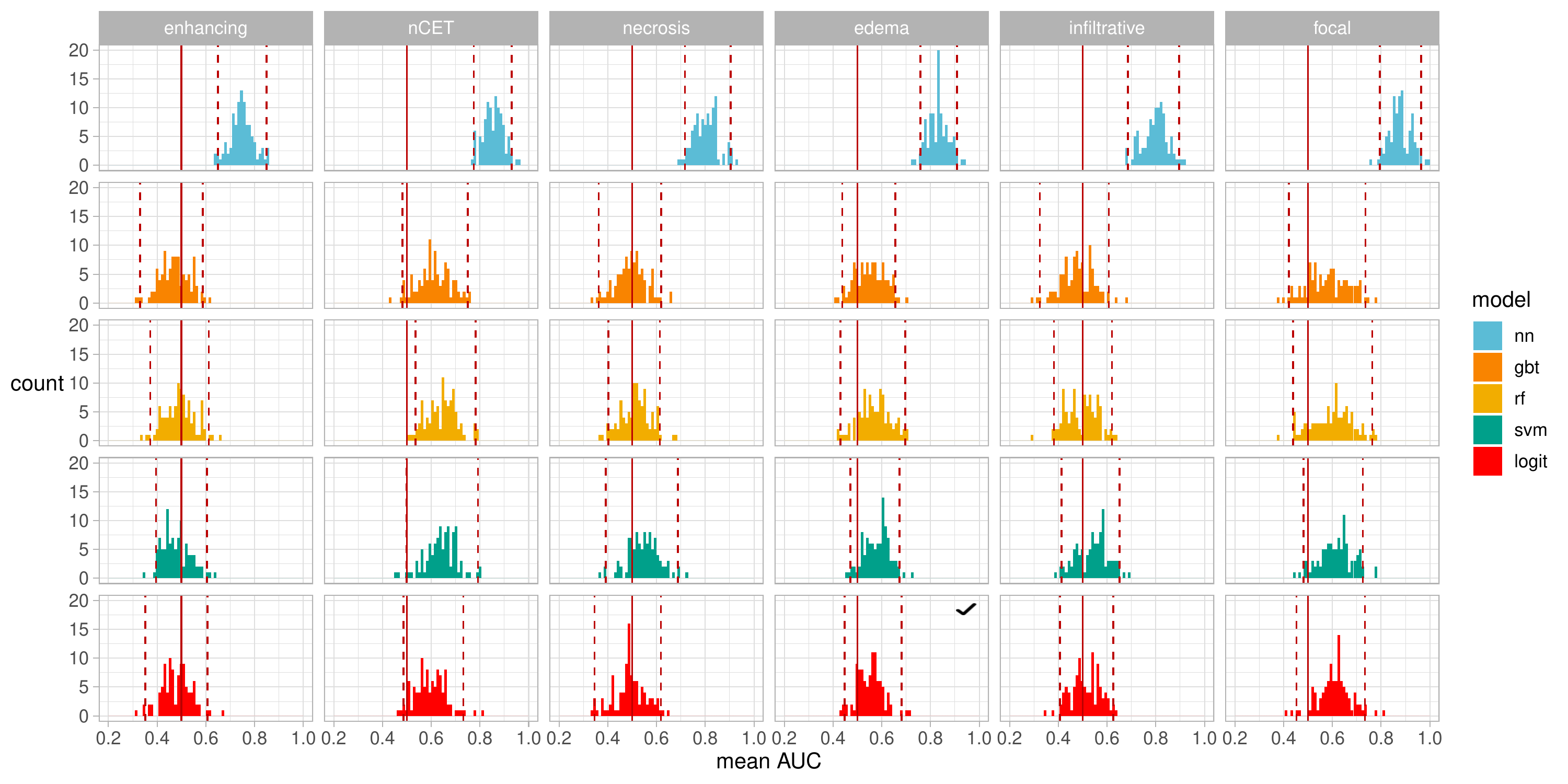}
\end{figure}

\begin{figure}[h!]
	\caption{Distribution of training and validation of neural networks in the 100 bootstrapped datasets. The 10-fold cross-validation means: 
		(\textbf{a}) number of  epochs  
		(\textbf{b}) loss,
		(\textbf{c}) AUC, 
		(\textbf{d}) average precision. Values were recorded at the epoch where highest performance was reached in early stopping. Grey intervals show the standard deviation of a value within a 10-fold cross-validation. Bootstraps were sorted by validation values for visualization purposes. }
	\label{supfig:boot_nn_metrics}
	\includegraphics[width=\textwidth]{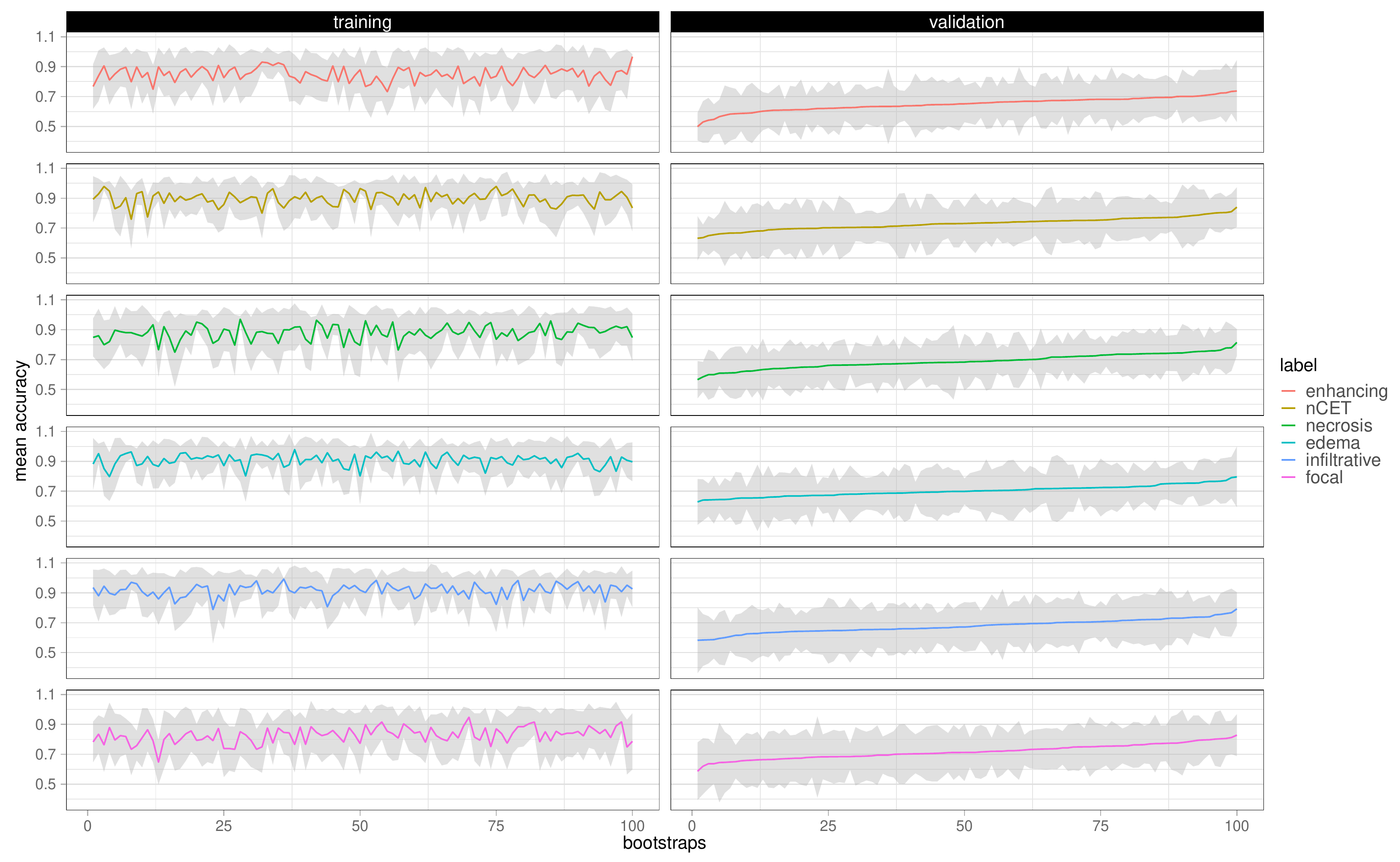}
	\centering
\end{figure}	

\clearpage
\section*{Gene masking}

 As gene set size increased, both AP and AUC increased. Although the full input, i.e., 12,042 gene expressions, was much larger, the models mapped multiple subsets of the input without losing a large proportion of classification performance. This was likely reflected the correlative nature between gene expressions, e.g., previous work with neural networks from Chen et al. was able to use landmark genes to predict another 9520 target genes \cite{Chen2016}.
 
\subsection*{Subtype neural network}

\begin{figure}[h!]
	\caption{Subtype gene set masking with subytpe model. 
		(\textbf{top}) Model's probabilities and, 
		(\textbf{bottom}) model performances. Subtypes and their gene sets were taken from \cite{Verhaak2010}. Random genes were randomly sampled and did not overlap with subtype genes.}
	\label{supfig:subtype_verhaak}
	\includegraphics[width=\textwidth]{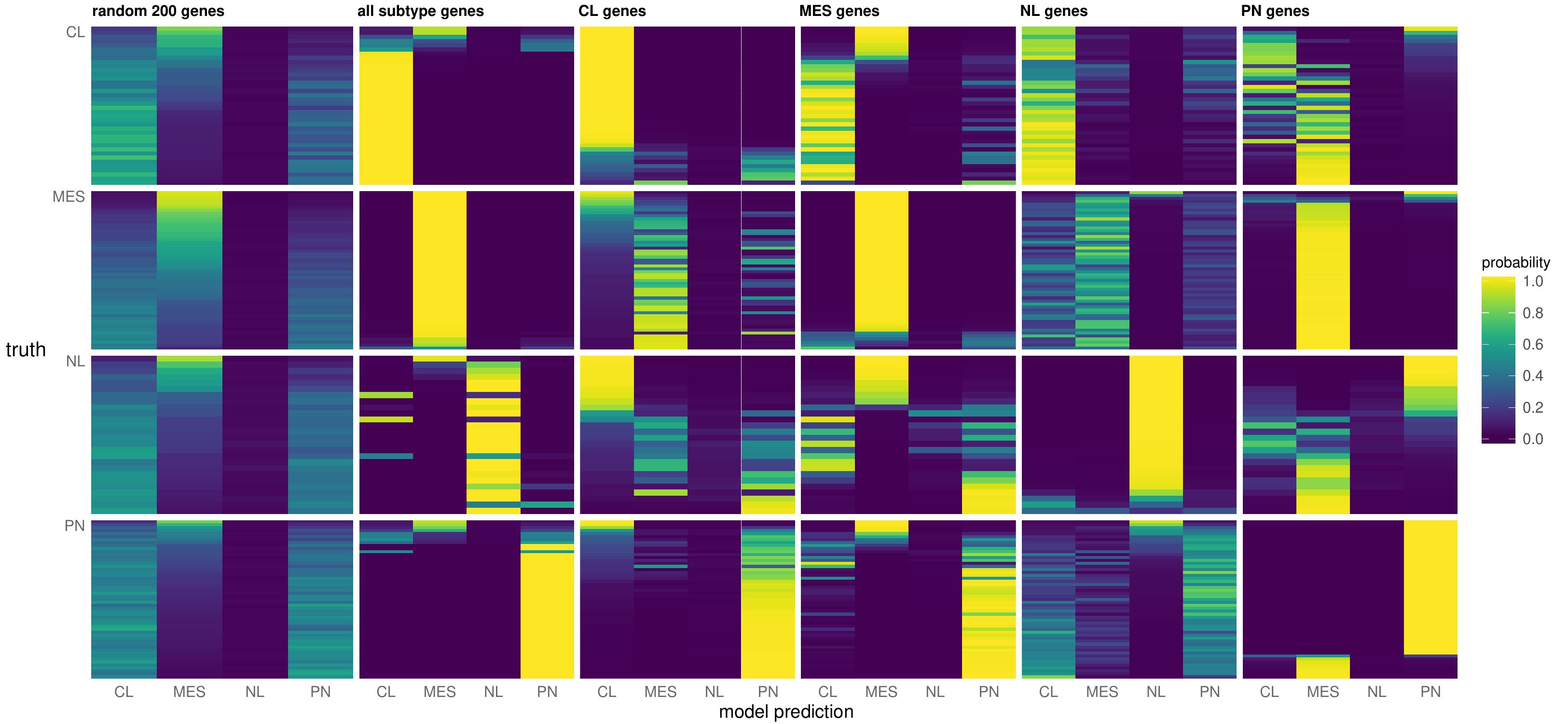}
	\includegraphics[width=.8\textwidth]{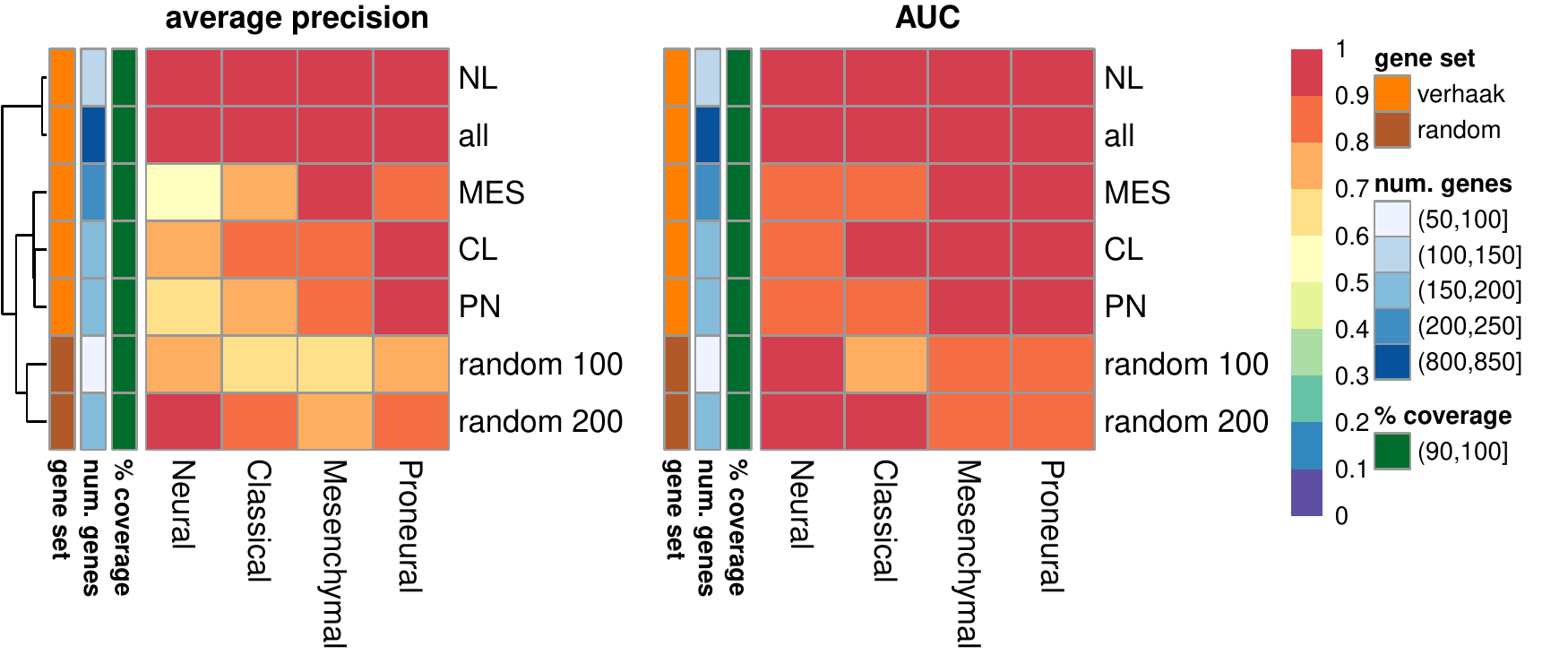}
	\centering
\end{figure}


\begin{figure}[h!]
	\caption{Perturbation of the subtype model with gene sets describing cell types and phenotypes \cite{Zhang2016neuron, Darmanis2015, Patel2014}, downloaded from Puchalski et al. \cite{Puchalski2018}.
		(\textbf{top}) Model performance in gene set masking and
		(\textbf{bottom}) gene set enrichment in genes ranked by single gene masking.
		For comparison, see the paper's Fig. S6. Neither rows nor columns were clustered in order mtach the order in the paper. Cell type enrichment for subtypes are similar to the gene masking scores for neural, proneural and mesenchymal findings in the paper. A major difference includes the subtype neural network's ability to predict with high AUC and precision the mesenchymal subtype with the GBM core astrocyte gene set, but this was not shown in the paper. The study found neural and proneural subtypes are often enriched by astroctye gene sets; mesenchymal associated with endothelial cells; and proneural with oligodendrocytes and quiescent fetal neurons. However, the associations between the study's and the neural network's gene masking do not completely agree. This disagreement was likely due to different goals: the study measures gene set associations based on single sample gene set enrichment analysis and gene masking was based on classification performance. Although, the overlap does show consistency between GBM subtypes and brain cell types, suggested that cell types were both enriched and predictive of the subtypes.}
	\label{supfig:subtype_puchalski}
	\includegraphics[width=\textwidth]{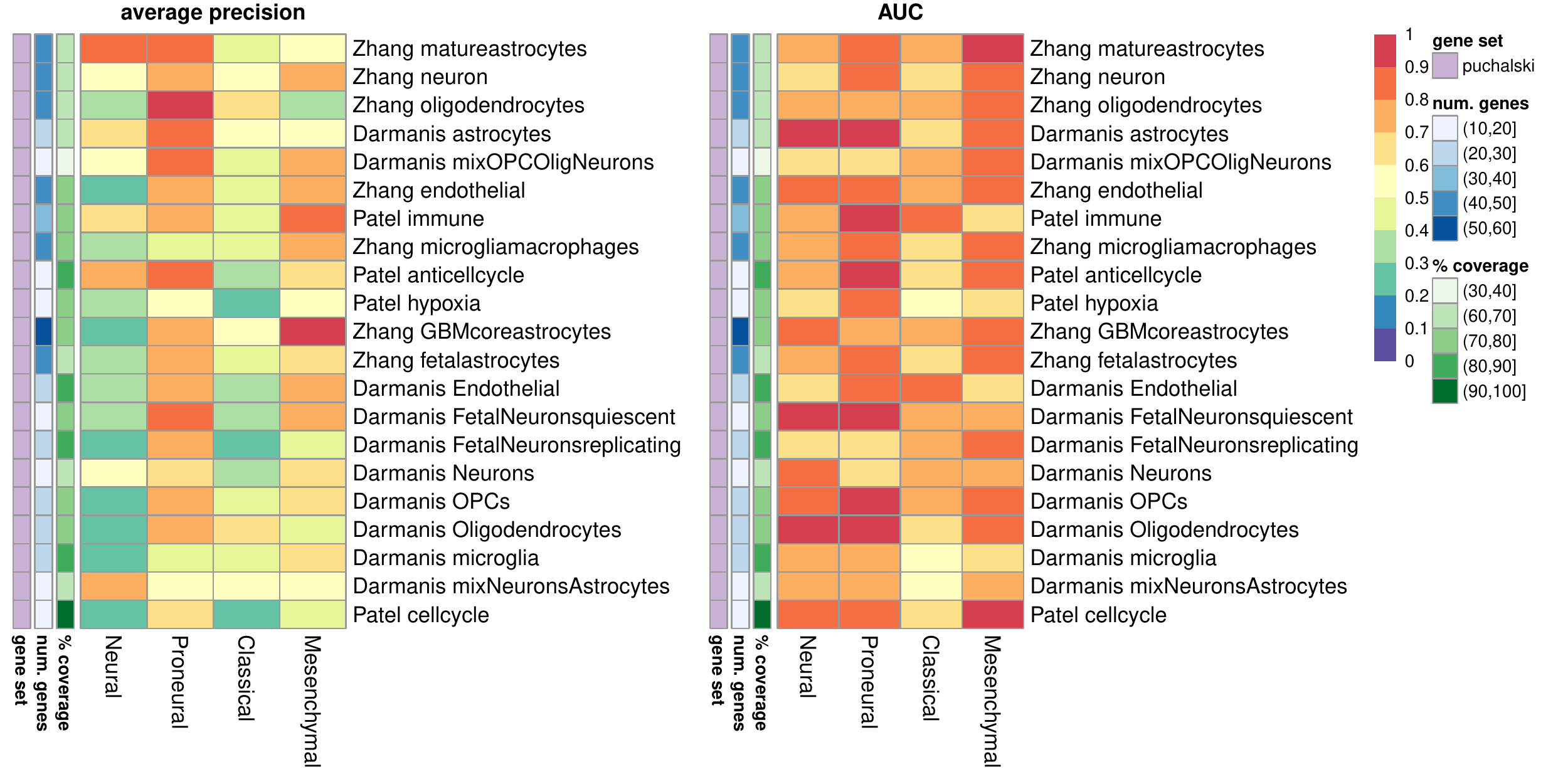}
	\includegraphics[width=\textwidth]{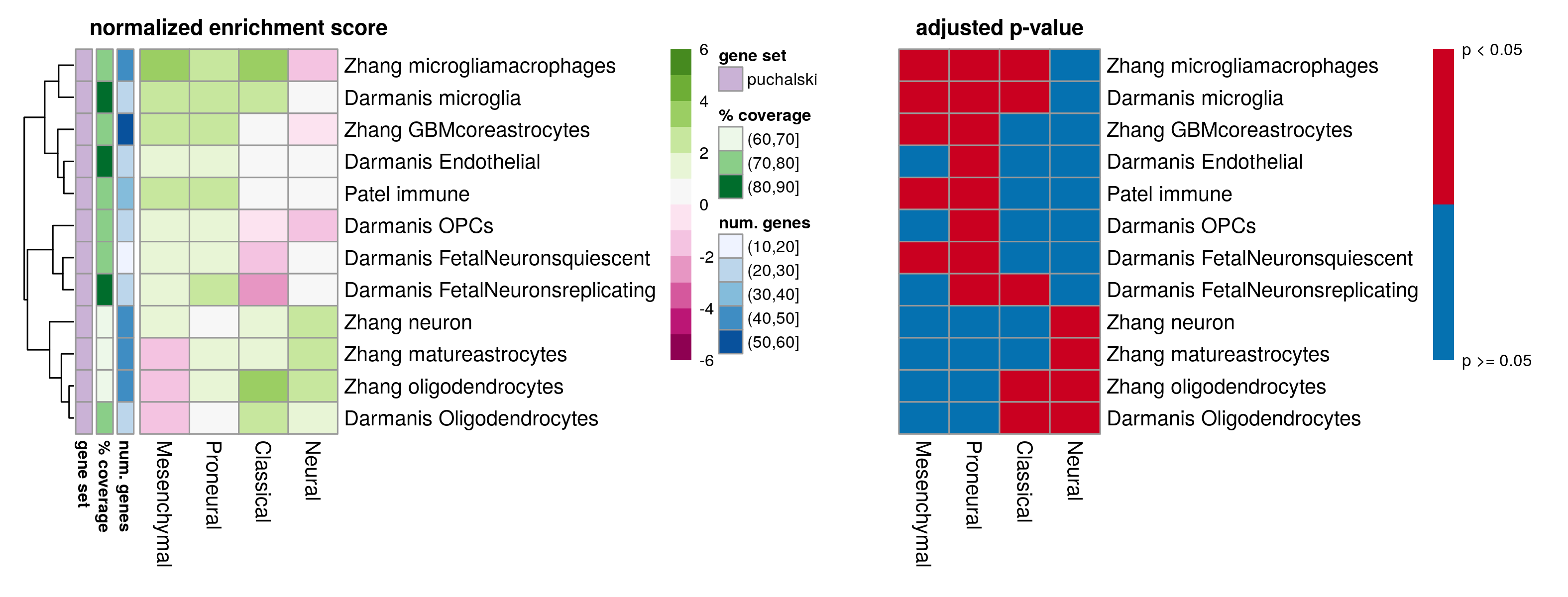}
	\centering
\end{figure}

\begin{figure}[h!]
	\caption{Perturbation of the subtype model with hallmark gene sets \cite{Liberzon2015}. Model performance in gene set masking. Shown are the top 20 hallmarks for each subtype ranked by average precision, totaling 32 gene sets. }
	\label{supfig:subtype_hallmark}
	\includegraphics[width=\textwidth]{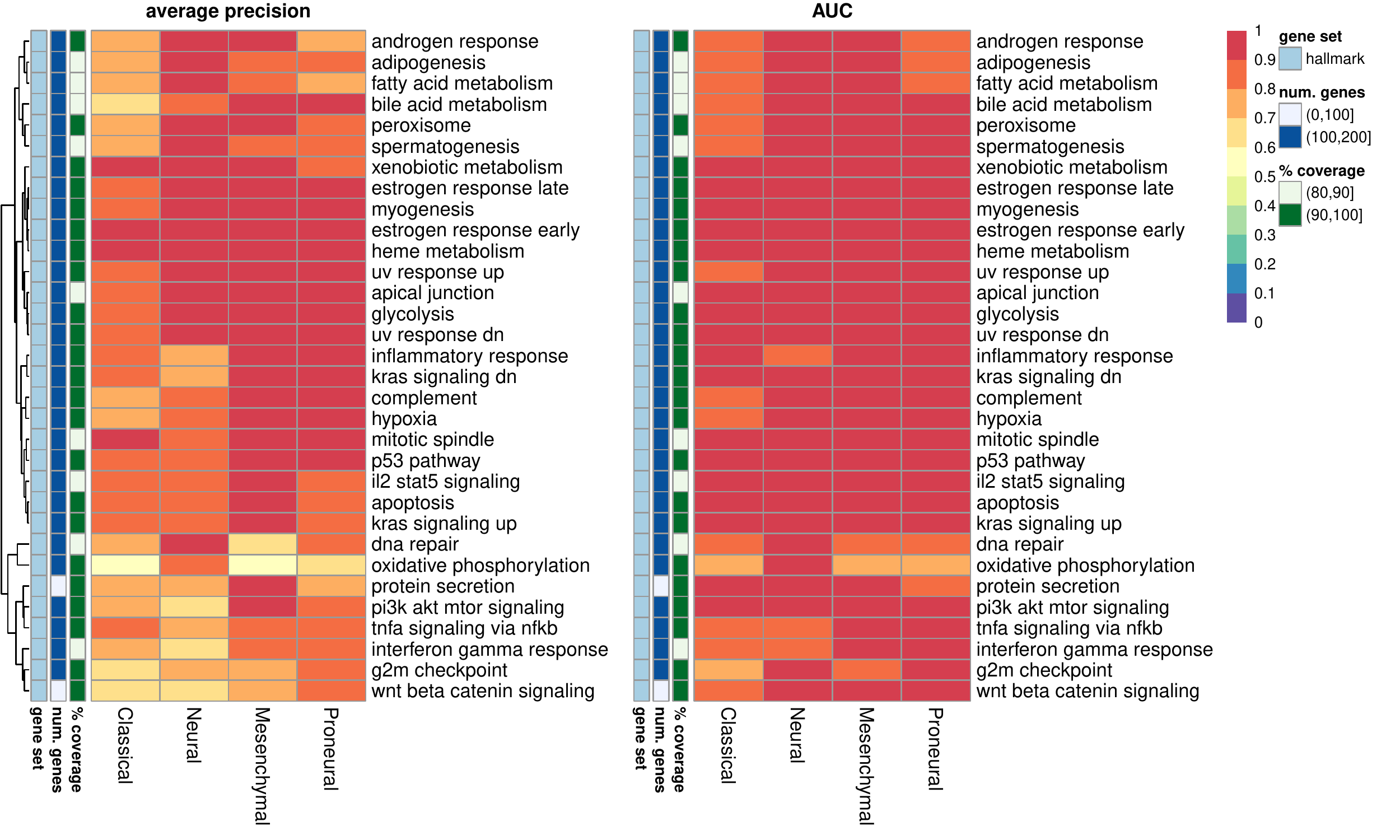}

	\centering
\end{figure}

\clearpage
\subsubsection*{Radiogenomic neural networks}

The 22 queried GBM genes from \cite{McLendon2008, Parsons2008} were \textit{AKT3}, \textit{CCND2}, \textit{CDK4}, \textit{CDK6}, \textit{CDKN2A}, \textit{CDKN2B}, \textit{CDKN2C}, \textit{EGFR}, \textit{ERBB2}, \textit{IDH1}, \textit{MDM2}, \textit{MDM4}, \textit{MET}, \textit{MYCN}, \textit{NF1}, \textit{PARK2}, \textit{PDGFRA}, \textit{PIK3CA}, \textit{PIK3R1}, \textit{PTEN}, \textit{RB1}, \textit{TP53}.

\begin{figure}[h!]
	\caption{Hallmark gene set masking in radiogenomic models ranked by AUC with corresponding average precision. Shown are the top 10 genes per label.}
	\label{supfig:vasari_hallmark_auc}
	\includegraphics[width=\textwidth]{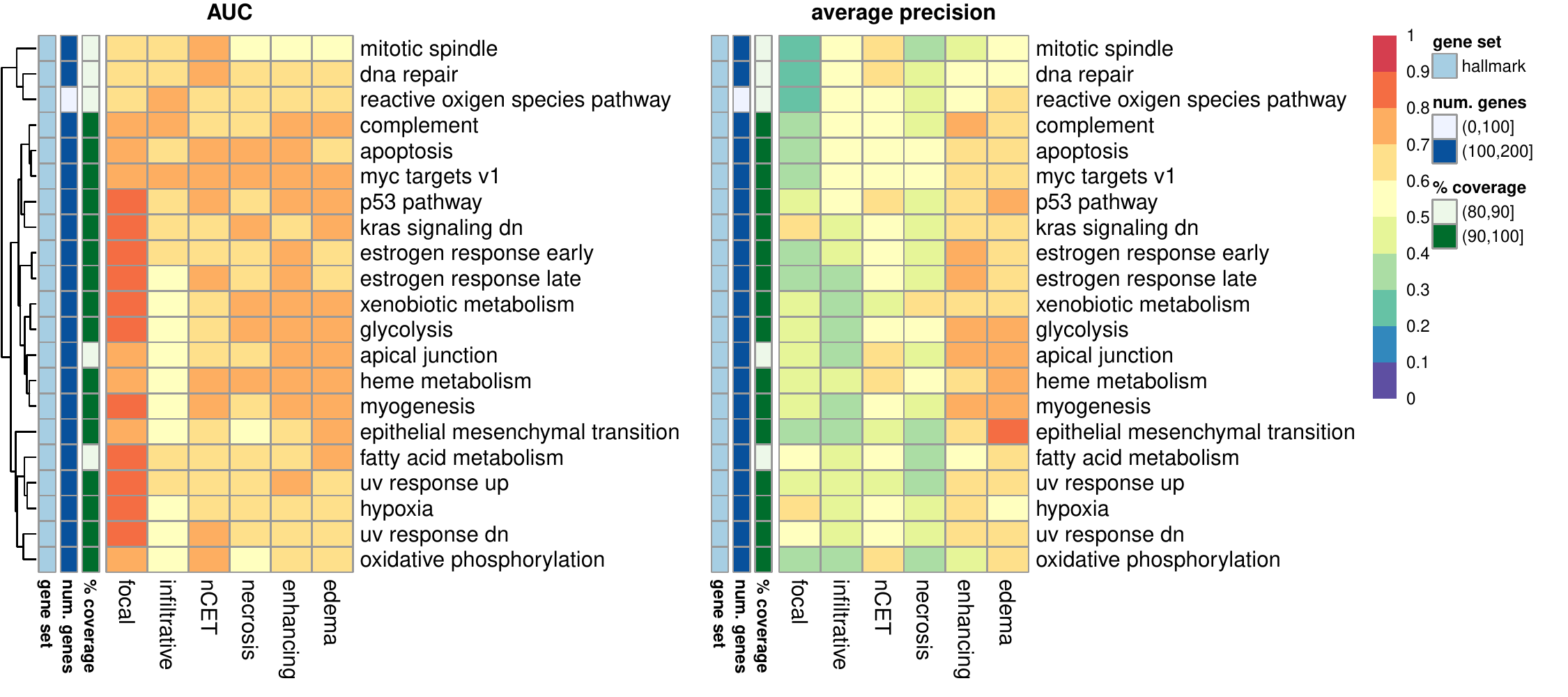}

	\centering
\end{figure}

\begin{figure}[h!]
	\caption{Single gene masking in radiogenomic models ranked by (\textbf{top}) AUC with corresponding average precision values on the right and (\textbf{bottom}) average precision with corresponding AUC values on the right. Shown are the top 10 genes per label.}
	\label{supfig:vasari_single}
	\includegraphics[width=.7\textwidth]{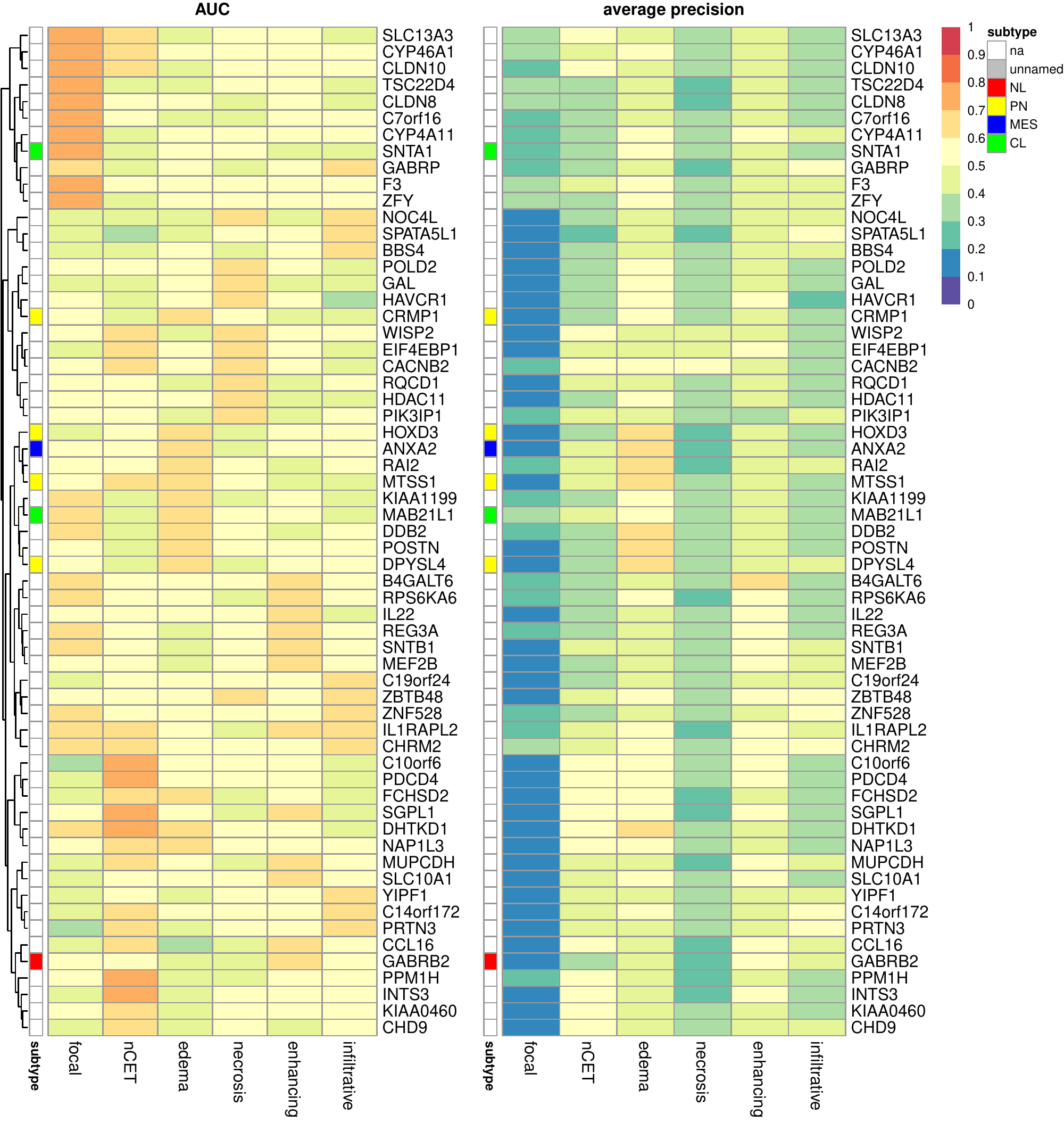}
	\includegraphics[width=.7\textwidth]{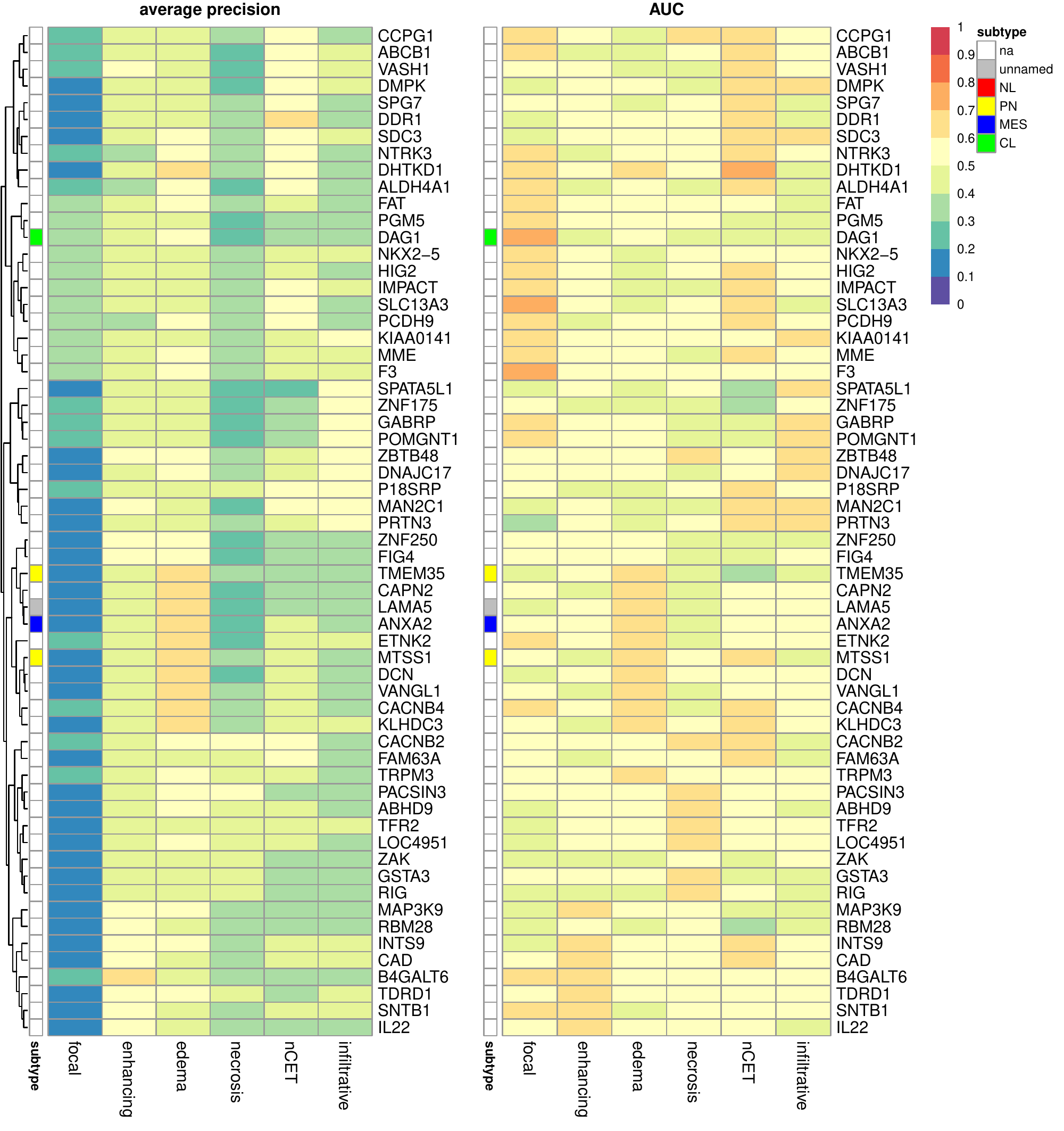}
	\centering
\end{figure}

\begin{figure}[h!]
	\captionsetup{font=small}
	\caption{Gene masking using gene sets associated with GBM genomic abnormalities \cite{McLendon2008, Parsons2008}. Shown are the top 15 gene sets ranked by average precision (AP) for
		(\textbf{a}) enhancing,
		(\textbf{b}) edema,
		(\textbf{c}) nCET,
		(\textbf{d}) necrosis,
		(\textbf{e}) focal, and
		(\textbf{f})  infiltrative neural networks. 
		Only the collections from GO, motif, and canonical pathways were considered. There were 22 genes queried, where gene sets may involve more than one of the queried genes.}
	\label{fig:vasari_gene_query}
	\includegraphics[width=\textwidth]{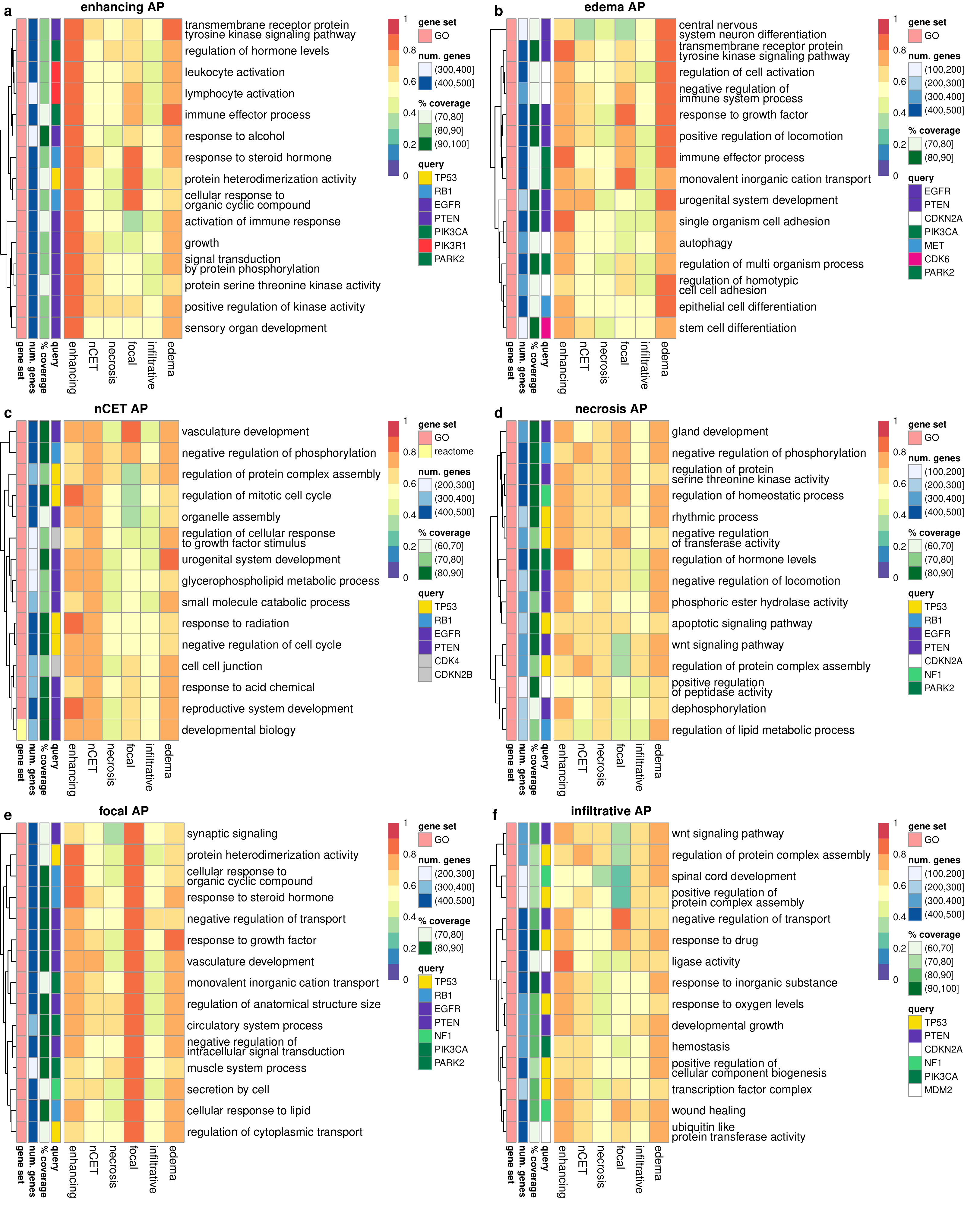}
	\centering
\end{figure}

\begin{figure}[h!]
	\captionsetup{font=small}
	\caption{Comparison of enhancement associations found in previous studies. Radiogenomic models were masked with gene sets queried from the Molecular Signature Database (MSigDB) based on key words from published findings.
		(\textbf{a}) Enhancement was found to be associated with hypoxia, ECM, and angiogenesis gene modules from Diehn et al. (n=22) \cite{Diehn2008}. There were 15 gene sets returned from querying hypoxia, ECM, and angiogenesis.
		(\textbf{b}) Jamshidi et al. found associations between 17 Biocarta pathways and enhancement [n=23] \cite{Jamshidi2014}. 
		(\textbf{c}) Jamshidi et al. also stated \textit{C1orf172}, \textit{CAMSAP2}, \textit{KCNK3}, \textit{LTBP1} genes were related to enhancement. There were 114 gene sets containing the four genes
		(\textbf{d}) Gutman et al. found a non-significant correlation between \textit{EGFR} copy number amplification and enhancement (n=75) \cite{Gutman2013}. Querying MSigDB for \textit{EGFR} resulted in 257 gene sets. The top ten gene sets ranked by the average precision (AP) in predicting enhancement were kept for (b-c).}
	\label{supfig:vasari_lit_enhan}
	\includegraphics[width=\textwidth]{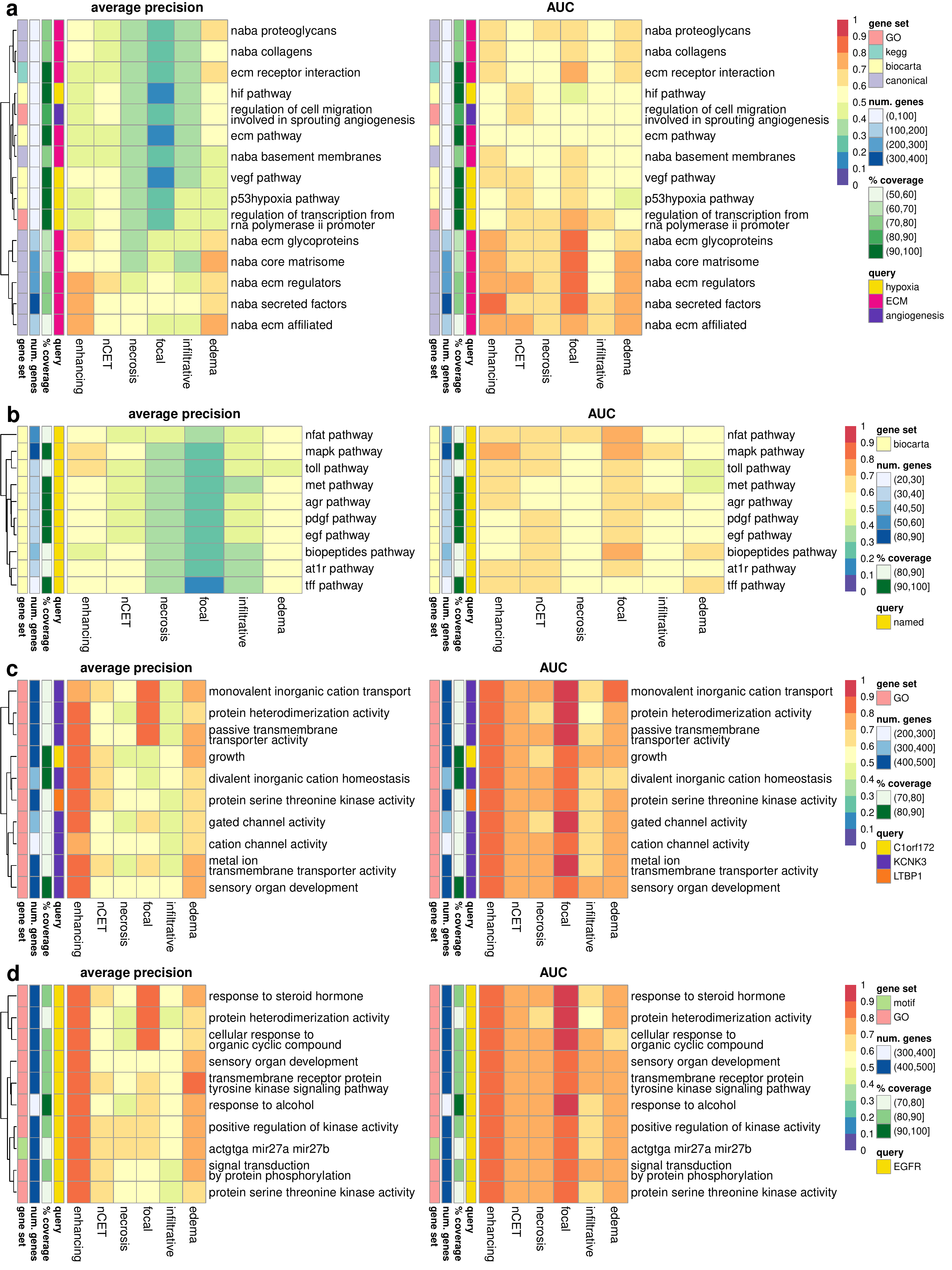} 
	\centering
\end{figure}

\begin{figure}[h!]
	\captionsetup{font=small}
	\caption{Comparison of edema associations found in Zinn et al. [n=78] \cite{Zinn2011}. Radiogenomic models were masked with gene sets queried from the MSigDB based on key words from published findings. Zinn et al. found concordant changes in mRNA and microRNA fold changes for patients with high edema or invasive traits: 
		(\textbf{a}) The top five upregulated microRNA and their associated genes and 
		(\textbf{b}) top five upregulated genes in patients with high FLAIR volumes, see Table 4 of the authors' paper. This resulted in 343 and 160 related MSigDB gene sets, respectively. Shown are the top ten gene sets ranked by average precision (AP) in predicting edema in our study.}
	\label{supfig:vasari_lit_edema}
	\includegraphics[width=\textwidth]{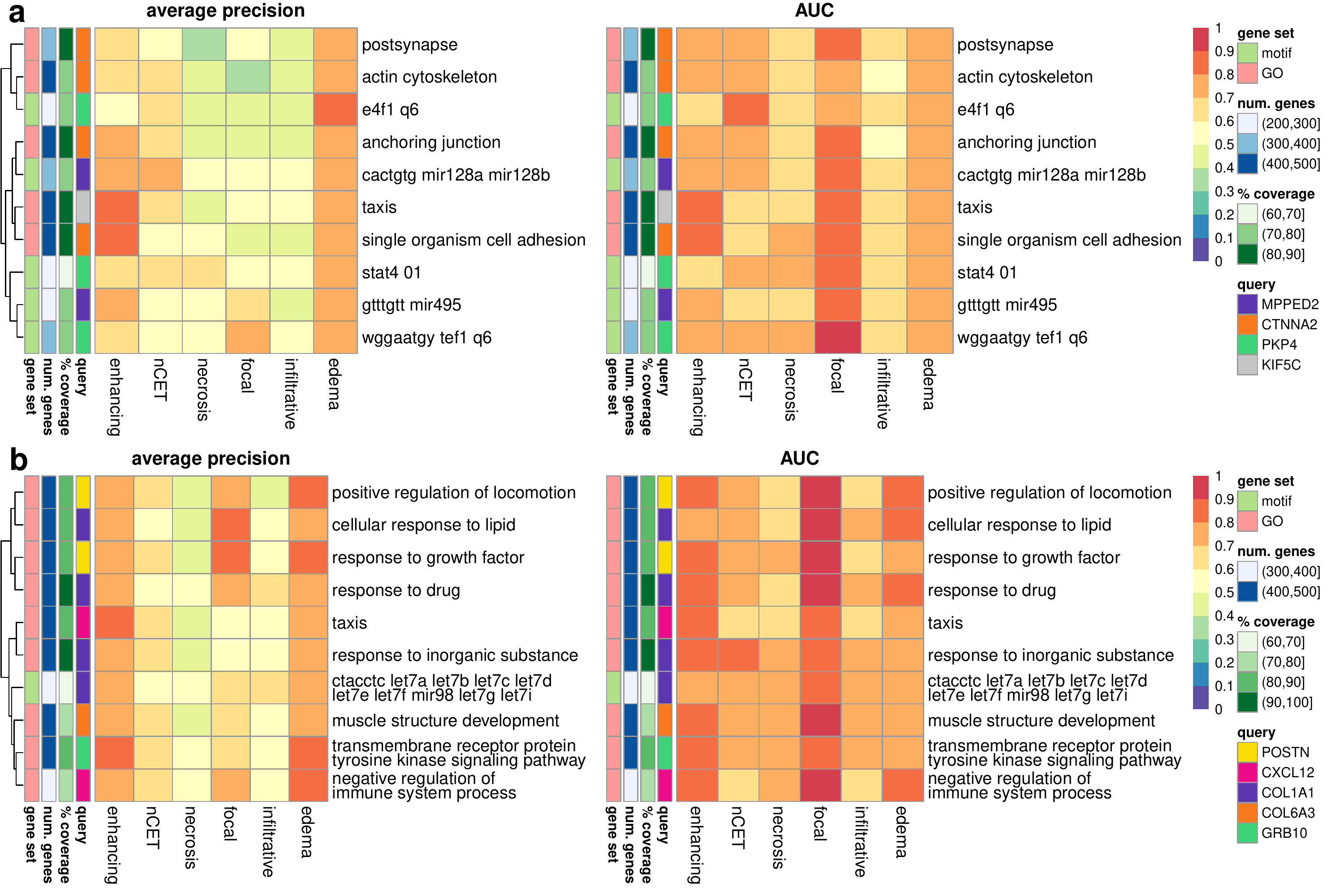}
	\centering
\end{figure}

\begin{figure}[h!]
	\captionsetup{font=small}
	\caption{Comparison of necrosis associations found in previous studies. Radiogenomic models were masked with gene sets queried from the MSigDB based on key words from published findings.
		(\textbf{a}) Gevaert et al. reported the boundary sharpness of necrosis region was associated with \textit{GAP43}, \textit{WWTR1}, IL4 pathway, and cell membrane genes [n=55]\cite{Gevaert2014}. This returned 378 gene sets from MSigDB.
		(\textbf{b}) Jamshidi et al. found four Biocarta pathways (named gene sets) and (\textbf{c}) two genes associated with necrosis \cite{Jamshidi2014}.
		(\textbf{d}) Gutman et al. found necrosis was correlated with the deletion of \textit{CDKN2A}, but was not significant \cite{Gutman2013}. An MSigDB query for the gene returned 183 gene sets.
		The top ten gene sets ranked by the average precision in predicting necrosis were kept in (a,b,d).}
	\label{supfig:vasari_lit_necro}
	\includegraphics[width=\textwidth]{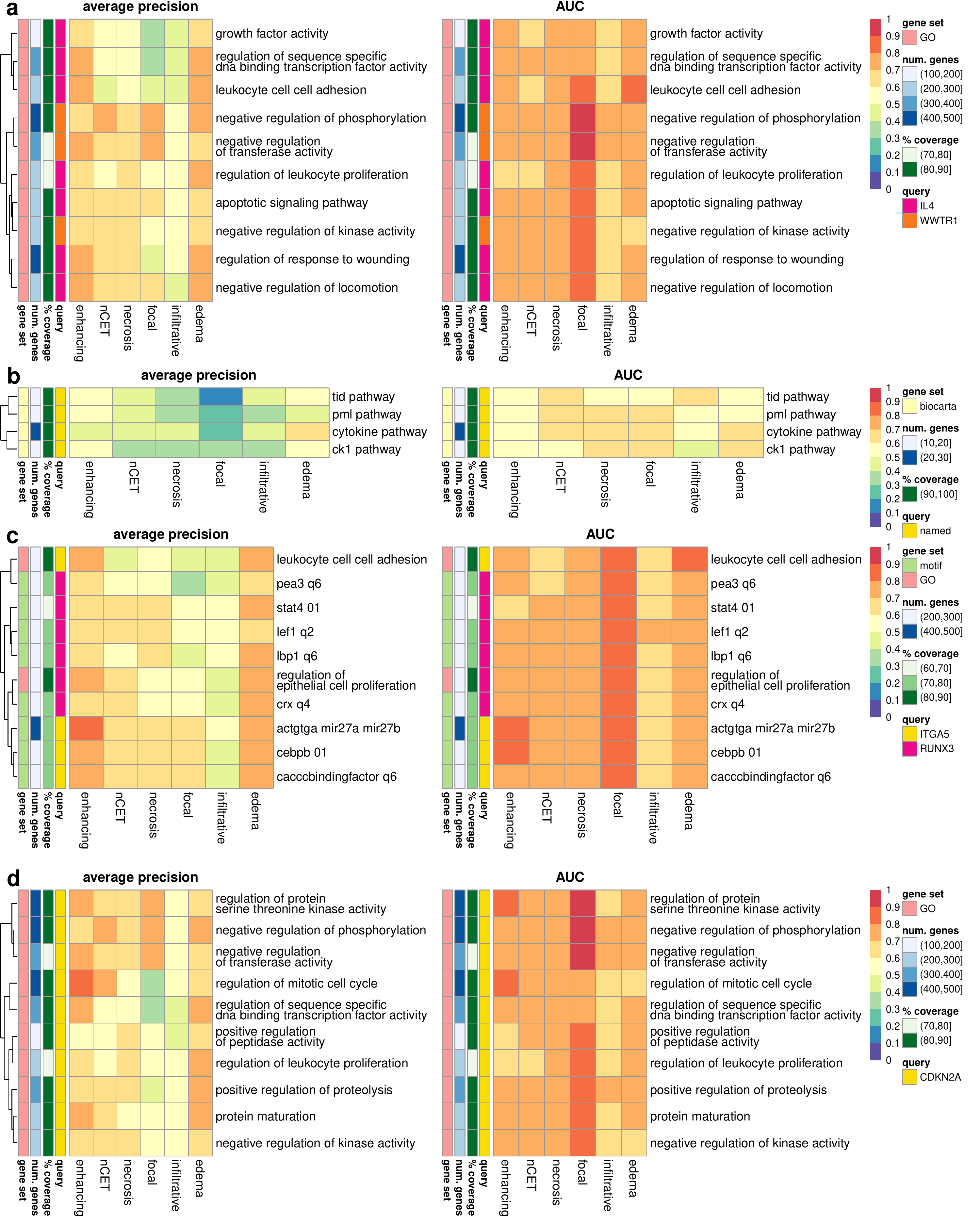}
	\centering
\end{figure}

\begin{figure}[h!]
	\captionsetup{font=small}
	\caption{Comparison of expansiveness versus infiltrative associations found in previous studies. Radiogenomic models were masked with gene sets queried from the MSigDB based on key words from published findings.
		(\textbf{a}) Colen et al. found invasive tumors to be associated with \textit{MYC}, leading to NFKBIA inhibition [n=104] \cite{Colen2014}. Querying MSigDB for \textit{MYC} and NFKBIA resulted in 316 gene sets. 
		(\textbf{b}) Diehn et al. looked at infiltrative vs. edematous T2 abnormality and found an association with an immune cell gene module \cite{Diehn2008}. An MSigDB query for `immune' returned 87 gene sets. Shown are the top ten gene sets ranked by average precision (AP) in predicting infiltrative tumors in our study.}
	\label{supfig:vasari_lit_f10}
	\includegraphics[width=\textwidth]{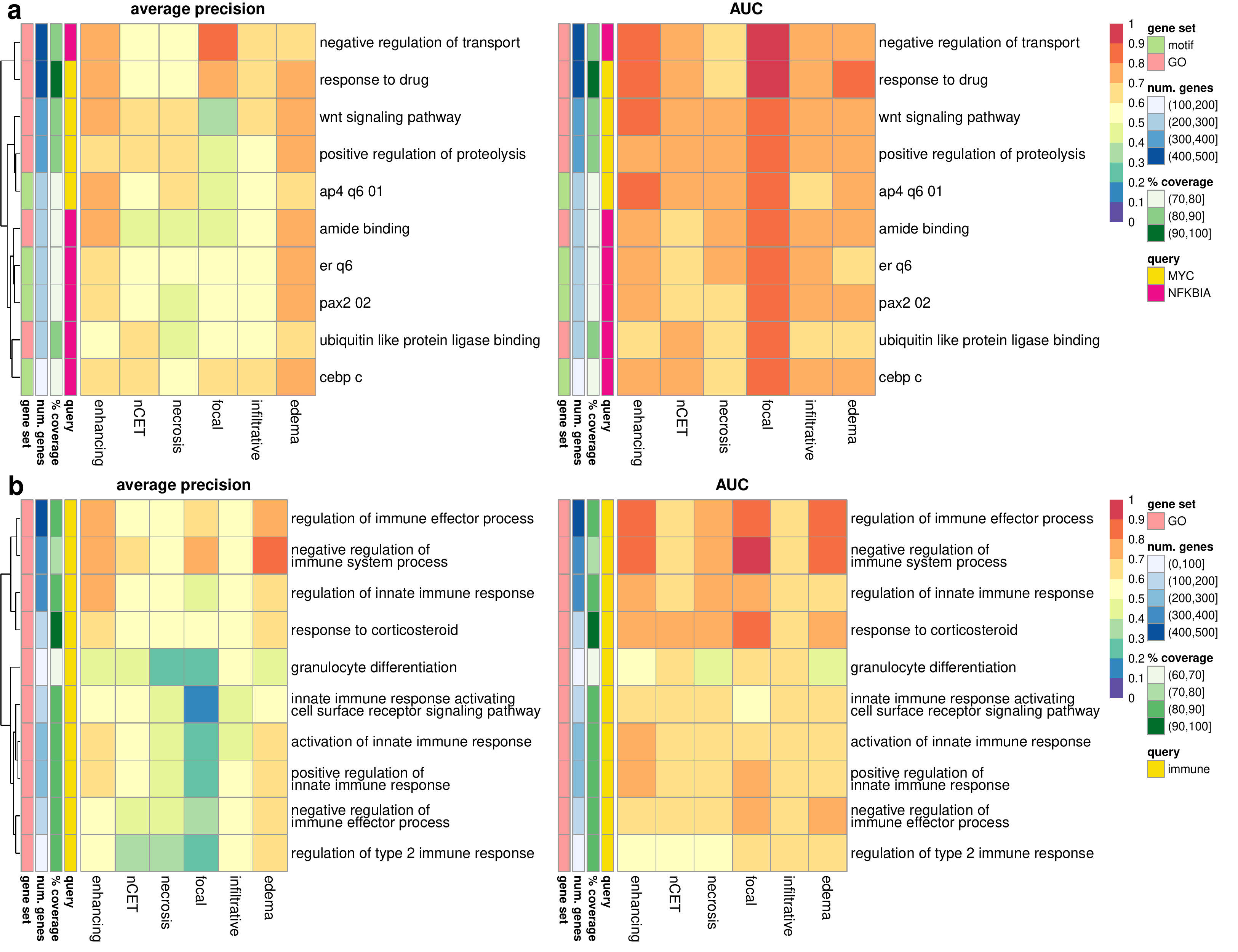}
	\centering
\end{figure}


\begin{figure}[h!]
	\caption{Perturbation of radiogenomic models with subtype gene sets from \cite{Verhaak2010}. 
		(\textbf{top}) Model performance in gene set masking, and
		(\textbf{bottom}) gene set enrichment in ranked genes after single gene masking. The authors found mesenchymal tumors had less nCET and proneural tumors had less enhancement. However, the radiogenomic models found neural genes were more predictive of nCET than mesenchymal genes (0.64 vs. 0.55 AUC, 0.52 vs. 0.46 AP) and other subtype genes were more predictive than proneural genes in predicting enhancement.}
	\label{supfig:vasari_verhaak}
	\includegraphics[width=.8\textwidth]{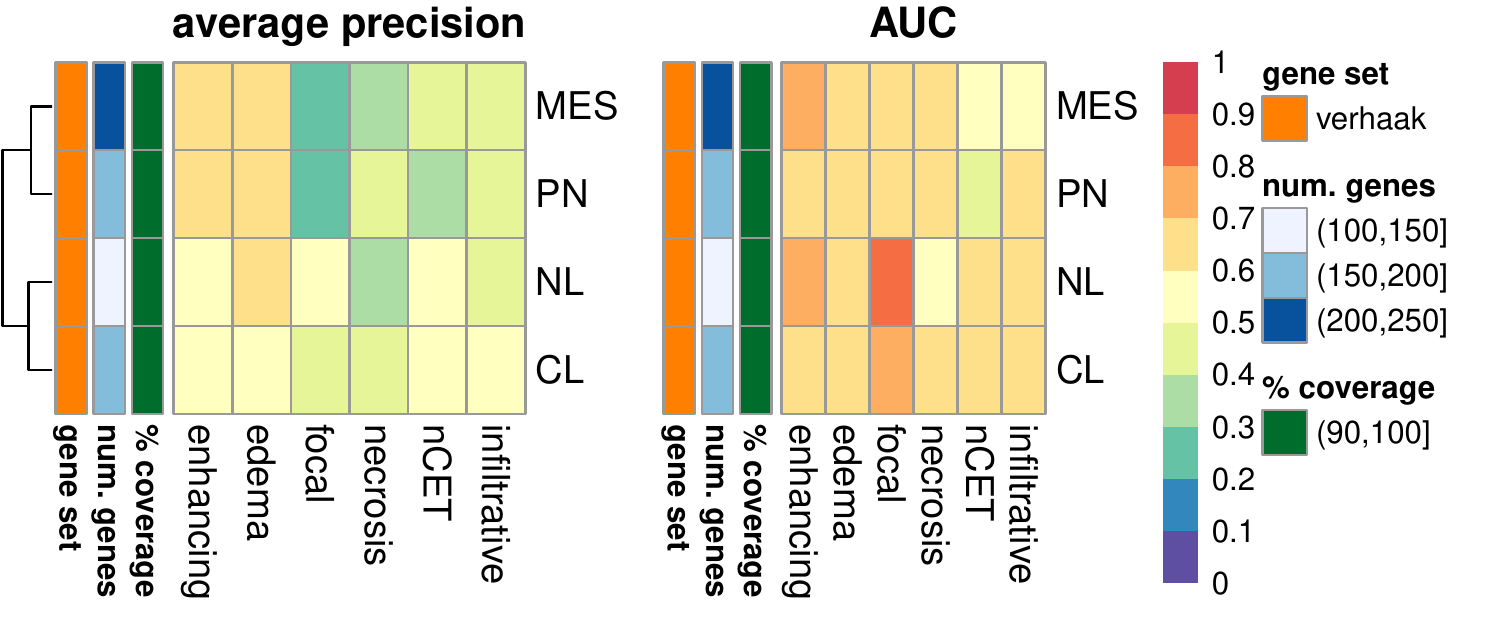}
	\includegraphics[width=.8\textwidth]{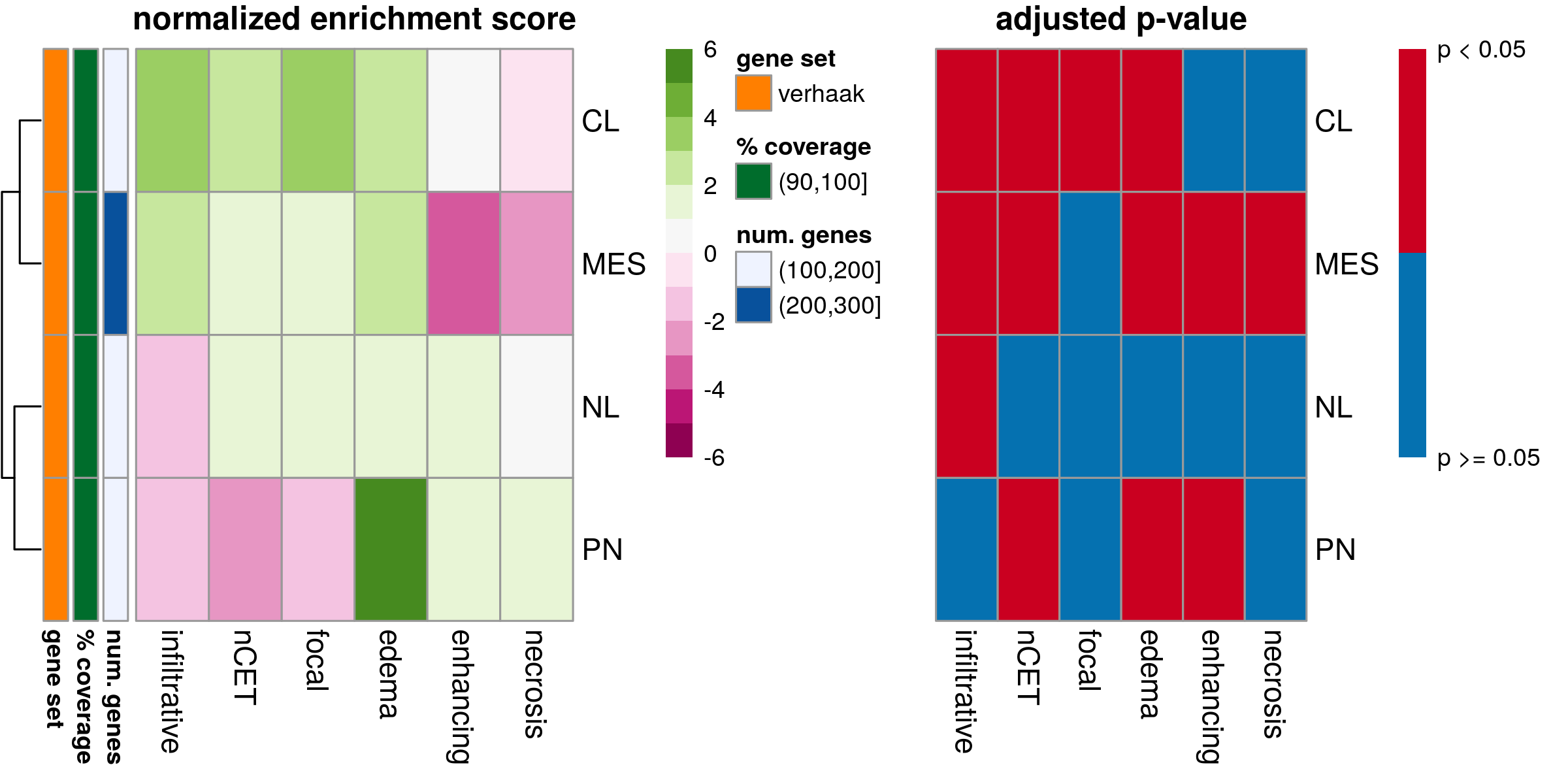}
	\centering
\end{figure}

\begin{figure}[h!]
	\caption{Perturbation of radiogenomic models with gene sets from brain cell types and phenotypes \cite{Zhang2016neuron, Darmanis2015, Patel2014}, downloaded from Puchalski et al. \cite{Puchalski2018}. 
	(\textbf{top}) Gene set masking, where the top 20 ranked by average precision were kept; shown are 21 gene sets.
	(\textbf{bottom}) GSEA of genes ranked in single gene masking. Shown are gene sets with at least one significant enrichment.}
	\label{supfig:vasari_puchalski}
	\includegraphics[width=\textwidth]{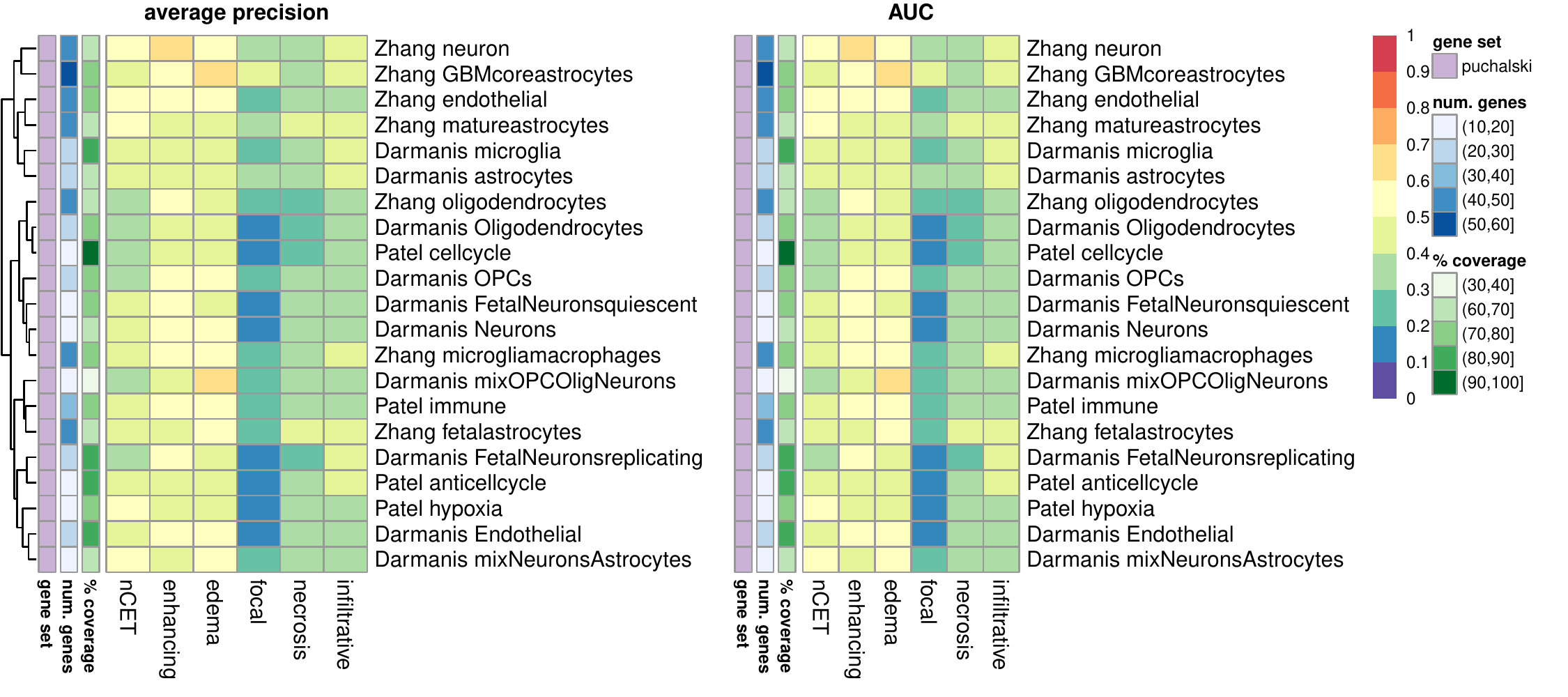}
	\includegraphics[width=\textwidth]{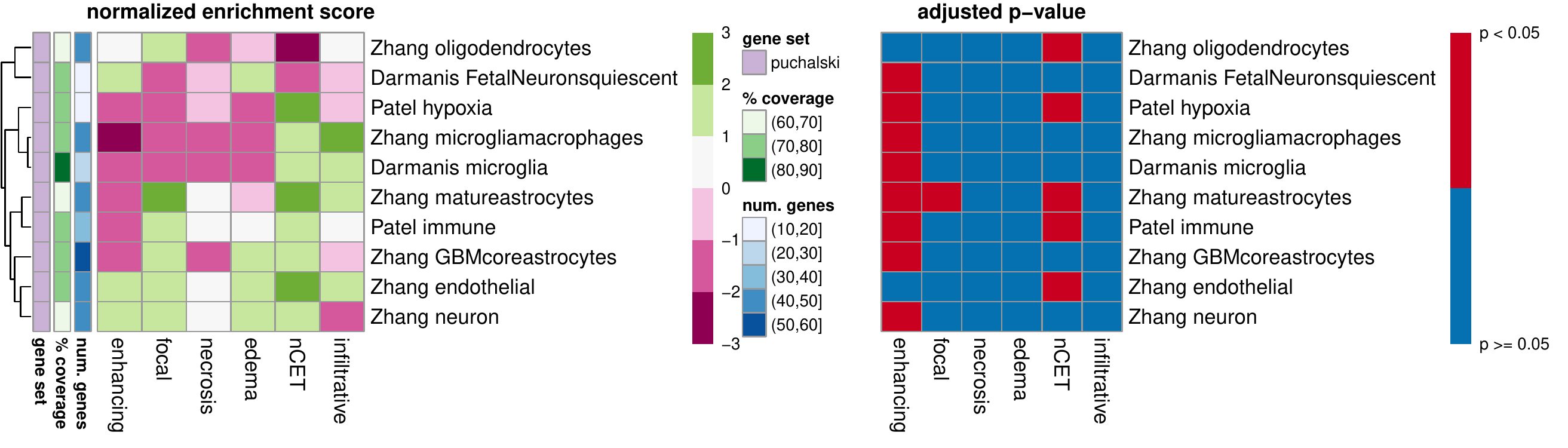}
	\centering
\end{figure}


\begin{figure}[h!]
	\caption{Gene masking in radiogenomic models with canonical gene sets from MSigDB. 
		(\textbf{top}) Gene set masking, where the top 5 ranked by average precision were kept; shown are 21 gene sets.
		(\textbf{bottom}) GSEA of genes ranked in single gene masking. Shown are gene sets with at least one significant enrichment.}
	\label{supfig:vasari_canonical}
	\includegraphics[width=\textwidth]{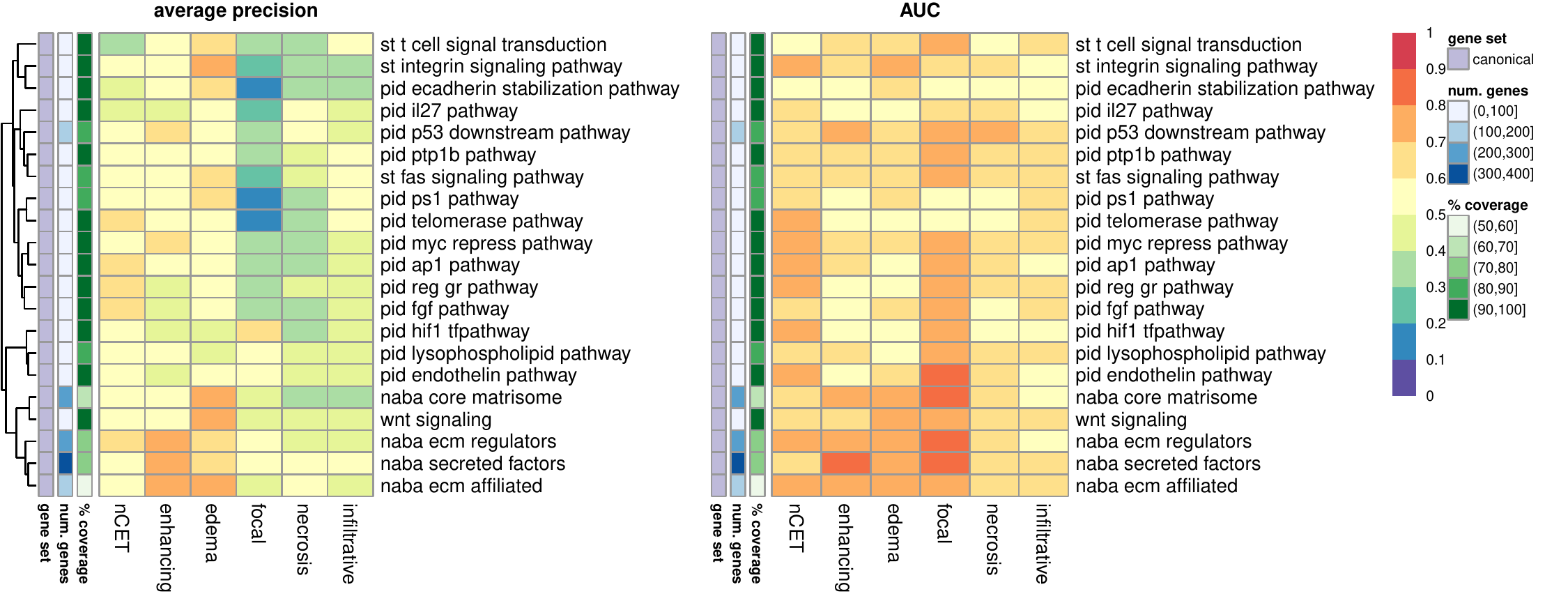}
	\includegraphics[width=\textwidth]{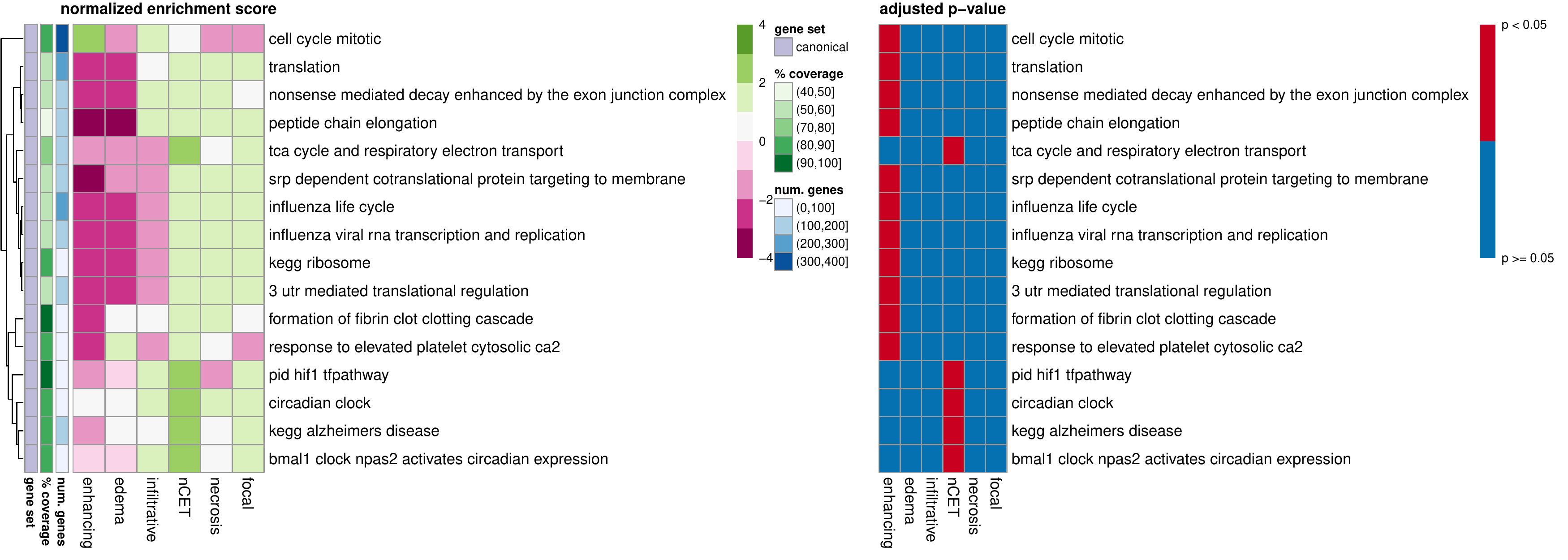}
	\centering
\end{figure}

\begin{figure}[h!]
	\caption{Gene masking in radiogenomic models with motif gene sets from MSigDB. 
		(\textbf{top}) Gene set masking, where the top 5 ranked by average precision were kept; shown are 29 gene sets.
		(\textbf{bottom}) GSEA of genes ranked in single gene masking. Shown are gene sets with at least one significant enrichment.
		}
	\includegraphics[width=\textwidth]{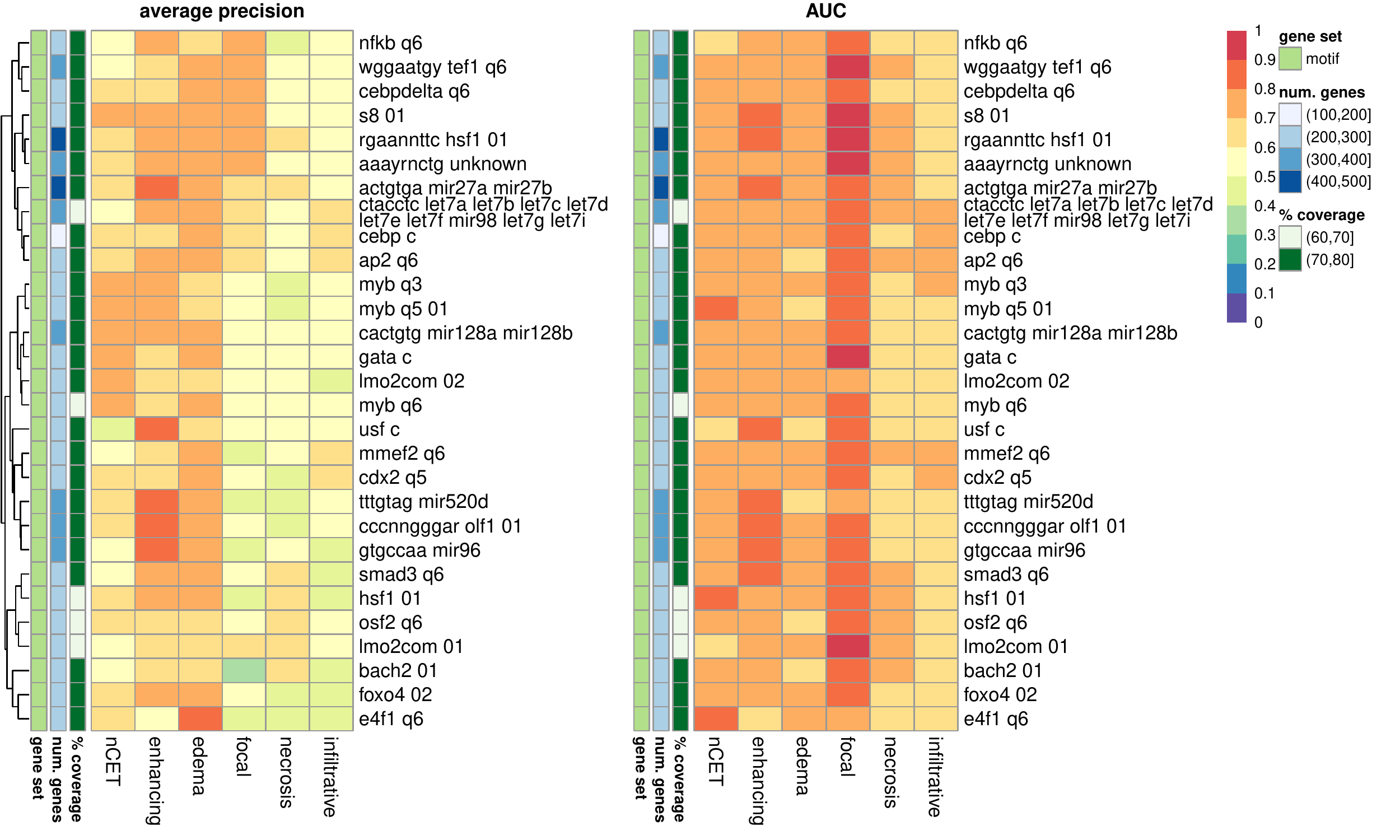}
	\includegraphics[width=\textwidth]{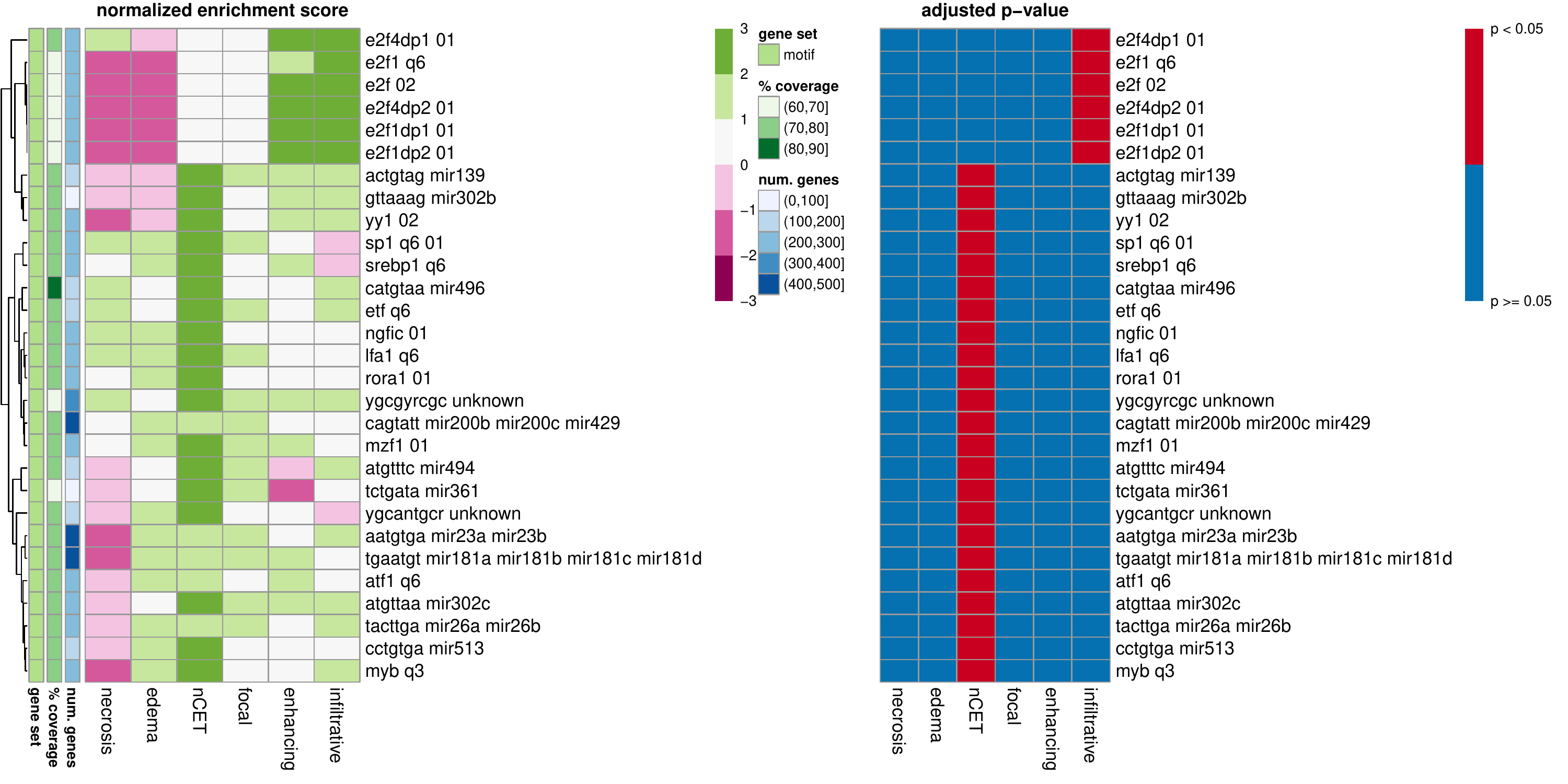}
	\label{supfig:vasari_motif}
	\centering
\end{figure}

\begin{figure}[h!]
	\caption{Gene masking in radiogenomic models with chromosome gene sets from MSigDB. 
	(\textbf{top}) Gene set masking, where the top 5 ranked by average precision were kept; shown are 25 gene sets.
	(\textbf{bottom}) GSEA of genes ranked in single gene masking. Shown are gene sets with at least one significant enrichment.}
	\label{supfig:vasari_chromosome}
	\includegraphics[width=.7\textwidth]{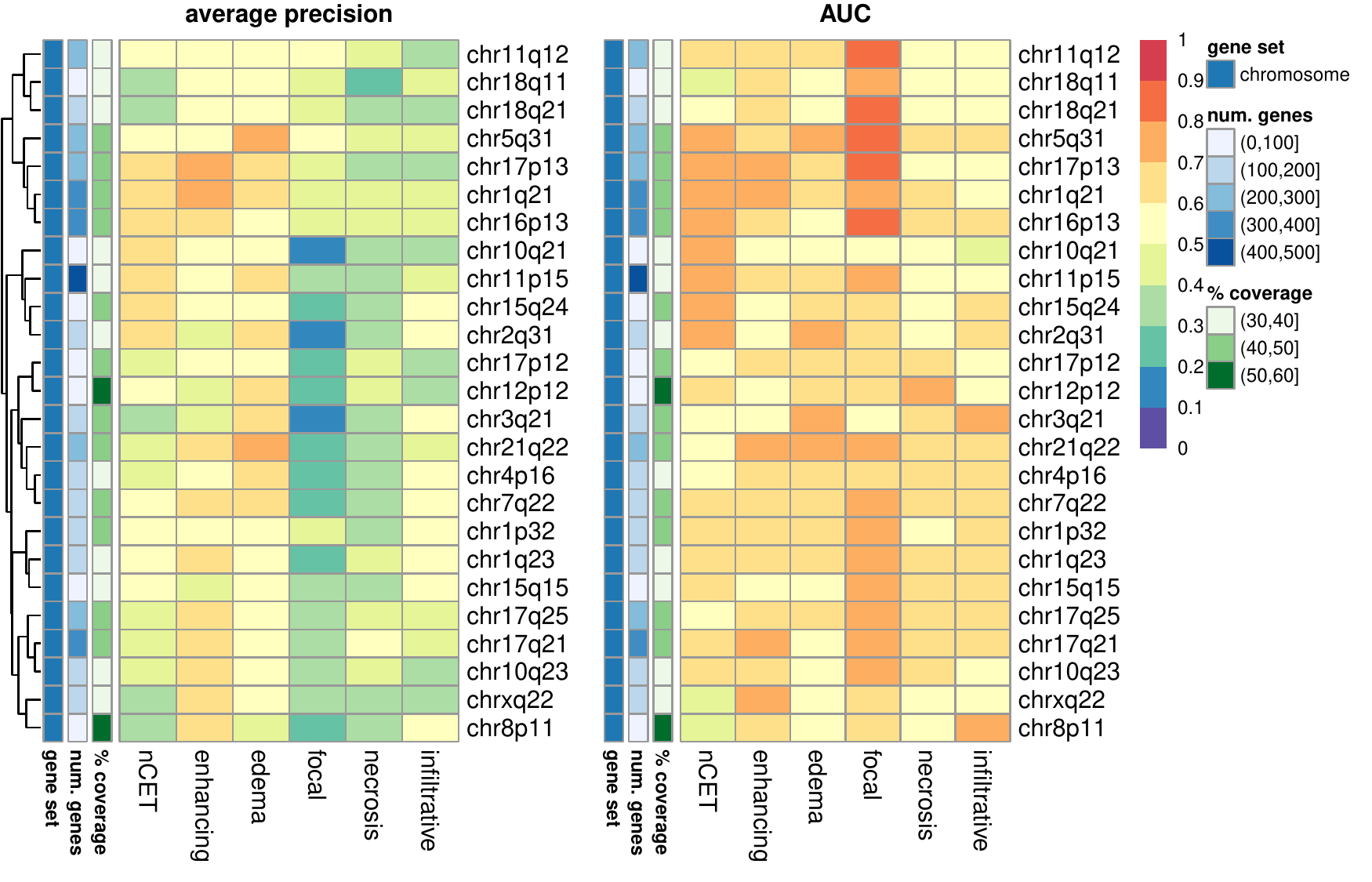}
	\includegraphics[width=.7\textwidth]{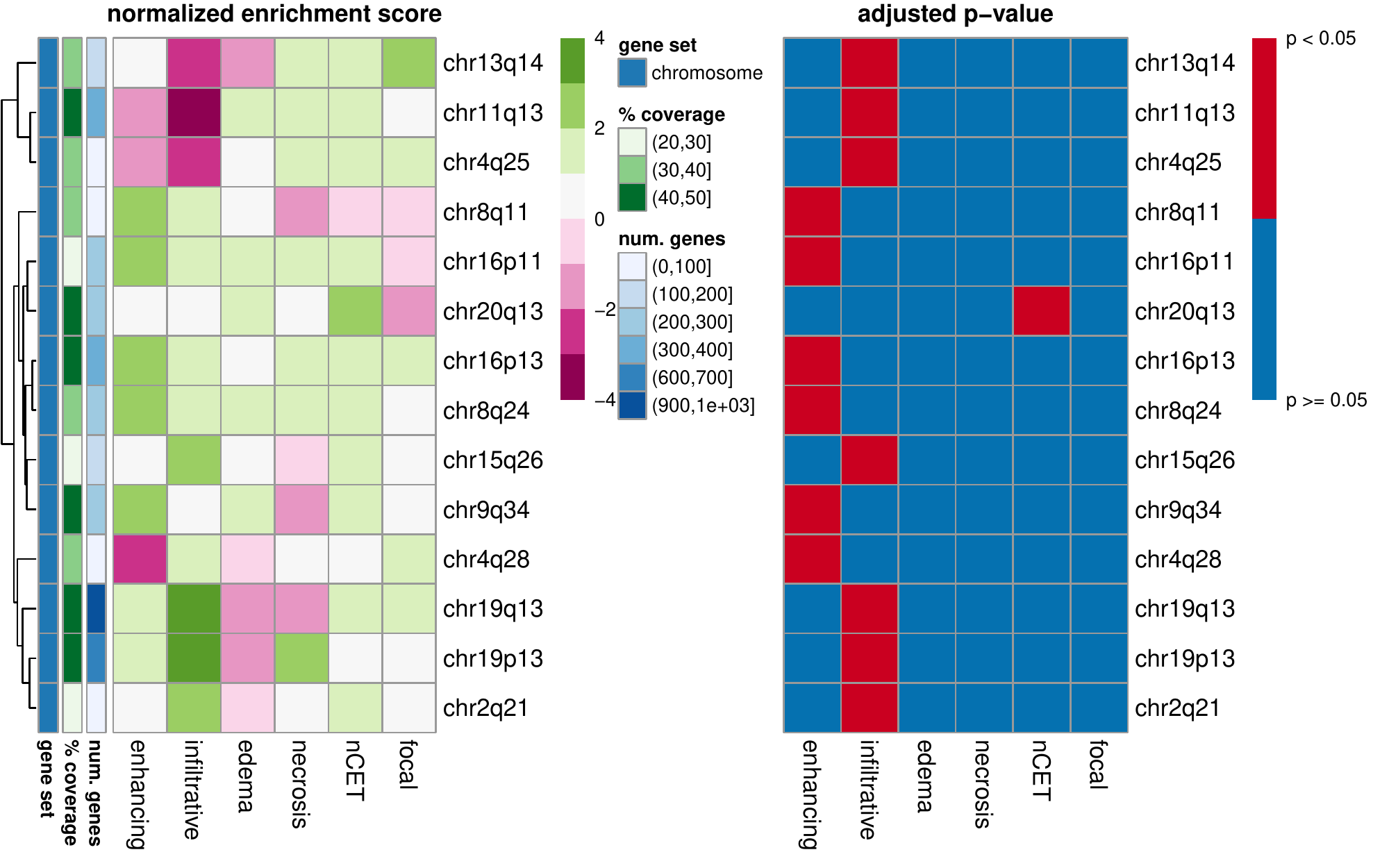}
	\centering
\end{figure}

\begin{figure}[h!]
	\caption{Gene masking in radiogenomic models with oncogenic signatures gene sets from MSigDB. 
	(\textbf{top}) Gene set masking, where the top 5 ranked by average precision were kept; shown are 25 gene sets.
	(\textbf{bottom}) GSEA of genes ranked in single gene masking. Shown are gene sets with at least one significant enrichment.}
	\label{supfig:vasari_onco}
	\includegraphics[width=\textwidth]{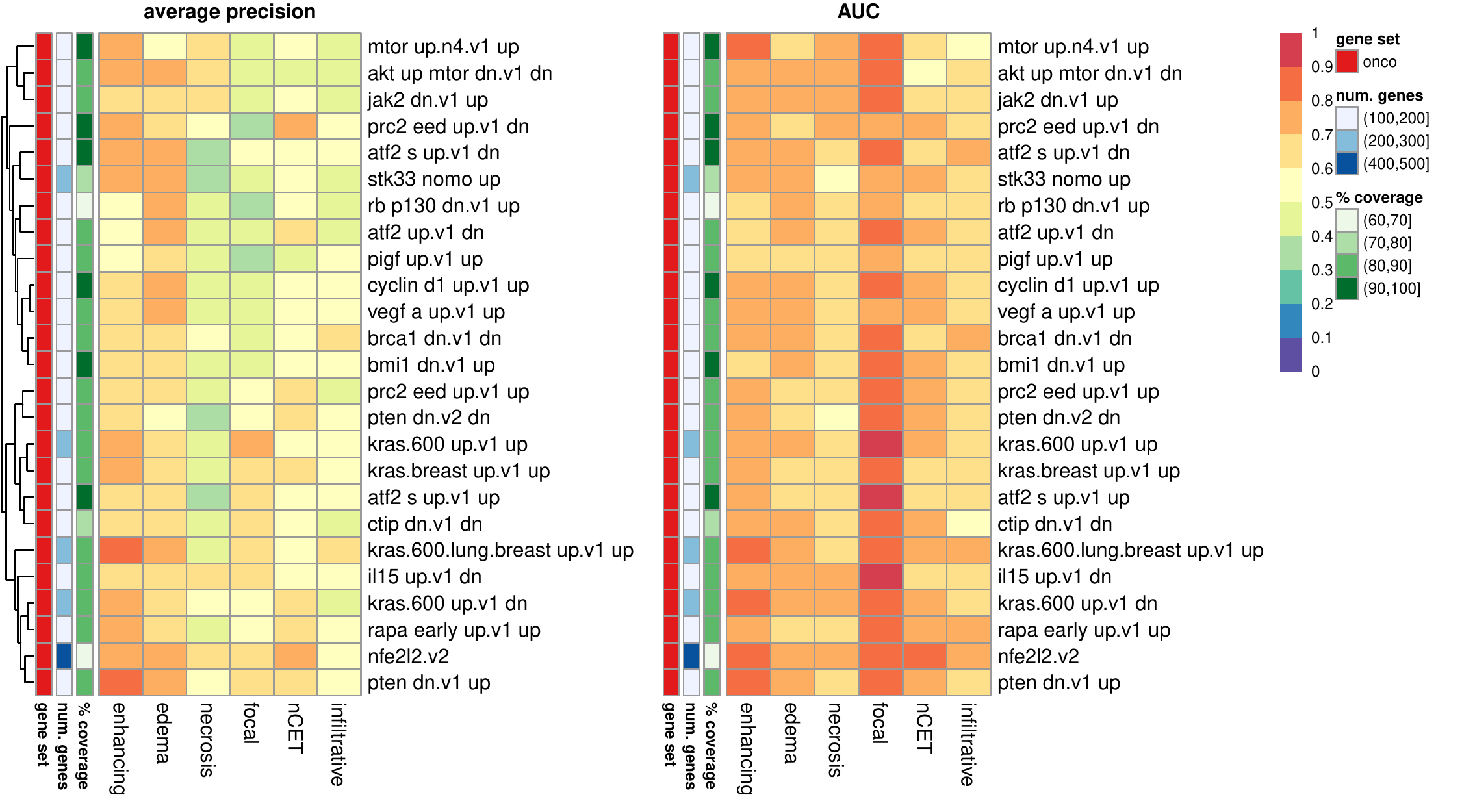}
	\includegraphics[width=\textwidth]{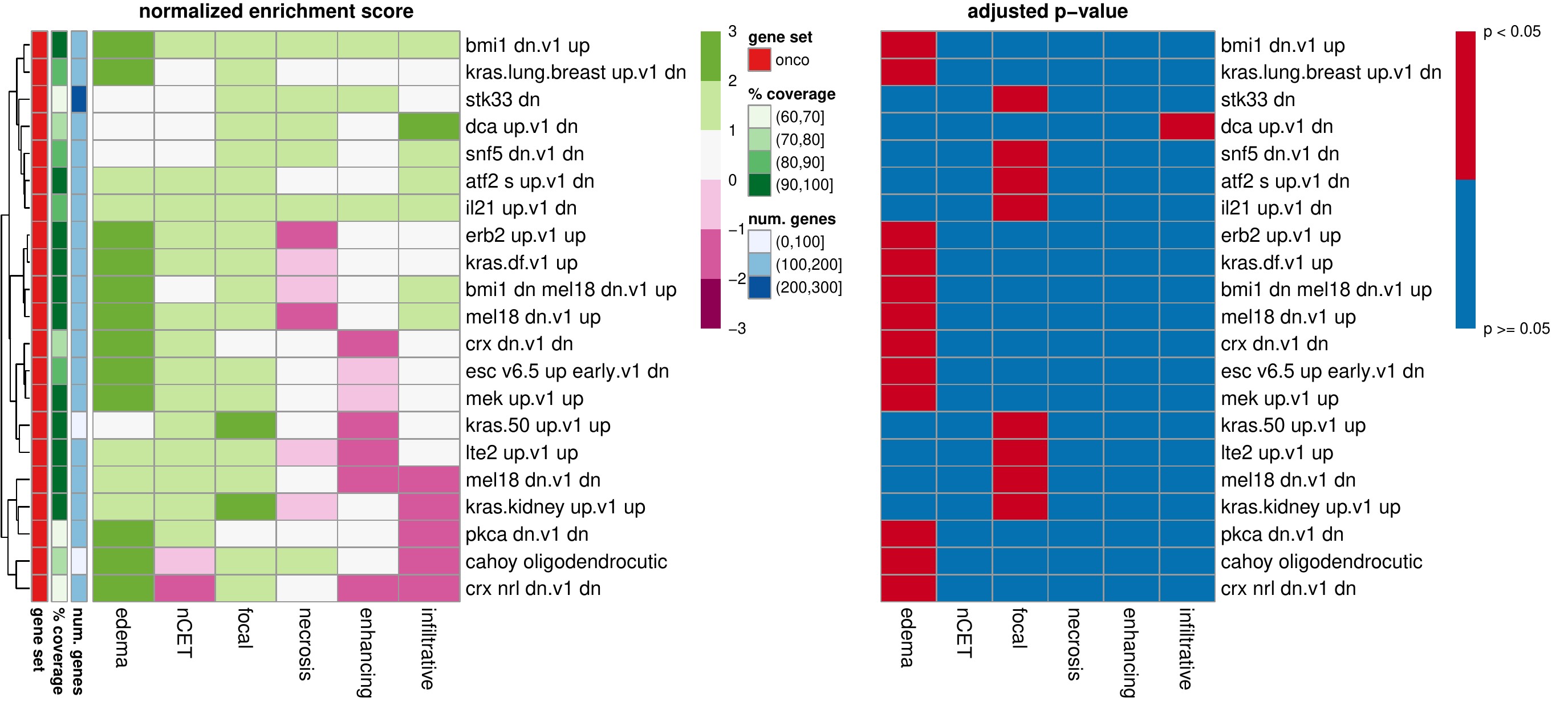}
	\centering
\end{figure}

\begin{figure}[h!]
	\caption{Gene masking performance compared to 
		(\textbf{top}) gene set size and 
		(\textbf{bottom}) coverage. Coverage was the percent of genes in the gene set that was in input gene expressions. Average precision (left column) and AUC (right column) were used.}
	\label{supfig:vasari_gs_info}
	\includegraphics[width=.65\textwidth]{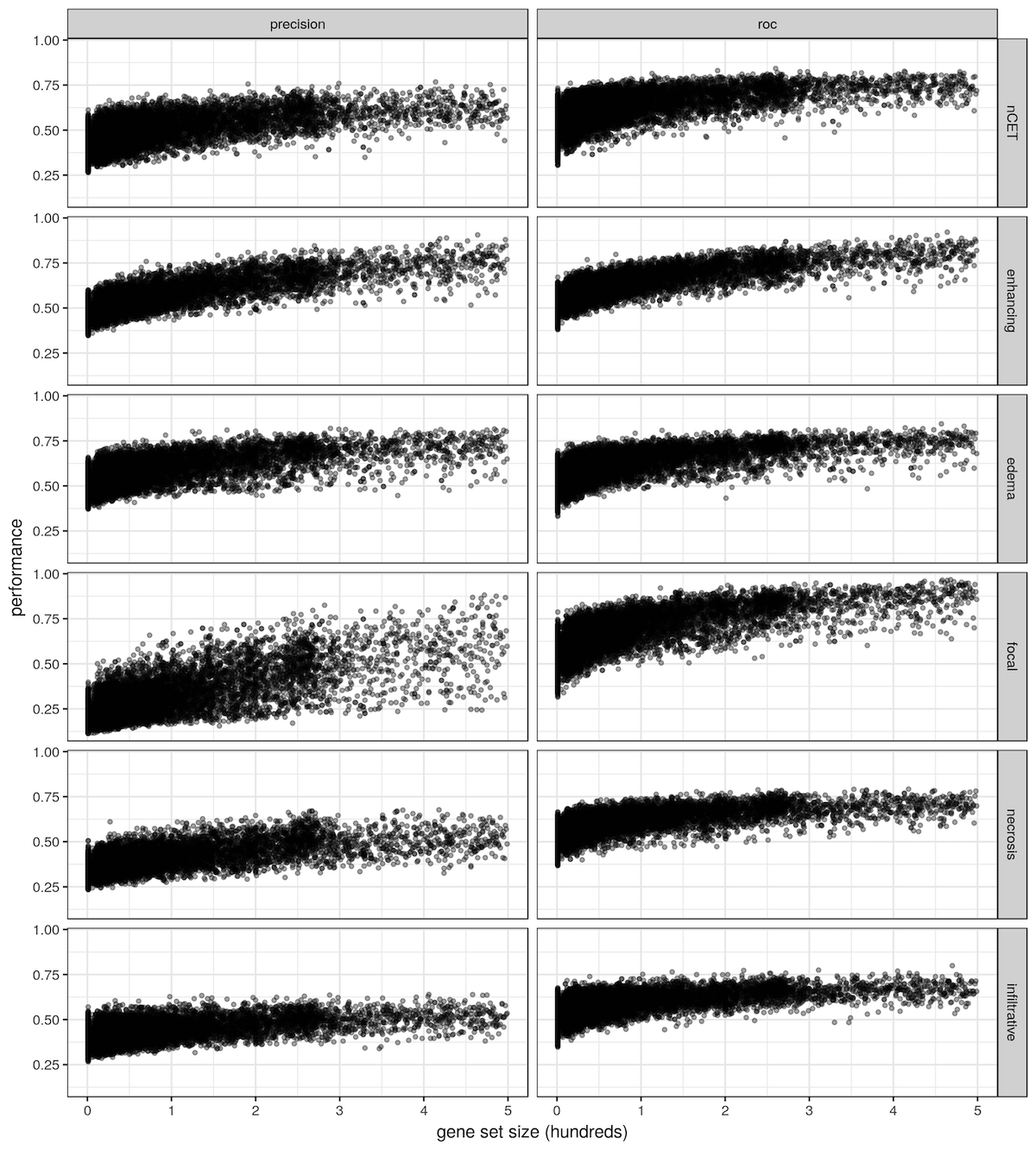}
	\includegraphics[width=.65\textwidth]{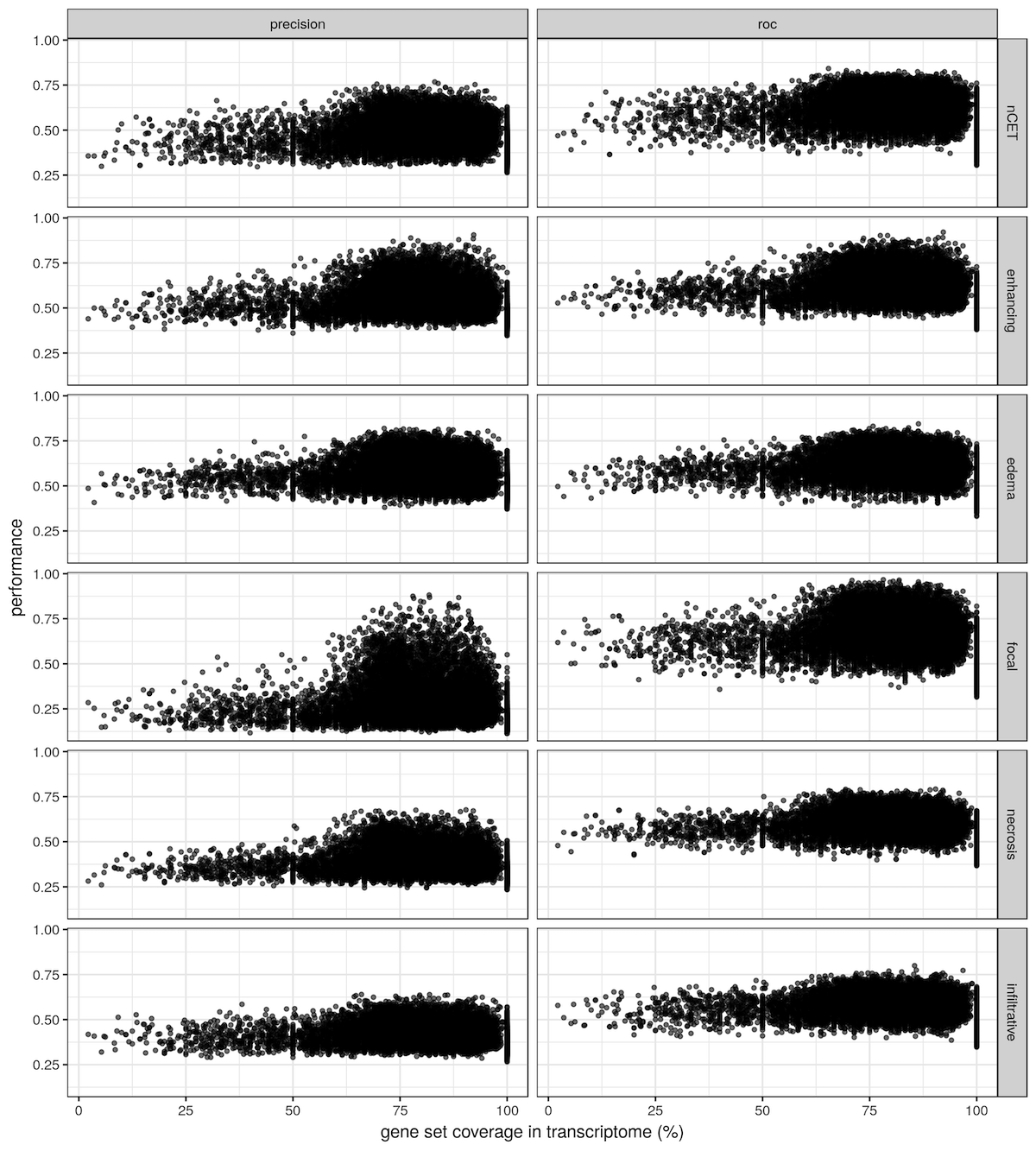}
	\centering
\end{figure}
\begin{figure}[h!]
	\caption{The trend between gene masking performance and gene set size broken down by gene set category: 
	(\textbf{top}) average precision and
	(\textbf{bottom}) AUC.}
	\label{supfig:vasari_gs_size}
	\includegraphics[width=\textwidth]{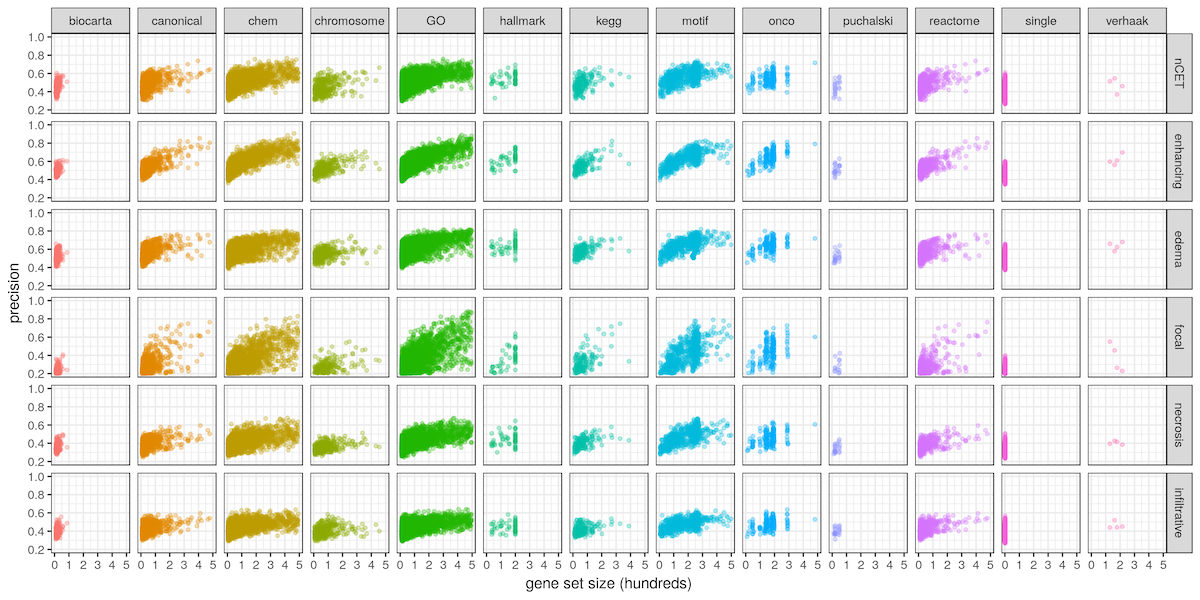}
	\includegraphics[width=\textwidth]{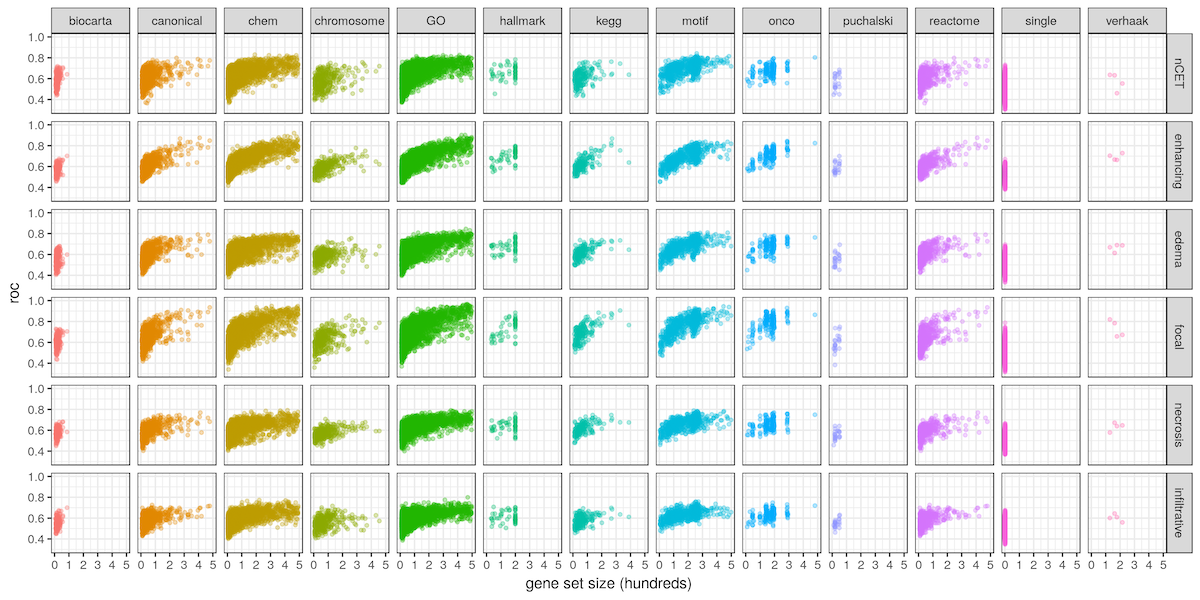}
	\centering
\end{figure}

\begin{figure}[h!]
	\caption{The trend between gene masking performance and gene set coverage broken down by gene set category: (\textbf{top}) average precision and
		(\textbf{bottom}) AUC.}
	\label{supfig:vasari_gs_cover}
	\includegraphics[width=\textwidth]{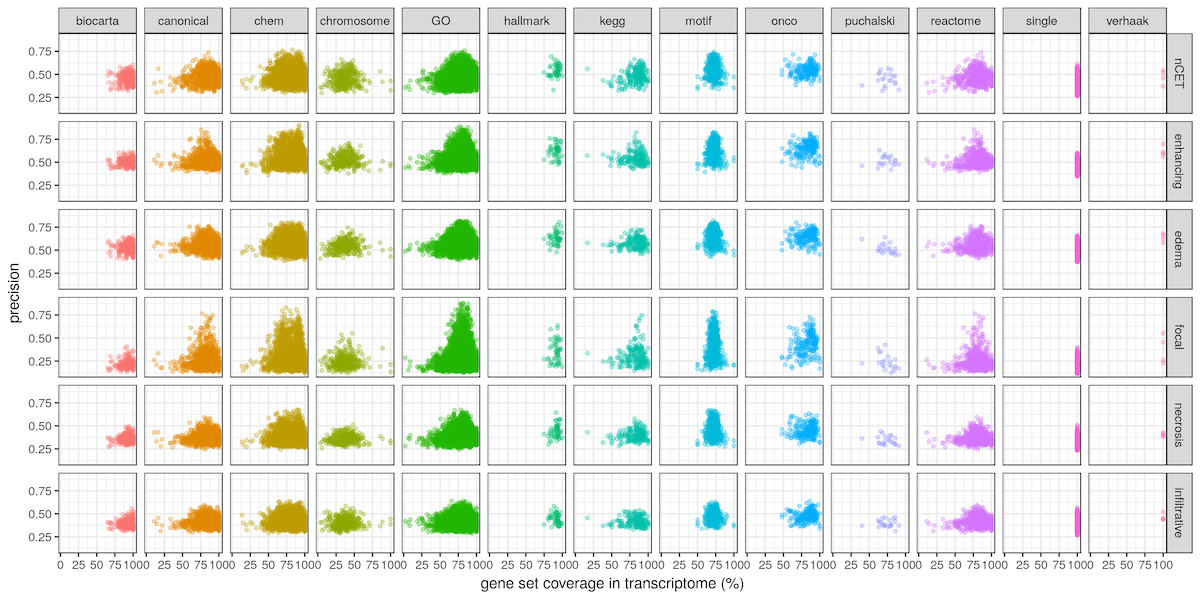}
	\includegraphics[width=\textwidth]{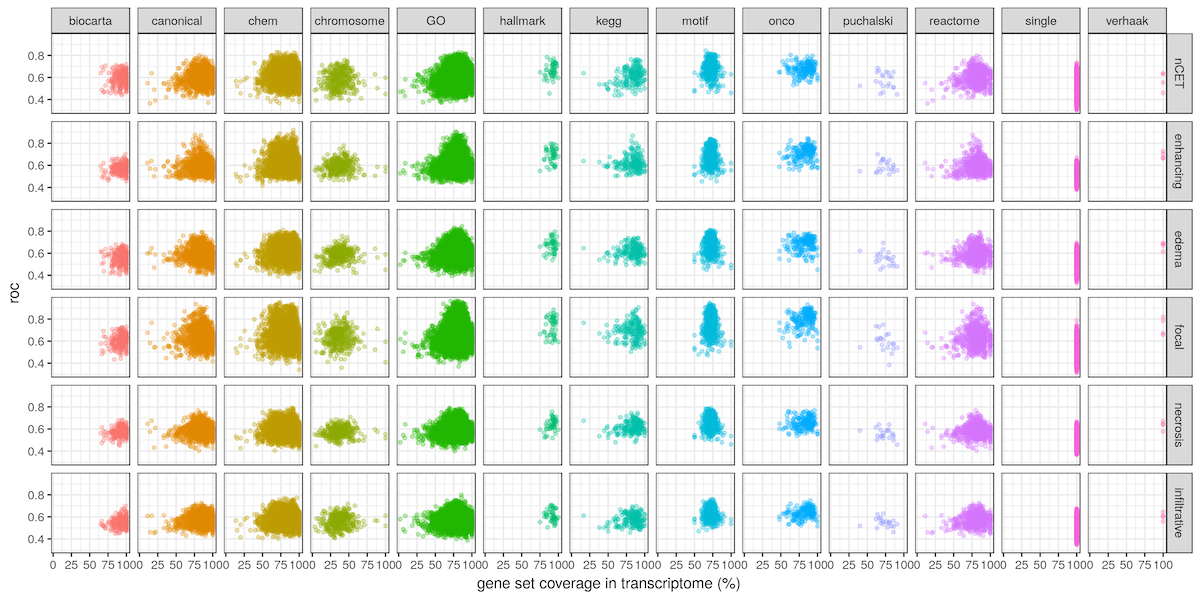}
	\centering
\end{figure}

\clearpage
\section*{Gene saliency}
\subsection*{Subtype neural network}
\begin{figure}[h!]
	\caption{Gene saliency in the subtype neural network. 
		(\textbf{a}) Clustering of the enrichment scores between each patients (columns) and each subtype gene set (rows). Enrichments were significant at an adjusted p-value $<$ 0.05. 
		(\textbf{b}) The number of patients with significant enrichment in each subtype gene set based on the bottom heatmap in (a) and broken down by their true subtypes.
		Of the patients with the mesenchymal subtype, more patients (33 patients) were enriched with mesenchymal genes than any other subtype genes. To a lesser degree, the same was seen among patients with the classical subtype, where 32 patients were enriched with classical genes. In contrast, 49 patients in the proneural subtype were enriched with classical genes while 48 were enriched with proneural genes. These aforementioned associations between gene saliency and subtype overlap with the associations from single gene masking in Fig. \ref{fig:subtype_single}b. As was seen in gene masking, genes from other subtypes were influential in predicting any individual subtype. In particular, neural patients' salient genes were enriched more often with mesenchymal genes (22 patients) than with neural genes (14 patients). Likewise, proneural patients were similarly enriched with both classical and proneural genes.}
	\label{supfig:subtype_subtype_saliency}
	\centering
	\includegraphics[width=\textwidth]{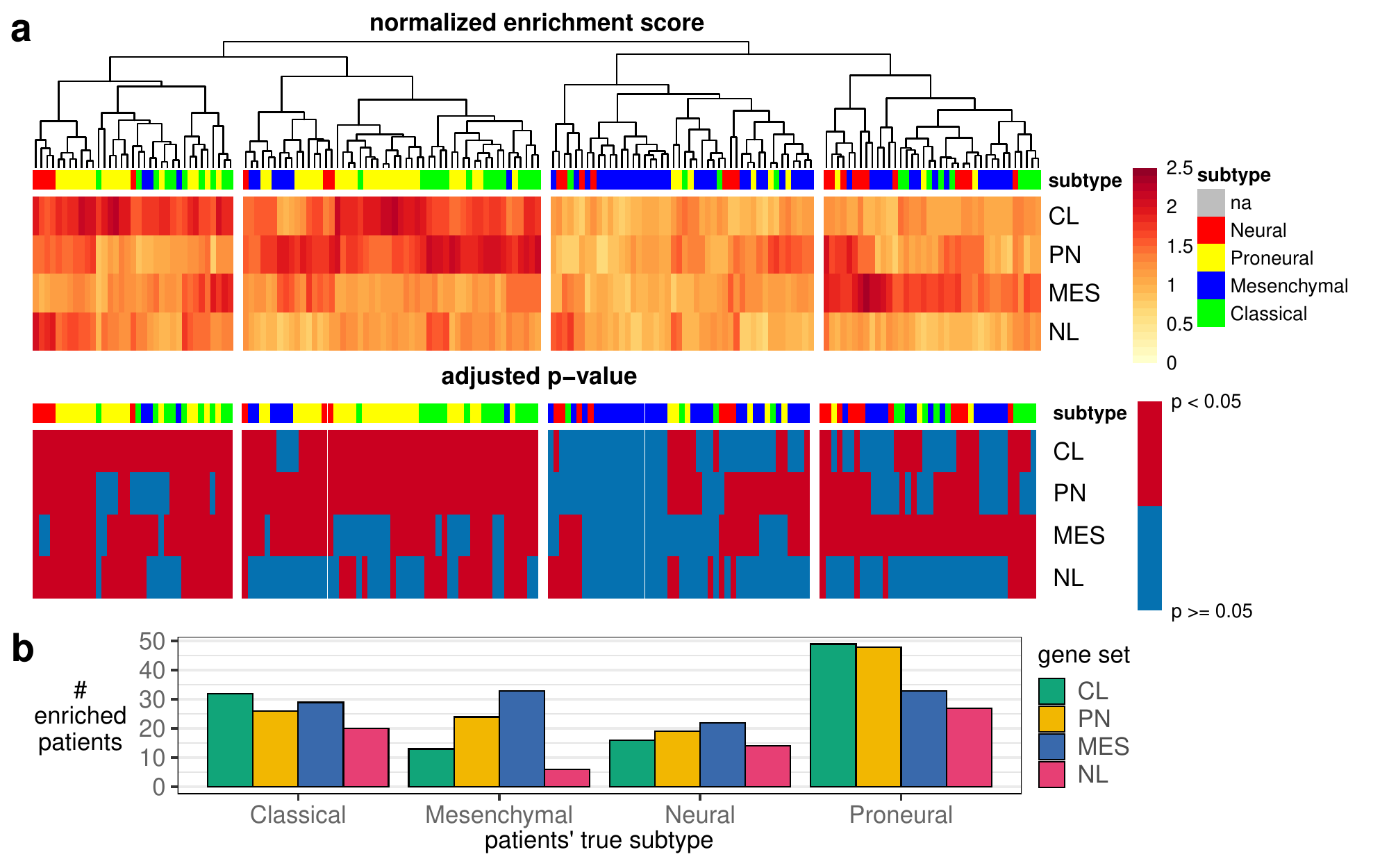}
\end{figure}

\clearpage
\subsubsection*{Radiogenomic neural networks}
\begin{figure}[h!]
	\caption{Patients with significant enrichment radiogenomic models based on gene saliency. Shown are other MSigDB gene set collections with at least 10 enriched patients. The chemical and genetic perturbations and computational collections were thresholded at 20 patients; all others were thresholded at 10 patients.}
	\label{supfig:vasari_barcounts}
	\includegraphics[width=\textwidth]{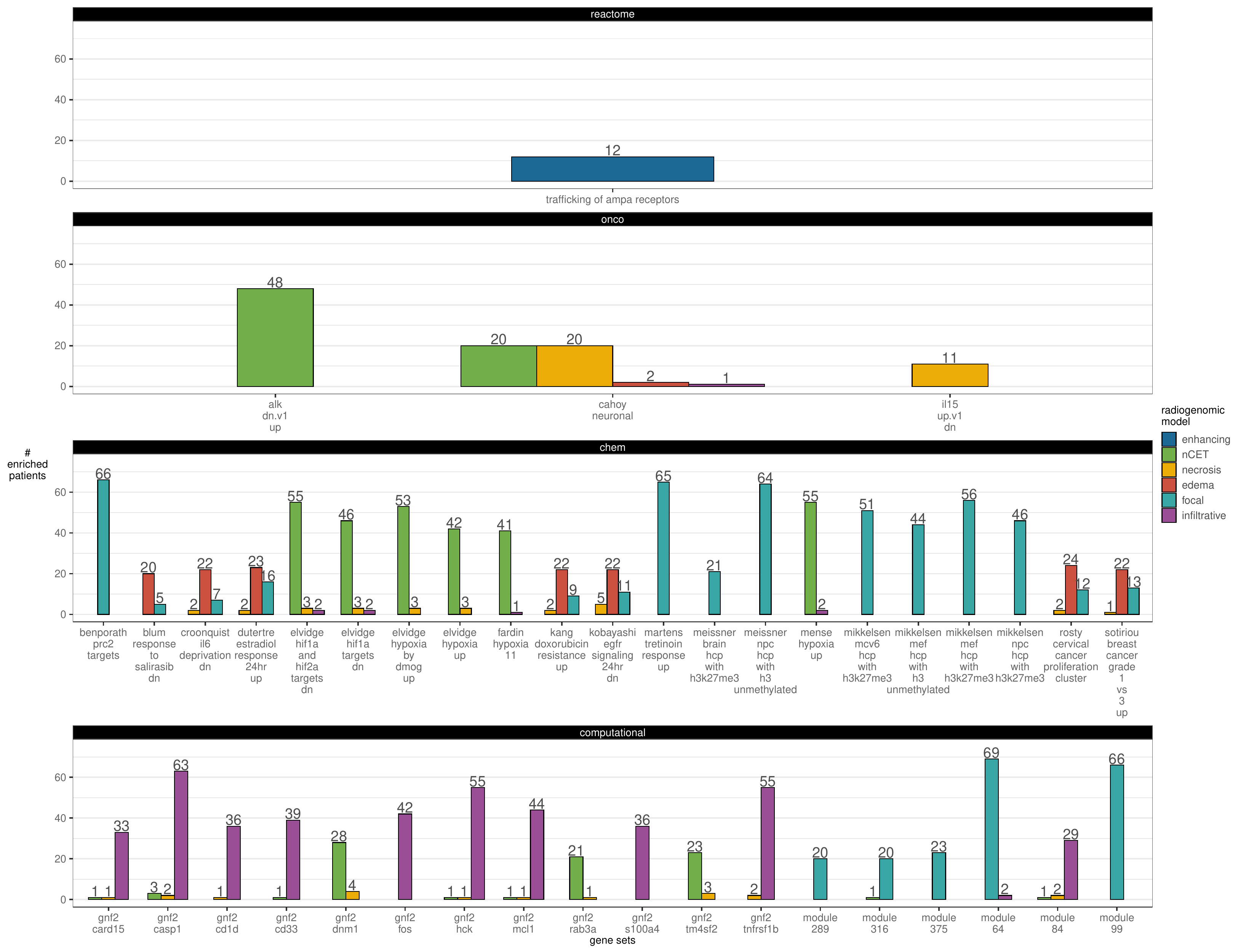}
	\centering
\end{figure}

\clearpage
\section*{Survival}

Ethnicity was removed since 98\% of patients with radiogenomic data were non-Hispanic.

Overall survival (OS) was defined in 'patient' file. For patients without death events, days-to-last-followup were obtained by finding the maximum day among any reported days-to-event in the `patient', `radiation', `drug', and `followup' files.

Progression-free survival (PFS) outcomes were defined by the `followup' file, which included the event types: locoregional disease, metastatic, progression of disease, or recurrence.
Unknown event types with days-to-event data were also removed unless other files stated patients had therapy regimes treated for progression, e.g., an unknown progression event type on day 554 and received radiation on day 576 to treat progression. For these cases (n=3), days-to-event data was kept, while event types were set to progression. 

\begin{figure}[h!]
	\caption{Survival estimates between TCGA-GBM (patients with transcriptome data),  VASARI subset (patients with at least one MRI trait), and radiogenomic subset (patients with all MRI traits) in 
		(\textbf{a}) overall survival (OS) and 
		(\textbf{b}) progression-free survival (PFS).  There were no differences in OS or PFS, see Supp. Table \ref{suptab_cohorts_surv}. Note: 27 of the overall cohort of 528 patients had missing survival and/or progression data, see Supp. Table \ref{tab:demo}. Similarly, 166 of 175 patient with any MRI trait also had survival information. Of the 166 patients, only 127 had all six MRI traits.}
	\label{supfig_cohort_sv_km}
	\includegraphics[width=\textwidth]{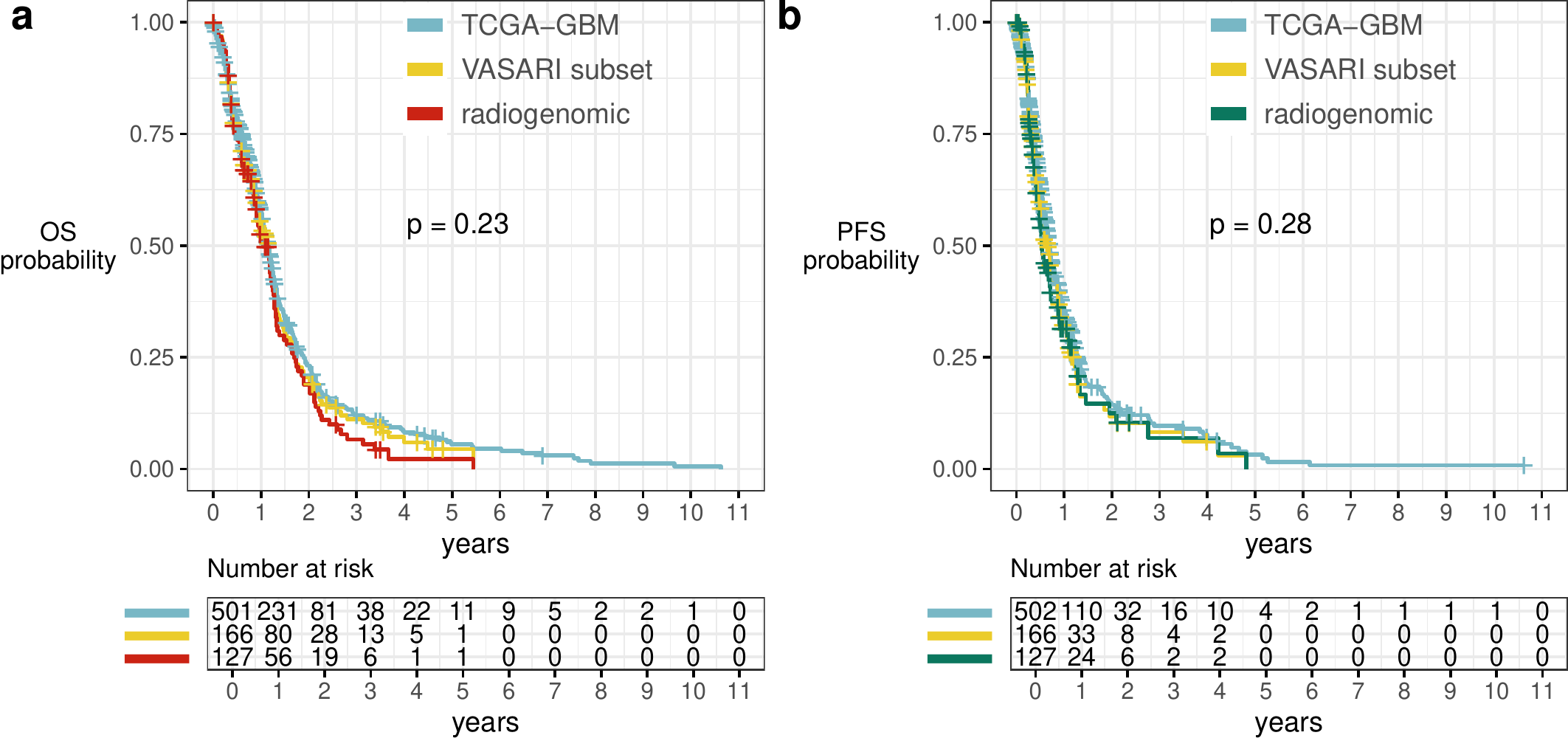}
	\centering
\end{figure}

\begin{figure}[h!]
	\caption{Survival estimates between patients split by their MRI traits. (\textbf{a}) Overall survival (OS) and (\textbf{b}) progression-free survival (PFS) differences.}
	\label{supfig_img_split_km}
	\includegraphics[width=.62\textwidth]{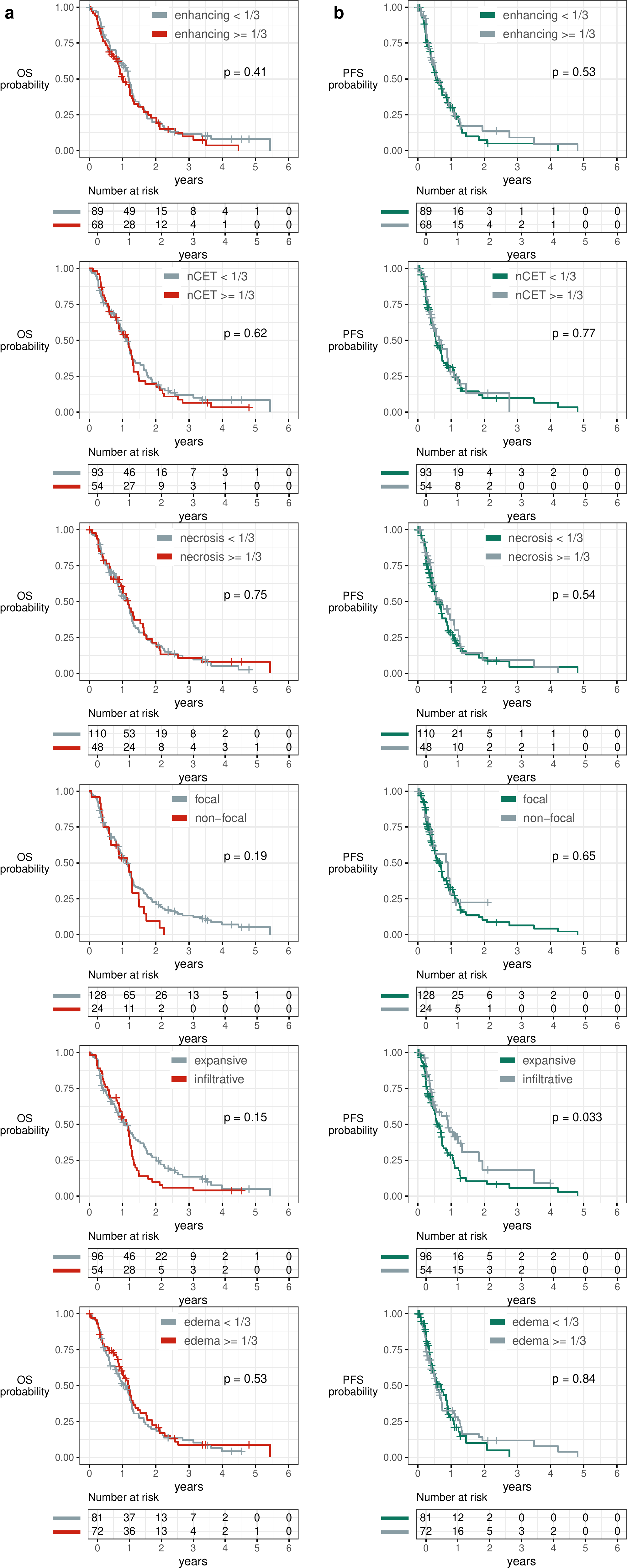} 
	\centering
\end{figure}

{\setlength{\tabcolsep}{6pt}
	\begin{table} [h!]
		\small
		\centering
		\caption{Summary measures of survival curves corresponding to cohorts in Supp. Figure \ref{supfig_cohort_sv_km}. }
		\label{suptab_cohorts_surv}	
		\begin{tabular}{lll lll lll lll l} 
			\toprule
		 \textbf{cohort} & \textbf{n} & \textbf{events} & \textbf{median (years)} & \textbf{95\% CI (years)} \\
			\midrule
			Overall survival \\
			\midrule
			TCGA-GBM 		 & 501  &  388 & 1.16 & 1.05, 1.24 \\
			VASARI subset  & 166 & 138 & 1.16 & 0.94, 1.26 \\
			radiogenomic   & 127 & 107 &  1.05 &    0.90, 1.27 \\
			 
			\midrule
			Progression-free survival \\
			\midrule
			TCGA-GBM   & 502    & 331   & 0.70   &  0.64, 0.77 \\
			VASARI subset    & 166    &  115   & 0.66   & 0.52, 0.85 \\
			radiogenomic    & 127     & 88   & 0.54     &  0.46, 0.73 \\
			
			\bottomrule	
		\end{tabular}
	\end{table}
}

\begin{figure}[h!]
	\caption{Frequency of enriched patients in each of the 54 radiogenomic traits. n = number of enriched patients}
	\label{supfig_rg_traits}
	\includegraphics[width=\textwidth]{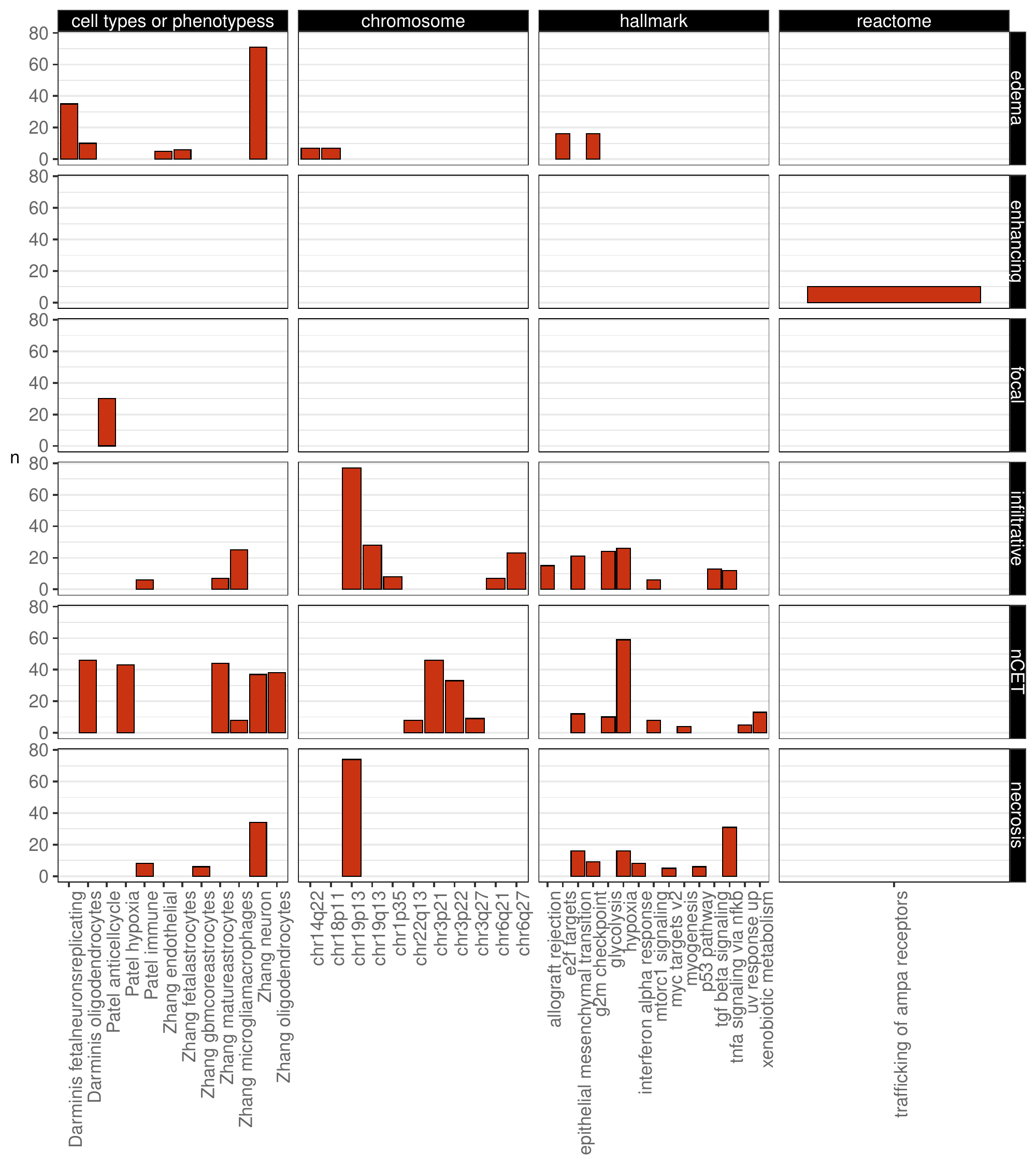}
	\centering
\end{figure}
\end{document}